\documentclass[11pt,a4paper]{article}
\pdfoutput=1
\usepackage[pdftex]{graphics}
\usepackage{jheppub}
\usepackage{amsmath,amssymb,amsfonts}
\usepackage{enumitem}
\usepackage{multirow}
\usepackage{array,booktabs}
\usepackage{tikz}
\usepackage{slashed}
\usepackage{verbatim}
\usepackage{float}
\usepackage{graphicx}
\usepackage{subcaption}
\usepackage{empheq}
\usepackage{cancel}

\newcommand{\eq}{\begin{equation}}
\newcommand{\eqe}{\end{equation}}
\newcommand{\eqa}{\begin{eqnarray}}
\newcommand{\eqae}{\end{eqnarray}}

\newcommand{\bn}{\begin{enumerate}}
\newcommand{\en}{\end{enumerate}}
\newcommand{\bl}{\begin{align}}
\newcommand{\el}{\end{align}}
\def\ie{\begin{equation}\begin{aligned}}
\def\fe{\end{aligned}\end{equation}}

\def\n{\nu}

\def\grad{\nabla}

\def\jmath{{j}}
\def\bl#1\el{\begin{align} #1 \end{align}}
\def\bg#1\eg{\begin{gather} #1 \end{gather}}

\def\bld#1\eld{\begin{aligned} #1 \end{aligned}}
\def\bgd#1\egd{\begin{gathered} #1 \end{gathered}}

\usetikzlibrary{snakes}
\usetikzlibrary{shapes.misc}
\usetikzlibrary{decorations.markings}

\tikzset{line/.style={line width=0.25mm},
curve/.style={line,smooth,tension=1},
->-/.style={decoration={
  markings,
  mark=at position #1 with {\arrow[>=stealth]{>}}},postaction={decorate}},
-<-/.style={decoration={
  markings,
  mark=at position #1 with {\arrow[>=stealth]{<}}},postaction={decorate}},
}

\title{De-projecting the EFThedron}
\author[1]{Li-Yuan Chiang}
\author[1,2]{Yu-tin Huang}
\author[1]{Laurentiu Rodina}
\author[1]{He-Chen Weng} 

\affiliation[1]{Department of Physics and Center for Theoretical Physics, National Taiwan University, Taipei 10617, Taiwan}
\affiliation[2]{Physics Division, National Center for Theoretical Sciences, Taipei 10617, Taiwan}
 
\emailAdd{yutinyt@gmail.com}

\abstract{ The space of Wilson coefficients of EFT that can be UV completed into consistent theories was recently shown to be described analytically by a positive geometry, termed the EFThedron.  However, this geometry, as well as complementary numerical methods of semi-definite programming, have so far focused on the positivity of the partial wave expansion, which allows bounding only ratios of couplings. In this paper we describe how the unitarity upper bound of the partial waves can be incorporated. This new problem can be formulated in terms of the well known $L$-moment problem, which we generalize and solve from a geometrical perspective. We find the non-projective generalization of the EFThedron has an infinite number of non-linear facets, which in some cases have remarkably simple descriptions. We use these results to derive bounds on single couplings, finding that the leading derivative operators are bounded by unity, when normalized by the cut-off scale and loop factors. For general operators of mass dimension $2k$ we find the upper bound is heavily suppressed at large $k$, with an $1/k$ fall-off.  
}

\begin{document}
\maketitle

\section{Introduction}
One of the original motivations for the modern revival of the conformal bootstrap (see \cite{Rychkov:2016iqz,Simmons-Duffin:2016gjk,Poland:2018epd,Poland:2022qrs} for reviews) was the question whether there exists an upper bound on the interaction strength of a relativistic quantum field theory~\cite{Caracciolo:2009bx}. An effective field theory version of this question can be framed as follows: given a dimension $2k{+}4$ operator, can we show that its corresponding Wilson coefficient $g$ is bounded by (naive dimensional analysis of~\cite{Manohar:1983md})
\eq
\frac{g}{(4\pi)^2}\leq \frac{\mathcal{O}(1)}{M^{2k}}\,,
\eqe
where the coefficient is normalized by a standard loop factor and $M$ is the cut-off scale for UV physics. Note that in principle this statement is a tautology since any EFT coupling saturating the $\mathcal{O}(1)$ bound at the cut-off will already be strongly coupled below the cut-off scale, and not a valid description. However, a direct derivation of the bound from axioms of quantum field theory can be viewed as a proof that dimensional analysis of EFT couplings is a theorem. Furthermore, as the couplings can be equally expressed as dispersive integrals linear in the S-matrix, such upper bounds are then applicable for observables whose low energy limit reduces to the couplings~\cite{Caron-Huot:2020cmc}.

The fact that higher dimension operators are constrained by their dispersive representation was known long ago as ``positivity bounds" of leading higher dimension operators~\cite{Adams:2006sv} (see~\cite{Pham:1985cr, Ananthanarayan:1994hf}), where the forward limit dispersion relations related the Wilson coefficient to a definite positive integral. In recent years, progress has been made in extending beyond the forward limit, including to scattering of particles with spin or mass~\cite{Bellazzini:2014waa,Bellazzini:2016xrt,deRham:2017avq,deRham:2017zjm,deRham:2018qqo,Bellazzini:2019bzh,Alberte:2019zhd,Bellazzini:2020cot,Tolley:2020gtv,Caron-Huot:2020cmc,Arkani-Hamed:2020blm,Green:2019tpt,Huang:2020nqy,Sinha:2020win,Wang:2020xlt,Trott:2020ebl,Wang:2020jxr,Hebbar:2020ukp,EliasMiro:2021nul,Alberte:2021dnj,Chowdhury:2021ynh,Raman:2021pkf,Haldar:2021rri,Zahed:2021fkp,Du:2021byy,Li:2021lpe,Zhang:2021eeo,Albert:2022oes,Bern:2021ppb,Henriksson:2021ymi,Davighi:2021osh,Melville:2022ykg}.  Importantly, the linchpin of gravitational EFTs, the graviton pole reflecting its long-range nature, was successfully incorporated through impact parameter space dispersion relations~\cite{Caron-Huot:2021rmr, Caron-Huot:2022ugt, Henriksson:2022oeu,Haring:2022cyf}. See also \cite{Kruczenski:2022lot,deRham:2022hpx} for recent reviews.

To set up the problem we will be exploring in this paper, let us begin with the partial wave representation of the four-scalar amplitude (see \cite{Correia:2020xtr} for a detailed discussion)
\eq
M(s,t)=s^{\frac{D{-}4}{2}}\sum_{\ell=even} n^{D}_{\ell} f_\ell(s) P^{D}_{\ell} \left(1+\frac{2t}{s}\right)\,,
\eqe
where $P^{D}_{\ell}$ are the $D$-dimensional Gegenbauer polynomial and
\eq\label{nfactor}
n^d_{\ell}=\frac{(4\pi)^{\frac{D}{2}}(d{+}2\ell{-}3)\Gamma[d{+}\ell{-}3]}{\pi \Gamma[\frac{d{-}2}{2}]\Gamma[\ell{+}1]}\,.
\eqe
At low energy, where the theory admits an EFT description, the amplitude is given by a polynomial in $s$ and $t$, 
\eq
M^{\rm IR}(s,t)=\sum_{k,q} \; g_{k,q}\,s^{k{-}q}t^q\,.
\eqe
These Taylor coefficients are linearly related to the Wilson coefficients of the on-shell action, with $2k$ counting the total number of derivatives. The coefficients can be analytically defined from the full amplitude via
\eq
g_{k,q}\equiv \frac{1}{2\pi i(q!)}\partial^q_t\oint_{s=0}\frac{ds}{s^{k{-}q{+}1}} M(s,t)
\eqe
This can be viewed as an on-shell definition of the EFT coefficients. Here we are neglecting the effects of massless loops, which can be incorporated by defining the coefficients on ``arcs" on the $s$-plane~\cite{Bellazzini:2020cot, Bellazzini:2021oaj}. Through the dispersion relation, these EFT couplings are related the imaginary part of the amplitude:
\eqa\label{Dispers}
g_{k,q}&=&\frac{1}{2\pi (q!)}\partial^q_t\left(\int_{M^2}^\infty \frac{ds'}{s'^{{k{-}q{+}1}}}{+}\int_{-M^2{-}t}^\infty \frac{ds'}{s'^{{k{-}q{+}1}}}\right) {\rm Im}[M(s',t)]\nonumber\\
&=&\frac{1}{2\pi}\sum_{\ell}\;n^{D}_{\ell} \lambda^{D}_{\ell,k,q} \,\int_{M^2}^\infty \frac{ds'}{s'^{{k{+}1}}} \; \rho_\ell(s') \;, 
\eqae
where unitarity bounds the spectral parameters to $0\leq\rho_{\ell}(s)\equiv {\rm Im}[f_\ell(s)]\leq2$ and the coefficients $\lambda^{D}_{\ell,k, q}$ receive contributions from both $s$- and $u$-channel thresholds and are given by:
\eq\label{lambdadef}
\lambda^{D}_{\ell,k, q}\equiv \frac{1}{q!}\partial^q_{t}\left[ \left(1{-}\frac{1}{(k{-}q)!}\partial^{k{-}q}_{s} \frac{1}{s{+}t{+}1}\right)P^{D}_{\ell} (1{+}2t)\right]\bigg|_{s,t{=}0}\,.
\eqe
Note that $\lambda_{\ell,k,q}$ is always a function of the spin Casimir $\mathcal{J}^2=\ell(\ell{+}D{-}3)$. In the remainder of the paper we will consider only $D=4$. Bounds in other dimensions can be just as easily derived. We next set the cut-off $M^2=1$ and define $z=1/s'$  to write
 \eq\label{ref1}
g_{k,q}=\sum_{\ell} v_{\ell,k,q}  \int_0^1 \rho_\ell(z) z^{k{-}1}  dz, \quad 0\le \rho_\ell(z)\le2\,,
\eqe
where $v_{\ell,k,q}=16(2\ell{+}1)\lambda_{\ell,k,q}$. This form of the dispersion relation will be the central object of our study. Finally, extra constraints on the couplings can be derived by requiring the dispersion relation to be consistent with permutation invariance. This translates to null constraints, expressing the fact that various couplings at equal mass dimension must be linearly related.

So far most of the focus in the literature has been on projective bounds, where one only bounds ratios of $g_{k,q}$. This is a reflection that when using either numerical semi-definite programming (SDP)~\cite{Simmons-Duffin:2015qma, Landry:2019qug}, pioneered for the EFT bootstrap in~\cite{Caron-Huot:2020cmc}, or analytical convex geometry (EFThedron)~\cite{Arkani-Hamed:2020blm,Chiang:2021ziz}, one only uses $0\leq\rho_{\ell}(s)$. To obtain non-projective bounds, imposing the upper bound on $\rho_{\ell}(s)$ is required. In~\cite{Caron-Huot:2020cmc}, the $\rho_{\ell}(s)\leq2$ was implemented for $g_{2,0}$ with a single null constraint at $k=4$.\footnote{Non-perturbative unitarity bounds have also been implemented in the gravitational EFT context using an ansatz approach~\cite{Guerrieri:2021ivu}, where the EFT is scaled with respect to $M_{pl}$, and thus those not assume a perturbative scale $M$. 
}

We will study this problem from a geometrical perspective.\footnote{Some of the results presented here have been implemented by the authors in the gravitation context in~\cite{Chiang:2022jep}, where a numerical approach using linear programming was also introduced. } Ignoring the upper bound on $\rho(z)$ first, the space of couplings can be viewed as a projective positive geometry, defined by the ``moment problem"~\cite{Bellazzini:2020cot, Chiang:2021ziz, Bellazzini:2021oaj}. In its most basic form, the moment problem asks what sequences of numbers $a_1,a_2,a_3,\ldots$ can be expressed as 
\eq\label{eq11}
a_k=\int_{-\infty}^\infty \rho(z)z^{k-1} dz, \quad \rho(z)\ge 0\,.
\eqe
The solution is that the Hankel matrices 
\eq\label{eq13}
H=\begin{pmatrix}a_1& a_2 &a_3&\\
a_2& a_3 &a_4&\cdots\\
a_3& a_4 &a_5&\\
&\vdots&&\ddots
\end{pmatrix}\,,
\eqe
must be positive semi-definite. Different integration domains for $z$ will translate to positivity of various generalizations of the Hankel matrix. The dispersion relation is a more general form of the moment problem, asking what sequences of numbers $a_{k_1,q_1},a_{k_2,q_2},\ldots$ satisfy
\eq\label{dbmom}
a_{k,q}=\sum_\ell \lambda_{\ell,k,q}\int_{0}^1 \rho_\ell(z)z^{k-1} dz, \quad \rho(z)\ge 0\,.
\eqe
In \cite{Arkani-Hamed:2020blm} this problem was treated as a product geometry between a single moment problem and a cyclic polytope, and was termed the EFThedron, adding yet another facet to our understanding of scattering amplitudes through geometry \cite{Herrmann:2022nkh}. In \cite{Chiang:2021ziz} the complete solution to this problem was formulated by treating it as a double moment problem, whose solution involves a generalization of the Hankel matrices, called moment matrices, together with positivity constraints enforcing $\ell\in \mathbb{N}$. 
One can alternatively view eq.(\ref{dbmom}) not as a double moment, but as a Minkowski sum of individual single moments, one for each spin. Since the boundaries of single moments are usually much easier to establish, the problem is reduced to computing the Minkowski sum of individually simple geometrical objects. This is the approach we will take in this paper.

\begin{figure}[H]
     \centering
         \centering
 \includegraphics[width=0.5\textwidth]{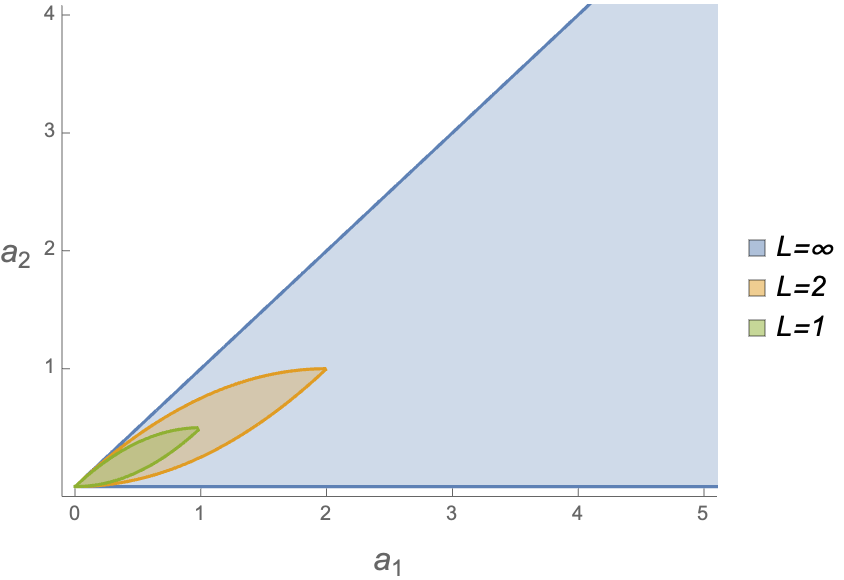} 
 \caption{The allowed space for single $L$-moments, for $L=1,2,\infty$. For $L=\infty$ only the projective bounds $0\le a_2/a_1\le 1$ are left.}
\label{fig:plotex2i}
 \end{figure}

We will introduce a systematic way of implementing the unitarity upper bound in the dispersion relation, revealing a fascinating geometric structure, of which the EFThedron was only a limiting case. At the core of this new geometry is the bounded version of eq.(\ref{eq11}), where now we also have the constraint $\rho(z)\le L$. This generalization is known as the $L$ moment problem, and was first solved a century ago \cite{Hausdorff:1923uf}, with generalizations and new approaches developed since (see eg. \cite{Kren1977TheMM,PUTINAR1990288,article}). The solution can be expressed in terms of more general Hankel-type matrices, but the boundaries themselves are parameterized by simple functions. For example, in 2D, for a space of moments $(a_{k_1},a_{k_2})$, the allowed region is bounded by two curves
\eqa\label{inteq1}
\nonumber  \textrm{lower bdy: }(a_{k_1},a_{k_2})&=&\left(L\frac{m^{k_1}}{k_1},L\frac{m^{k_2}}{k_2}\right)\,,\\
\textrm{upper bdy: }(a_{k_1},a_{k_2})&=&\left(L\frac{1-m^{k_1}}{k_1},L\frac{1-m^{k_2}}{k_2}\right)\,,
\eqae
where $m$ is a parameter $\in[0,1]$. As we take $L\rightarrow \infty$ we are left only with projective boundaries. We sketch this in Figure \ref{fig:plotex2i}. 

\begin{figure}[H]
     \centering
     \begin{subfigure}[b]{0.44\textwidth}
         \centering
\includegraphics[width=\textwidth]{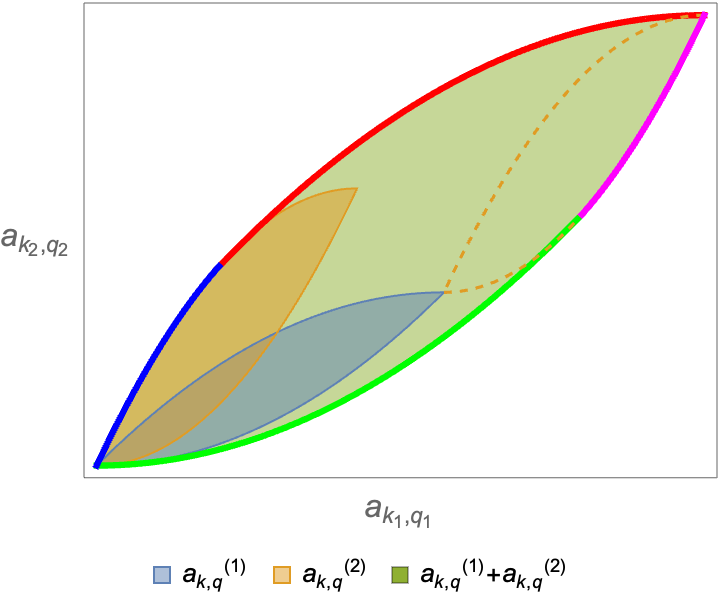}
\caption{Structure of the Minkowski sum of two single moments}
\label{fig:M000a}
\end{subfigure}
\hfill
  \begin{subfigure}[b]{0.44\textwidth}
         \centering
 \includegraphics[width=\textwidth]{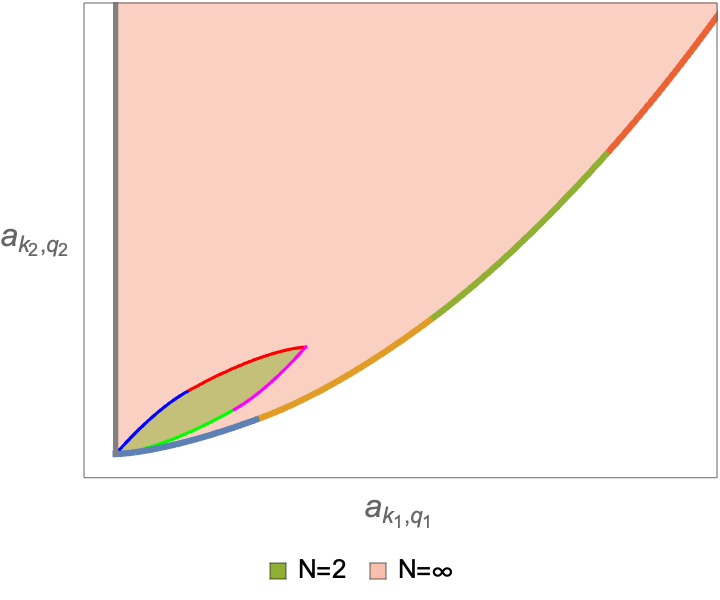}
 \caption{Minkowski sum for two and an infinite number of moments}
\label{fig:plot5500}
 \end{subfigure}
 \caption{Minkowski sums of single $L$-moments}
 \end{figure}

The next step is to compute the dispersion relation as the Minkowski sum of single $L$-moments, one for each spin and rescaled by the $\lambda_{\ell,k,q}$ factors. Since the boundary of a Minkowski sum $A+B$ is contained in the Minkowski sum of the boundaries of $A$ and $B$, we simply take a sum of boundary parametric curves such as eq.(\ref{inteq1}) and extremize with respect to the parameters.  For example, the Minkowski sum of two moments is shown in Figure \ref{fig:M000a}.
Remarkably, we find that the Minkowski sum of an infinite number of moments converges in many of our cases of interest, leading to the non-projective EFThedron. This geometric object has an infinite number of distinct facets, each resulting from the Minkowski sum of different subsets of spins. We sketch an example in Figure \ref{fig:plot5500}.

\begin{figure}
\centering
\begin{minipage}{.47\textwidth}
  \centering
  \includegraphics[width=\linewidth]{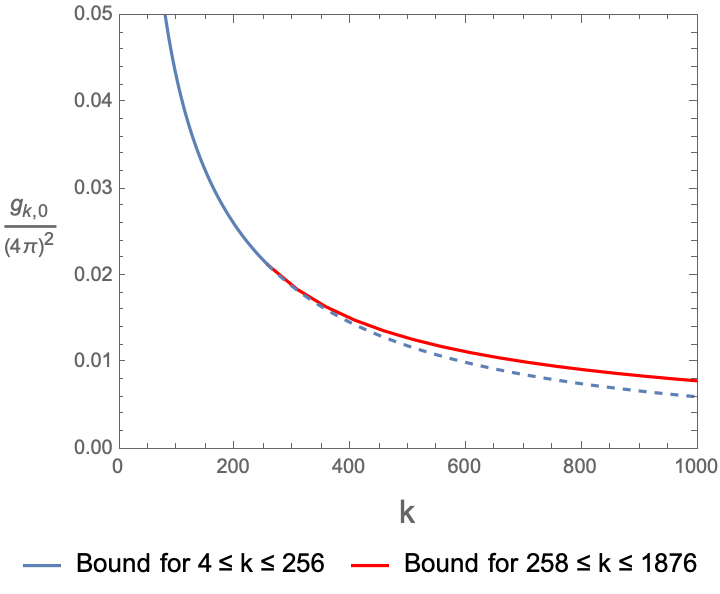}
  \captionof{figure}{Upper bound on $g_{k,0}$, as a function of $k$, with $k=4$ null constraint. Blue corresponds to eq.(\ref{genbound}), valid for $k\le256$, while red is valid for $258\le k\le 1876$; both exhibit $1/k$ dropoff.}
  \label{fig:test1a}
\end{minipage}%
\hfill
\begin{minipage}{.47\textwidth}
  \centering
  \includegraphics[width=0.98\linewidth]{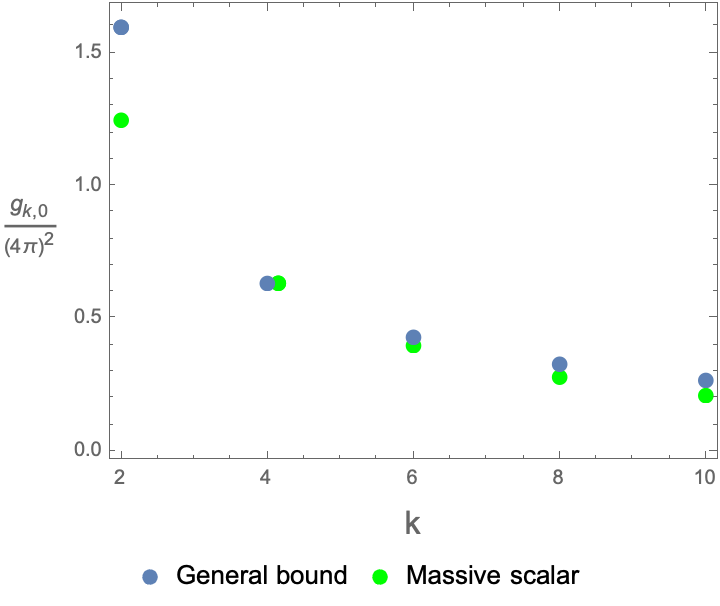}
  \captionof{figure}{Eq.(\ref{genbound}) compared to the massive scalar amplitude, after normalizing by matching at $g_{4,0}$; the scalar amplitude exhibits a stronger, approximately $1/k!$ dropoff.}
  \label{fig:test2a}
\end{minipage}
\end{figure}

\begin{figure}[H]
     \centering
\includegraphics[scale=0.77]{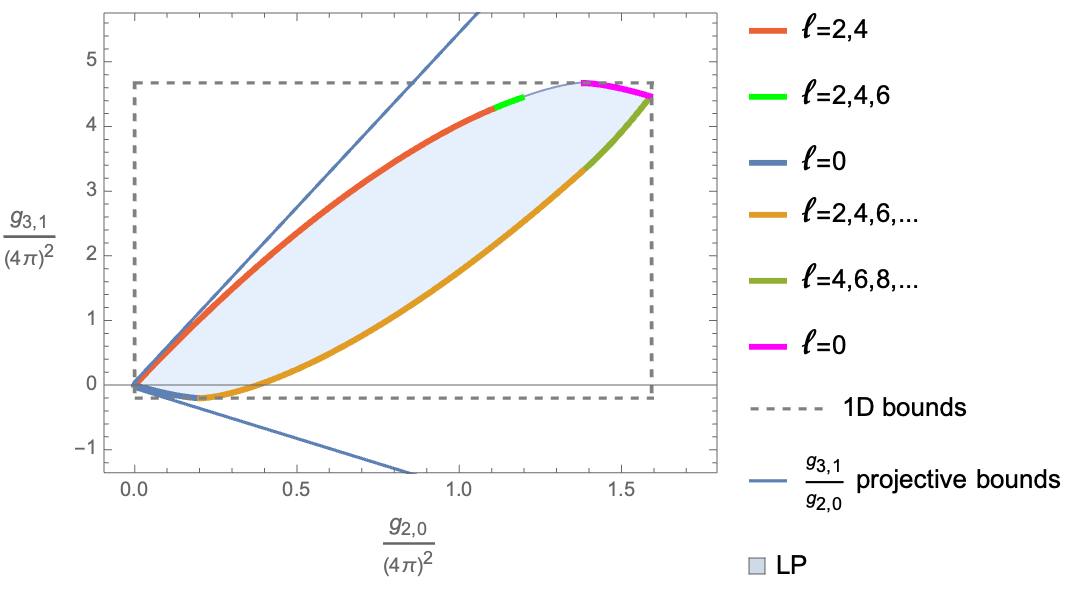}
 \caption{Allowed region for the space $(g_{2,0},g_{3,1})$ subjected to the null constraint at $k=4$. The various boundaries contain contributions from different subsets of spins. In contrast, the projective bounds only imply bounds on the ratio $g_{3,1}/g_{2,0}$.}
\label{fig:gz00i}
 \end{figure}

Using this analytic geometry, we can now derive bounds for individual couplings. For example, for $g_{k,0}$ with $k\in even$, imposing permutation invariance at eighth derivative order ($k=4$) we reproduce the result for $g_{2,0}$ in~\cite{Caron-Huot:2020cmc}, while for $4\leq k\leq 256$ we find
\eq\label{genbound}
 \frac{g_{k,0}}{(4\pi)^2}\le\frac{2}{k} \left(30-18\left(\frac{35}{36}\right)^\frac{k}{4}\right)\frac{1}{\pi ^2 M^{2k}}\,.
\eqe
with small corrections for higher $k$, illustrated in Figure \ref{fig:test1a}. While the leading derivative operators are bounded by $\mathcal{O}(1)$, subsequent operators are in fact suppressed as $1/k$ for $k\gg1$. Such results are stronger than the combination of non-projective bounds for a single operator and projective bounds for the remaining, which would only imply $g_{k,0}\le g_{k,2}$. In Figure \ref{fig:test2a} we compare this bound with the EFT coefficients arising from the UV amplitude of a massive scalar at one loop. 
The massive scalar satisfies the bound due to combinatorics from expanding the loop integrand at low energies. Similar to the projective geometry, in higher dimensional spaces we can derive analytic expressions for the boundaries, which we compare to numerical results from linear programming (LP), as well as previously known projective bounds. For example, with permutation invariance imposed at $k=4$, we obtain Figure \ref{fig:gz00i}.

The paper is organized as follows. In Section \ref{sec2} we find the non-projective polytope resulting from the Minkowski sum of finite segments. In addition to describing the space of couplings at equal mass dimension, we then show how this result can be used to solve the single $L$-moment problem. We present other aspects of the $L$ moment problem, including Hankel constraints, in Appendix \ref{aspectsL}. In Section \ref{sec3} we compute the Minkowski sum of single $L$ moments. This is carried out by extremizing the sum using Lagrange multipliers, but we also discuss an alternative solution in Appendix \ref{DiscreteM}. We solve this problem completely in 2D. For higher dimensions we present closed form solutions for particular cases.  In Appendix \ref{appgend} we discuss how this approach could be extended to more general cases. In Section \ref{sec4} we apply the results to obtain physical bounds on the scattering amplitude of four massless scalars. Although not directly applicable to our four scalar amplitude, in Appendix \ref{LSD} we show how the previous results can be used to impose the low spin dominance condition. We conclude in Section \ref{conc}.
\section{De-projection of convex hulls}\label{sec2}
As reviewed in the previous section, the dispersive representation of EFT coefficients generally takes the form
\eq\label{akq}
a_{k,q}=\sum_{\ell\in \mathbb{N}} v_{\ell,k,q} \int_0^1 \rho_\ell(z)z^{k-1} dz, \quad 0\le \rho_\ell(z)\le 2\,.
\eqe
When we neglect the upper bound on $\rho_\ell$, the space for allowed $a_{k,q}$ is projective and one can only bound the ratios of $a_{k,q}$. The corresponding projective space is characterized by the convex hull of product moments as described in ~\cite{Arkani-Hamed:2020blm,Chiang:2021ziz}.

To incorporate the effects of the upper bound, we will view the geometry as the Minkowski sum of convex hulls.  To illustrate this point, we will first consider a simplified version of eq.(\ref{akq}) where the coefficients $v_{\ell,k,q}$ are independent of $k$, and the vectors 
\eq
{\bf v}_{\ell}=\begin{pmatrix}
v_{\ell,q_1}\\
v_{\ell,q_2}\\
\vdots\\
v_{\ell,q_D}
\end{pmatrix}\,,
\eqe
 are cyclically ordered by $\ell$ in the sense that,
\eq\label{eq:cyclic}
\textrm{det}\begin{pmatrix}{\bf v}_{i_1}&{\bf v}_{i_2}&{\bf v}_{i_3}&\ldots\end{pmatrix}\equiv \langle i_1,i_2,i_3,\cdots i_D\rangle\ge 0, \quad \forall i_1<i_2<i_3\cdots <i_D\,.
\eqe
We begin with the space of couplings with equal $k$ (``same mass-dimension"), 
 \eq
 (a_{k,q_1}, a_{k,q_2}, \cdots)\,,
 \eqe 
whose projective geometry is simply that of a cyclic polytope. We will show that the ``de-projected" space can be constructed by taking the Minkowski sum of line segments, resulting in a convex polytope whose vertices are given by linear sums of successive ${\bf v}_{\ell}$s.

This result will allow us to derive the boundaries for a closely related geometry: the $L$-moment problem. This concerns the space of $a_k$ given by 
\eq
a_k=\int_0^1 \rho_\ell(z)z^{k-1} dz, \quad 0\le \rho_\ell(z)\le L\,.
\eqe
The limit $L\rightarrow \infty$ corresponds to the Hausdorff moment problem, whose solution is given by the positivity of Hankel matrices constructed from $a_k$. We will derive the solutions for finite $L$.

\subsection{Minkowski sums }\label{minksum}
\begin{figure}
\begin{center}
\includegraphics[scale=.38]{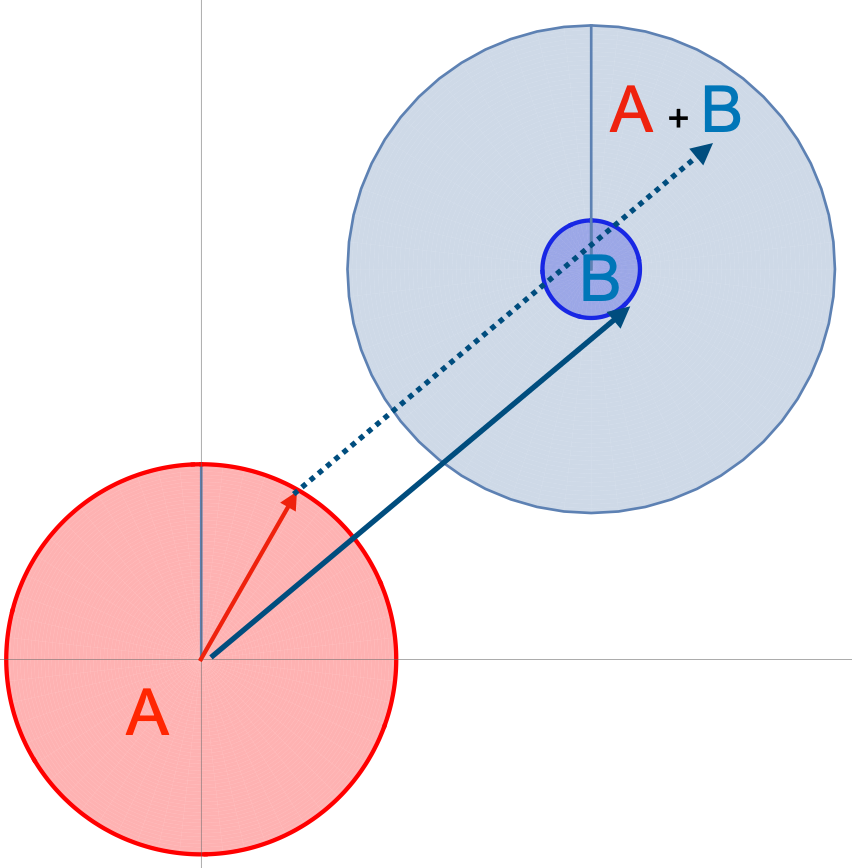}
\caption{The Minkowski sum of two disks $ A$ and $ B$. Note that the sum of two points on the boundary of individual region is not necessarily on the boundary of the Minkowski sum.}
\label{Minkfig1}
\end{center}
\end{figure}

In this section we review some basic notions related to Minkowski sums.
Let us consider two regions $A, B$ with each point inside the regions defined by a vector. The Minkowski sum of the two regions is then given by all possible pairs of vectors, one from $A$ and one from $B$, and taking their sum:
\eq
A{+}B=\{a{+}b,\forall a\in A, \forall b\in B\}\,.
\eqe
It can be understood as translating the region $B$ by all vectors $a \in A$, or vice versa. As an example consider $A$ and $B$ to be separate disks shown in Figure \ref{Minkfig1}. Their Minkowski sum is then displayed as the light blue region, which contains any vector in $B$ shifted by all vectors in $A$.

\begin{figure}
\begin{center}
\includegraphics[scale=.4]{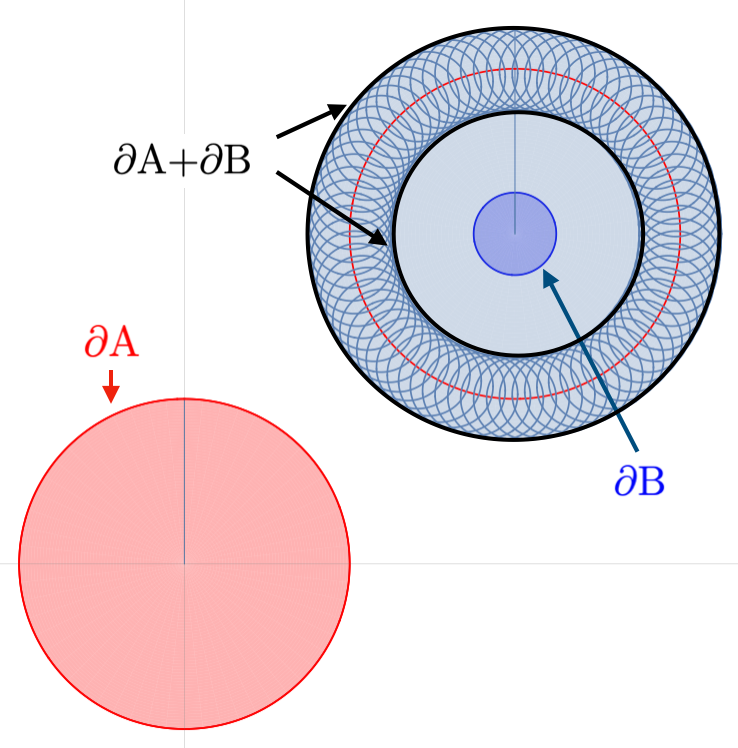}
\caption{The Minkowski sum of the boundary of the two disks $\partial A{+}\partial B$. We see that it gives a annulus with two boundaries, and only one is the true boundary of the Minkowski sum.}
\label{Minkfig2}
\end{center}
\end{figure}

We will be interested in characterizing this sum in terms of its boundaries $\partial (A{+}B)$, which are determined by the individual boundaries $\partial A$ and $\partial B$. More precisely, any vector on $\partial (A{+}B)$ must be comprised by the sum of a vector on $\partial A$ and a vector on $\partial B$. However the reverse is not necessarily true, as illustrated in the above figure, where a sum of boundary vectors sits inside the Minkowski sum. In fact, the boundary of the Minkowski sum $A{+}B$ is a \textit{subset} of boundaries for the Minkowski sum $\partial A{+}\partial B$. Indeed for our example in Figure \ref{Minkfig2}, the  Minkowski sum of $\partial A$ (denoted by the red contour) and $\partial B$ (denoted by the blue contour) covers a subregion of $A{+}B$ which contains two boundaries, with the outer one being $\partial (A{+}B)$\,. Thus when one is interested in the boundary of the Minkowski sum, only the boundaries of the individual sets are relevant. 

\begin{figure}
\begin{center}
\includegraphics[scale=.4]{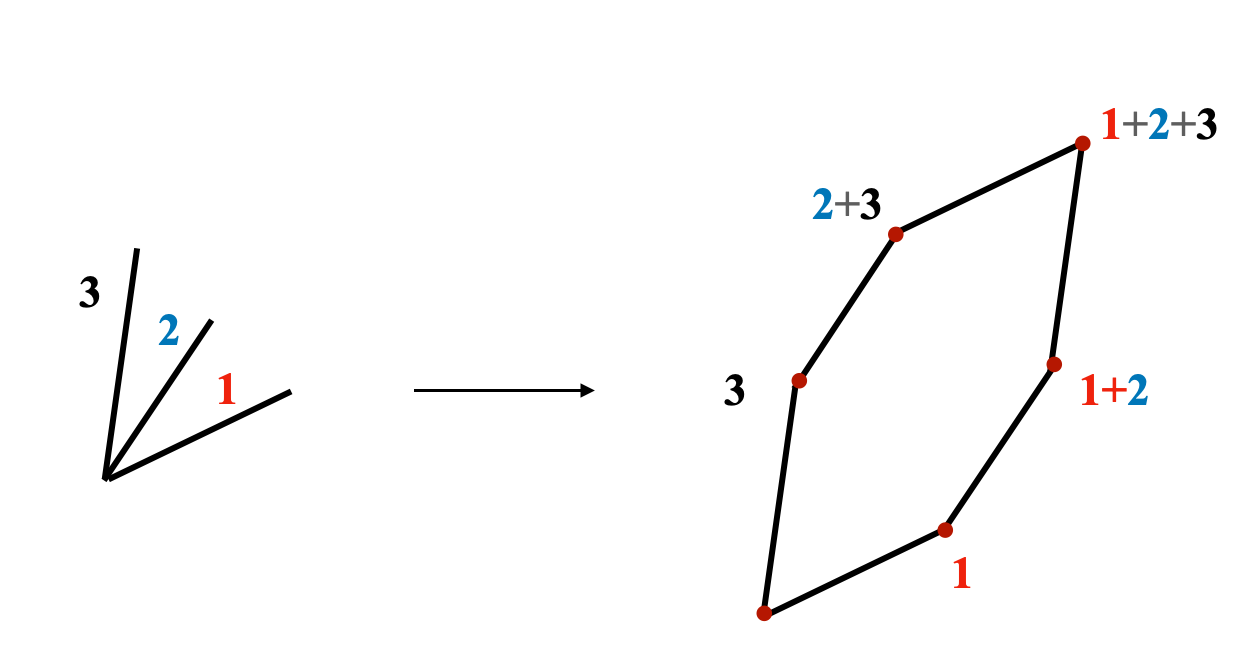}
\caption{The Minkowski sum of three segments labelled 1,2,3, ordered according to their slope.}
\label{PolySum}
\end{center}
\end{figure}

Let us take a simpler example which will be of immediate use. Consider the Minkowski sum of three segments labeled $1,2,3$ in Figure \ref{PolySum}.
The boundary of each segment is given by the end points. Thus following the discussion above, the boundary of the Minkowski sum is given by the Minkowski sum of the end points, i.e. three vectors. In this case, the vectors are labeled according to their slope, and thus the boundary can be sequentially constructed by starting with the vector with the smallest slope, and successively adding the next vector until all vectors are included. Then one starts subtracting the first vector and so forth until one is left with the vector of the largest slope. 

\subsection{De-projecting the cyclic polytope}\label{sec22}
Let us begin with 
\eq\label{Strawman}
a_{k,q}=\sum_{\ell} v_{\ell,q}\int_0^1 \rho(z)z^{k-1} dz=\sum_\ell  v_{\ell,q}a_{k}^{(\ell)}=\sum_\ell  a_{k,q}^{(\ell)}\,,
\eqe
where $\rho(z)$ is bounded as $0\leq \rho(z)\leq L$.  Note that each $a_{k}^{(\ell)}$ is a 1D segment, $0\le a_k^{(\ell)}\le \frac{L}{k}$. For later convenience we will parameterize the segments as
\eq
a_k^{(\ell)}=m_\ell \frac{L}{k} , \quad m_{\ell}\in[0,1]\,.
\eqe
The combination 
\eq
{\bf a}^{(\ell)}=
\begin{pmatrix}
a_{k,q_1}^{(\ell)}\\
a_{k,q_2}^{(\ell)}\\
\vdots\\
a_{k,q_D}^{(\ell)}
\end{pmatrix}=  m_\ell\frac{L}{k}\begin{pmatrix}
v_{\ell,q_1}\\
v_{\ell,q_2}\\
\vdots\\
v_{\ell,q_D}
\end{pmatrix}=  m_\ell {\bf v}_{\ell}\,,
\eqe
then represents a segment in $D$ dimensions, also parameterized by $m_{\ell}$. Eq.(\ref{Strawman}) is therefore
\eq\label{bod11}
{\bf a}\equiv \sum_\ell m_\ell {\bf v}_\ell
\eqe
that is, ${\bf a}$ corresponds to a Minkowski sum of segments, with endpoints $(0,0,\cdots,0)$ and ${\bf v}_\ell=\frac{L}{k}(v_{\ell,q_1}, v_{\ell,q_2}, \cdots, v_{\ell,q_D})$ in $D$ dimensions. As discussed previously, the boundary of the Minkowski sum ${\bf a}$ is found from the Minkowski sum of the boundaries of ${\bf a}^{(\ell)}$. Therefore the endpoints of the segments, represented by vectors ${\bf v}_\ell$ will be what is relevant to the problem.

\begin{figure}
   \centering
   \includegraphics[width=3in]{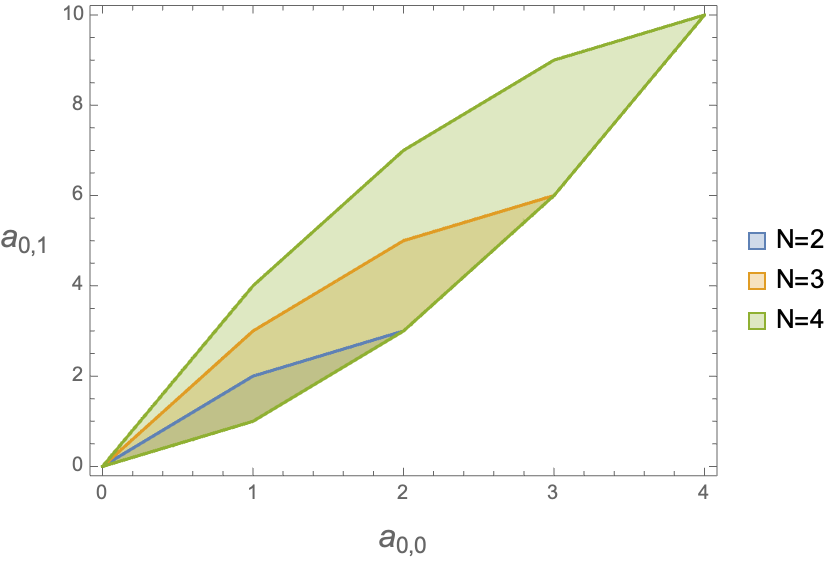} 
   \caption{2D polytope for $N=2,3,4$, with ${\bf v}_i=(1, i)$.}
   \label{ExamplePolytopes}
\end{figure}

\paragraph{Two dimensions}
Let us start in two dimensions, where the $N$ ordered vectors ${\bf v}_\ell$ satisfy
\eq\label{2DOrder}
\langle i_1, i_2\rangle >0, \quad \forall i_1< i_2\,.
\eqe 
As a consequence the relative angle of any two vectors is less than $\pi$, and they span at most a half space. From the example in the beginning of this section, it is straightforward to determine that the Minkowski sum is a polygon with $2N$ ordered vertices ${\bf V}_{i}=\sum_{j<i} {\bf v}_{j}$ and $\bar{\bf V}_{i}=\sum_{j> i} {\bf v}_{j}$, $1\le i\le N$. This region can be characterized by the inequalities
\eq\label{2Dbound}
{\rm det}\left(\begin{array}{ccc}1 & 1 & 1 \\ {\bf a} &{\bf V}_i &{\bf V}_{i+1}\end{array}\right),\quad {\rm det}\left(\begin{array}{ccc}1 & 1 & 1 \\ {\bf a} & \bar{{\bf V}}_i & \bar{{\bf V}}_{i+1}\end{array}\right)\geq0,\quad \, 1\leq i\leq N\,.
\eqe
where we have ${\bf V}_1=\bar{{\bf V}}_{N{+}1}={\bf 0}$. Note that the boundaries are non-linear in $ {\bf a}$, since the geometry is non-projective. As an example the Minkowski sum with $N=2,3,4$ vectors are displayed in Figure \ref{ExamplePolytopes}.

\begin{figure}
   \centering
   \includegraphics[width=4in]{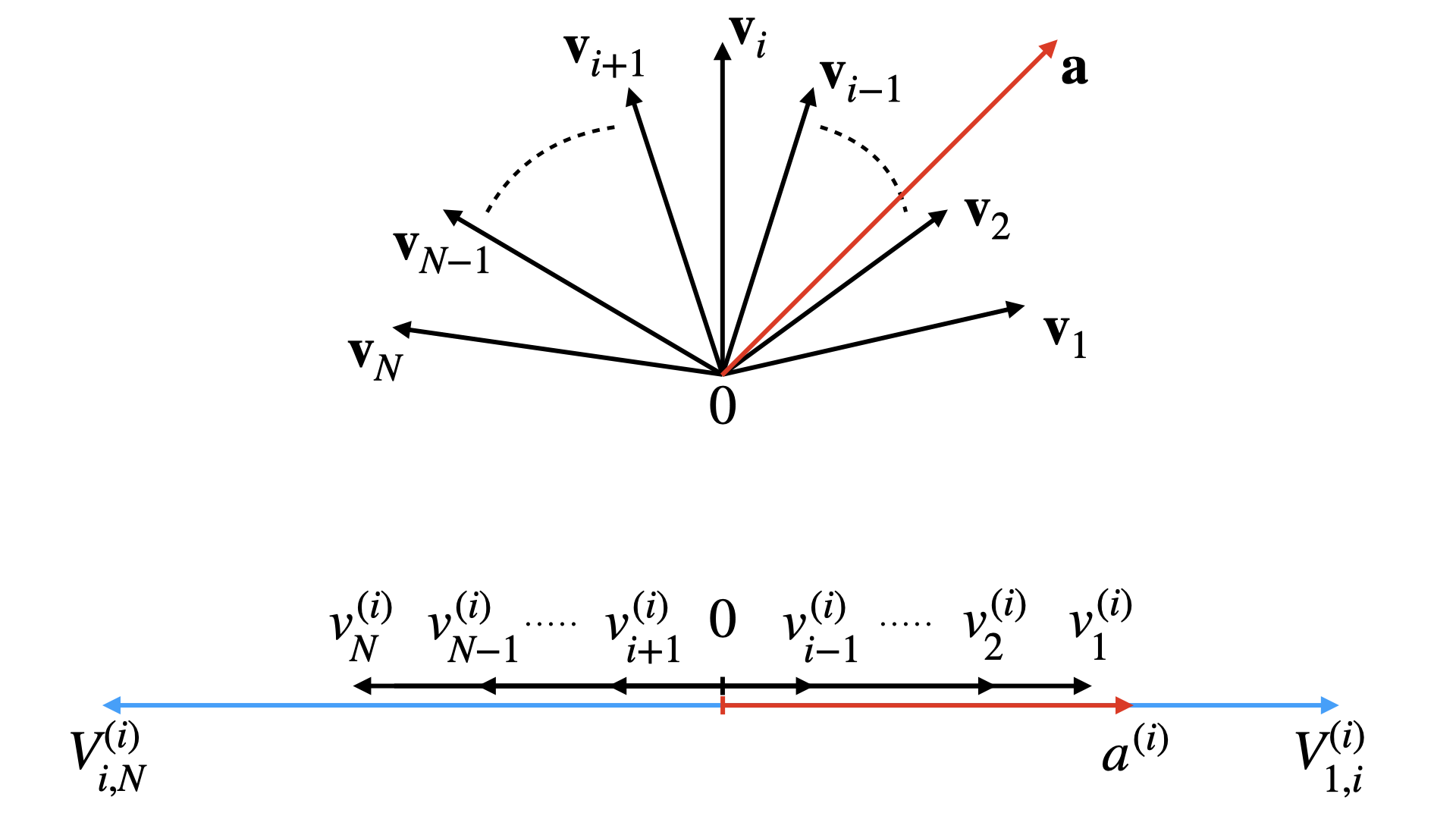} 
   \caption{Projecting vectors ${\bf v}$ to the line orthogonal to ${\bf v}_i$. The projection of any vector ${\bf a}$ in the Minkowski sum must lie between the end points $V^{(i)}_{i,N}$ and $V^{(i)}_{1,i}$.}
   \label{fig:plotprojA}
\end{figure}

In higher dimensions we can derive the boundaries by projecting the Minkowski sum to 1D. Each projection will give necessary constraints corresponding to  two of the co-dimension 1 boundaries. Let us use the 2D example as an illustration. We project all the vectors ${\bf v}_\ell$ to the one-dimensional line orthogonal to one of the vectors, say ${\bf v}_i$.  The end-points of the projected segments are now given by $0$ and 
\eq
v_j^{(i)}\equiv \frac{1}{|{\bf v}_i|}\langle j, i\rangle\,.
\eqe
If the vectors satisfy eq.(\ref{2DOrder}), the projections are oriented according to the ordering: segments corresponding to $1\le j\le i$ are in  the negative direction, while segments corresponding to $i\le j\le N$ are in the positive direction.\footnote{Note that for notational simplicity, we are  considering ${\bf v}_i$ to be part of both sets. Including it or not does not change any of the equations, since ${\bf v}_i^{(i)}=0$.} This projection is shown in Figure \ref{fig:plotprojA}. 
The end points of the 1D Minkowski sum are then simply the sum over the 1D vectors on either side respectively:
\eqa
\nonumber {V}^{(i)}_{1,i}&\equiv&\sum_{j\le i}{ v}_j^{(i)} =\sum_{j\le i}\frac{1}{|{\bf v}_i|}\langle j, i\rangle=\frac{1}{|{\bf v}_i|}\langle {\bf V}_{1,i}, i\rangle\,,\\
{V}_{i,N}^{(i)}&\equiv&\sum_{j\ge i}{v}_j^{(i)}= \frac{1}{|{\bf v}_i|}\langle {\bf V}_{i,N}, i\rangle\,,
\eqae 
where we have defined ${\bf V}_{i_1,i_2}=\sum_{j=i_1}^{i_2}{\bf v}_j\,$. The last form indicates that a 1D end point is the projection of one of the vertices of the whole 2D Minkowski sum, as it must.  Now a necessary condition for any point ${\bf a}$ to be inside the Minkowski sum is that its projection through ${\bf v}_i$ must lie within the $1D$ boundaries. This implies
\eq\label{UpLow}
\langle {\bf V}_{i,N}, i\rangle \le \langle {\bf a}, i\rangle \le \langle {\bf V}_{1,i},  i\rangle\,.
\eqe
We obtain two boundaries, which we denote as
\eqa\label{poly1bar}
P_i^{\textrm{lower}}&\equiv &
-\langle 
{\bf a}{-}{\bf V}_{1,i}, i\rangle=\textrm{det}\begin{pmatrix}
1&1&1\\
 {\bf a}&{\bf V}_{1,i}&{\bf V}_{1,i}
\end{pmatrix}_{3\times3}\geq0\,,\nonumber\\
{P}_i^{\textrm{upper}}&\equiv &\langle 
{\bf a}{-}{{\bf V}}_{i,N}, i\rangle=\textrm{det}\begin{pmatrix}
1&1&1\\
 {\bf a}&{{\bf V}}_{i,N}&{{\bf V}}_{i,N}
\end{pmatrix}_{3\times3}\geq0\,,
\eqae 
matching the boundaries in eq.(\ref{2Dbound}). Successively repeating the same operation for each vector ${\bf v}_{i}$ then yields the complete set of boundaries. Throughout this paper, we will often refer to boundaries as either upper or lower, labeled in terms of how they would appear in plots. There is however a more invariant definition: given a set of ordered segments, the vertices of the lower boundary are obtained from the successive sum starting from the beginning of the ordering, while the upper boundaries are from the successive sum starting from the end of the ordering. This is apparent from eq.(\ref{UpLow}). In higher dimensions the equivalent distinction will be such that the upper boundary is the one containing ${\bf V}_{i,N}$ (regardless of whether it also contains  ${\bf V}_{1,i}$).

To check these boundaries are correct and can also be saturated, we can substitute ${\bf a}$ by its general form in  eq.(\ref{bod11}), ${\bf a}=\sum_{j} m_j{\bf v}_j$, where $0\leq m_j\leq1$. Since ${\bf a}{-}{\bf V}_{1,i}=\sum_{j< i} (m_j{-}1){\bf v}_j {+}\sum_{i\geq j} m_j {\bf v}_j$, the determinant in eq.(\ref{poly1bar}) then becomes
\eq
P_{i}=\sum_{j<i}(1-m_j)\langle j, i\rangle-\sum_{j>i}m_j\langle j,i\rangle\,.
\eqe
Due to the cyclic condition, we have that $\langle j, i\rangle$ is positive for $j<i$, and negative for $j>i$, therefore the expression above is non-negative term by term. These are true boundaries since the equality $P_i=0$ can be saturated by taking $m_j=1$ for $j<i$ and $m_j=0$ for $j>i$, while $m_i$ can be any value between 0 and 1, parameterizing the edge between vertices $i$ and $i+1$.

\begin{figure}
   \centering
   \includegraphics[width=2.5in]{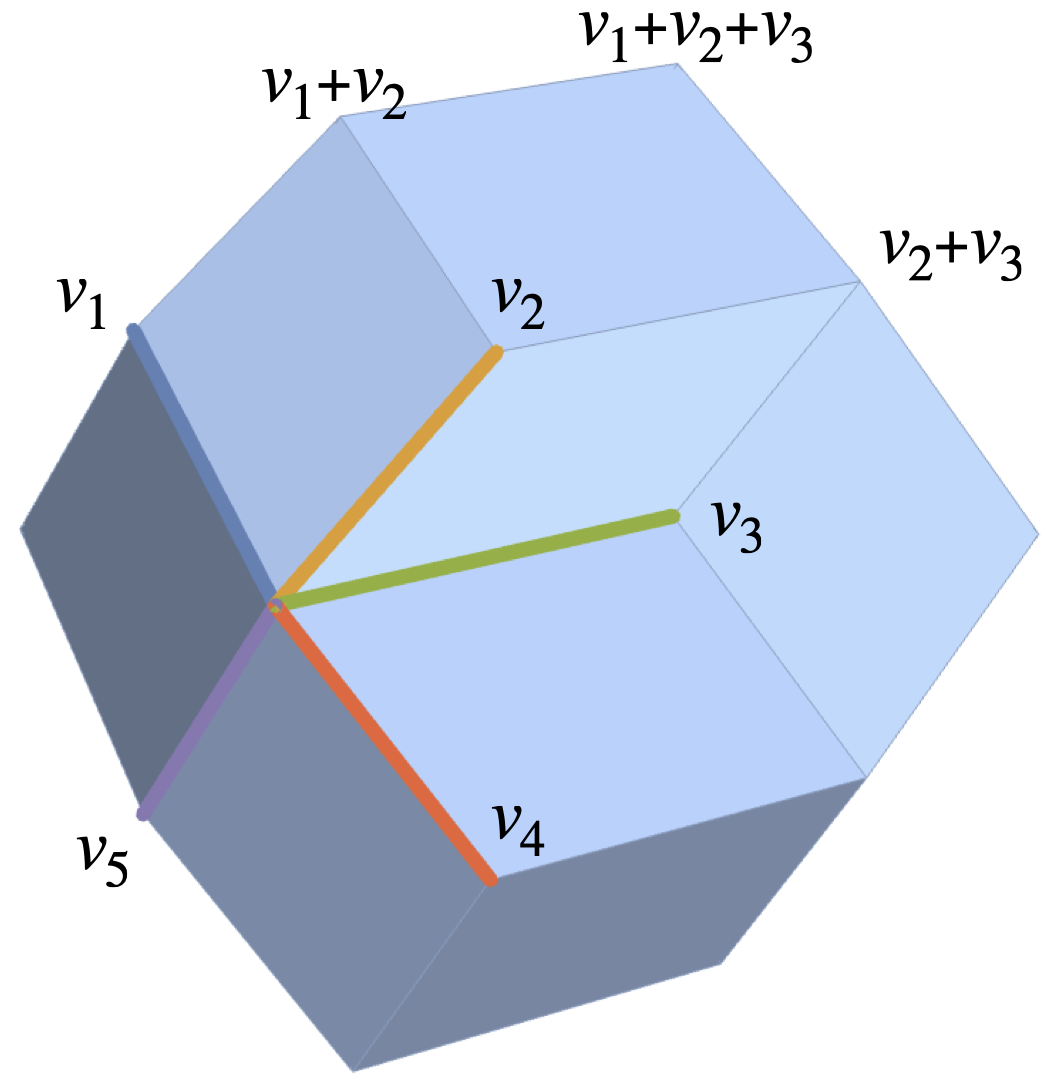} 
   \caption{3D polytope with $N=5$}
   \label{fig:plotprojB}
\end{figure}

\paragraph{3D}
We next consider three-dimensional vectors, which we again assume satisfy the ordering of eq.(\ref{eq:cyclic}). We will derive the boundary by considering two successive projections down to 1D, through two vectors ${\bf v}_{i_1}$ and ${\bf v}_{i_2}$, for $i_1<i_2$. Since the vectors are ordered, we deduce:
\eqa
\begin{array}{c} 
\langle j, i_1, i_2 \rangle>0\,, \quad j{<}i_1{<}i_2\,, \\ \langle j, i_1, i_2 \rangle<0\,, \quad i_1{<}j{<}i_2\,,  \\ \langle j, i_1, i_2 \rangle>0\,, \quad i_1{<}i_2{<}j \,.
\end{array} 
\eqae
That is, when projecting twice down to 1D, the remaining $N{-}2$ vectors will be split into two sets:  the vectors ${\bf v}_{i_1{+}1}, {\bf v}_{i_1{+}2},\ldots, {\bf v}_{i_2{-}1}$ yield points on one side, and ${\bf v}_{i_2{+}1}, {\bf v}_{i_2{+}2},\ldots, {\bf v}_{i_1{-}1}$ on the other.\footnote{The ordering is cyclic, thus ${\bf v}_{N{+}1}\equiv {\bf v}_{1}$. } The resulting end points of the one-dimensional Minkowski sum are then given by summing over the vectors in the respective sets. By requiring any point in the Minkowski sum to lie between the two endpoints we obtain
\eq
\langle {\bf V}_{i_1,i_2}, i_1, i_2\rangle\le \langle {\bf a}, i_1, i_2\rangle \le \langle {\bf V}_{1,i_1}{+}{\bf V}_{i_2,N}, i_1, i_2\rangle \,,
\eqe
or
\eqa\label{3dpol}
\nonumber P_{i_1,i_2}&\equiv&\langle 
{\bf a}{-}{\bf V}_{i_1,i_2}, i_1, i_2 \rangle \geq0,\,\\
\overline{P}_{i_1,i_2}&\equiv&{-}\langle 
{\bf a}{-}{\bf V}_{1,i_1}{-}{\bf V}_{i_2,N}, i_1, i_2\rangle \geq0\,.
\eqae
The complete boundary is then given by the collection of $P_{i_1,i_2},\overline{P}_{i_1,i_2}$ for all pairs of $\{i_1< i_2\}$. We present an example with $N=5$ vectors in Figure \ref{fig:plotprojB}. Note that in this case we refer to $\overline{P}$ as the upper boundary, as it contains ${\bf V}_{i_2,N}$.

\paragraph{General dimension}
In $D$ dimensions each boundary is labeled by $D{-}1$ indices, corresponding to the $D-1$ vectors we use to project. We denote this set as $I=(i_1,i_2,i_3,\ldots i_{D-1})$, $1{\le} i_1{<}i_2{<} \ldots {<}i_{D-1}{\le} N$. Now we must compute, for a given set of projections $I$, whether a vector ${\bf v}_j$ has positive or negative projection. Let 
\eqa\label{Idef}
\mathcal{I}^-=\{j|v_j^{(I)}\le 0\},\quad  \mathcal{I}^+=\{j|v_j^{(I)}\ge 0\},
\eqae
If the ${\bf v}$ are ordered, we have, for $D=\textrm{even}$:
\eqa\label{Isign1}
\nonumber \mathcal{I}^-&=&[1,i_1]\cup[i_2,i_3]\cup\ldots\cup [i_{D-2},i_{D-1}]\\
 \mathcal{I}^+&=&[i_1,i_2]\cup[i_3,i_4]\cup\ldots\cup [i_{D-1},i_{D}]
\eqae
and for $D=\textrm{odd}$:
\eqa\label{Isign2}
\nonumber \mathcal{I}^-&=&[i_1,i_2]\cup[i_3,i_4]\cup\ldots\cup [i_{D-2},i_{D-1}]\\
\mathcal{I}^+&=&[1,i_1]\cup[i_2,i_3]\cup\ldots\cup [i_{D-1},i_N]
\eqae
To maintain our convention that the upper boundary, corresponding to the $\overline{P}_I\ge 0$, is the one containing ${\bf v}_N$, we then define ${\bf V}_I$ and ${\bf \overline{V}}_I$ such that
for $D=\textrm{even}$:
\eqa\label{evenpol}
\nonumber \textrm{lower:}\quad {\bf V}_I&\equiv\displaystyle\sum_{i\in\mathcal{I}^+} {\bf v}_j= &{\bf V}_{1,i_1}+{\bf V}_{i_2,i_3}+\ldots {\bf V}_{i_{D-2},i_{D-1}}\,,\\
\textrm{upper:}\quad {\bf \overline{V}}_I&\equiv \displaystyle\sum_{i\in\mathcal{I}^-} {\bf v}_j=&{\bf V}_{i_1,i_2}+{\bf V}_{i_3,i_4}+\ldots {\bf V}_{i_{D-1},N}\,,
\eqae
and for $D=\textrm{odd}$:
\eqa\label{oddpol}
\nonumber\textrm{lower:}\quad {\bf V}_I&\equiv\displaystyle\sum_{i\in\mathcal{I}^-} {\bf v}_j=&{\bf V}_{i_1,i_2}+{\bf V}_{i_3,i_4}+\ldots {\bf V}_{i_{D-2},i_{D-1}}\,,\\
\textrm{upper:}\quad {\bf \overline{V}}_I&\equiv\displaystyle \sum_{i\in\mathcal{I}^+} {\bf v}_j=&{\bf V}_{1,i_1}+{\bf V}_{i_2,i_3}+\ldots {\bf V}_{i_{D-1},N}\,.
\eqae

  Then we have the constraints
\eqa\label{gendimpol}
P_I&=&(-1)^{D+1}\langle 
{\bf a}{-}{\bf V}_{I}, i_1, i_2,\ldots ,i_{D{-}1}\rangle \nonumber\,,\\
\overline{P}_I&=&(-1)^{D}\langle {\bf a}{-}{\bf \overline{V}}_{I}, i_1, i_2, \ldots , {i_{D{-}1}}\rangle\,.
\eqae

By considering all possible choices $1\le i_1{<}i_2{<}\ldots{<}i_{D-1}{\le}N$ we obtain the complete boundary. 

\paragraph{Non-cyclic vectors}\label{noncyc}
In the case where the ${\bf v}$s cannot be cyclically ordered, then we no longer know beforehand whether the 1D projection of any vector will be positive or negative.  Therefore, eqns.(\ref{Isign1}) and (\ref{Isign2}) no longer necessarily follow from the definition in eq.(\ref{Idef}). In such case one has to individually check, for every sequence $I$, which vectors ${\bf v}$ belong to which set, $\mathcal{I}^+$ or $\mathcal{I}^-$, and use this to compute ${\bf V}$ and ${\bf \overline{V}}$. In practice, since we are usually interested in intersecting the geometry with a null constraint, typically this means only a finite number of boundaries are relevant, so this is a trivial computational task.

In two dimensions only vectors spanning more than a half-plane cannot be cyclically ordered, and the problem can be solved straightforwardly. We first pick a half-plane,\footnote{In the physical problem, there will be an infinite number of vectors, but always accumulating in a single quadrant. In this case we must pick the half-plane including this quadrant.} and name the vectors in this half-plane ${\bf v}$, and the remaining ${\bf \tilde{v}}$. The Minkowski sum is now
\eq
{\bf a}=\sum_i m_i {\bf v_i}+\sum_i m_i {\bf \tilde{v}_i}\,.
\eqe
Next we add and subtract $\sum_i {\bf \tilde{v}}_i$. Since the sum is associative, we can group this as 
\eq
{\bf a}=\left(\sum_i m_i {\bf v_i}+\sum_i (m_i -1){\bf \tilde{v}_i}\right)+\sum_i{\bf \tilde{v}_i}\,.
\eqe
The combinations $ (m_i -1){\bf \tilde{v}_i}$ are now segments in the same half-plane as the ${\bf v}$, so the sum in parantheses can be computed as before, simply ordering the vectors according to their slope $v_{\ell,k_2,q_2}/v_{\ell,k_1,q_1}$, regardless of what half-plane they are in. Finally we translate the whole boundary by $\sum_i{\bf \tilde{v}_i}$.

\subsection{De-projecting the moment problem:  The $L$-moment}\label{sect2L}
Let us now consider a different problem, where our previous result will guide us to the answer. We want to find the allowed space of moments $(a_{k_1},a_{k_2},\ldots)$ that satisfy
\eq\label{Lcoupling}
a_k=\int_{0}^1 \rho(z)z^{k-1}dz,\quad 0\le \rho(z)\le L\,.
\eqe
This is known as the ``$L$-moment problem" \cite{akhiezer1934fouriersche, akhiezer1962some}. We will solve it in a straightforward fashion by taking the continuous limit of the polytope  discussed in the previous section. To relate the two we express the above integral as a Riemann sum
\eq
a_k=\int_{0}^1 \rho(z)z^{k-1}dz\;\rightarrow\; \sum_{i=1}^N \rho_i \frac{i^{k-1}}{N^{k}}\,,
\eqe
where $\rho_i\equiv\rho(i/N)\le L$, which again we can consider to be a segments of length $L$, parameterized by $m_i\in[0,1]$. We can then write the above as
\eq\label{minki}
a_k=\sum_{i=1}^N m_i  v_{i,k} 
\eqe
where $v_{i,k}=Li^{k-1}/N^k$. Expressed as a discrete sum the $L$ moment problem is equivalent to a Minkowski sum of segments, with endpoints at 0 and $L(v_{i,k_1},v_{i,k_2},\ldots)$. Furthermore,  the $v_{i,k}$ are ordered, as taking a collection of $N$ such vectors the resulting $N\times N$ determinant will be proportional to the Vandermonde determinant, and thus eq.(\ref{eq:cyclic}) holds. The results in the previous section therefore apply to this case, so the vertices of the Minkowski sum in eq.(\ref{minki}) are given by eq.(\ref{gendimpol}). As we take $N\rightarrow \infty $, the infinite number of polytope boundaries merge into a smooth curve and each vertex is a point on the curve. 

\begin{figure}[H] 
   \centering
   \includegraphics[height=1.8in]{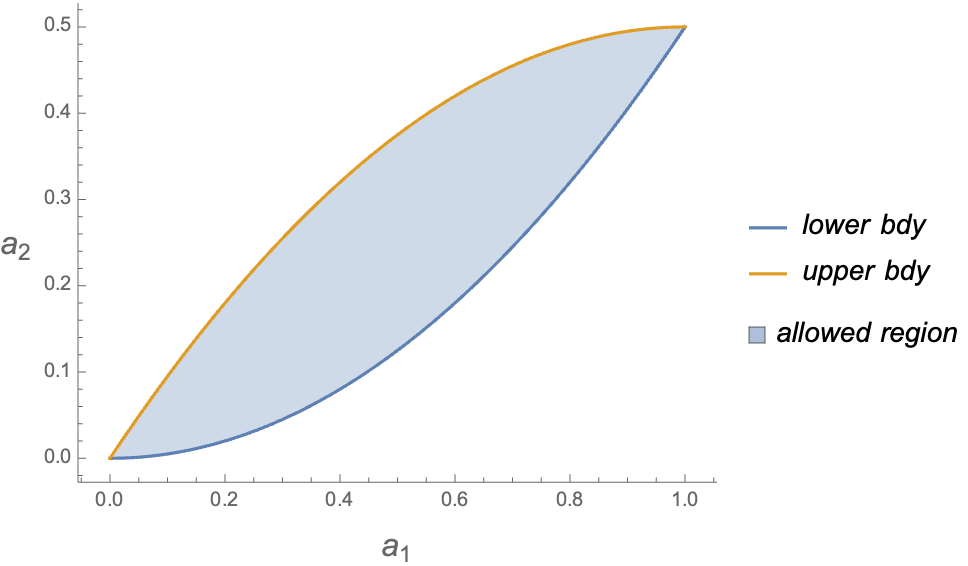} 
     \includegraphics[height=1.8in]{plots/plotex2.png} 
   \caption{$(a_1,a_2)$ allowed space and boundaries. In the limit $L=\infty$, only the projective boundaries $0\le a_2/a_1\le 1$ are left.}
   \label{fig:plotex3}
\end{figure}

\noindent \textbf{Two dimensions}
Let us take the two dimensional space $(a_{k_1}, a_{k_2})$ as an example. From the previous subsection, the vertices are 
\eq\label{sum}
{V}_{1,i|k}=L\sum_{j=1}^{i} v_{j,k}=L\sum_{j=1}^{i} \frac{j^{k-1}}{N^{k}}\,.
\eqe
Now we define $m\equiv \frac{i}{N}\in[0,1]$ and $z\equiv \frac{j}{N}\in[0,m]$, which in the $N \rightarrow \infty$ limit can be treated as continuous parameters. The sum in  eq.(\ref{sum}) simply becomes an integral
\eq
{ V}_{1,i|k}=L\sum_{j=1}^{i} \frac{j^{k-1}}{N^{k}}\rightarrow L \int_0^{m} z^{k-1} dz=L\frac{m^k}{k}\,,
\eqe
and the boundary vertex becomes a point on the parametric curve
\eq\label{pareq}
(a_{k_1}, a_{k_2})=({U}_{k_1}(m), {U}_{k_2}(m)),\quad  {U}_{k}(m)\equiv L\frac{m^{k}}{k}\,.
\eqe
Similarly for the vertices of type ${\bf V}_{i,N}$ in the continuous limit we obtain: 
\eq\label{parequ}
(a_{k_1}, a_{k_2})=(\overline{U}_{k_1}(m), \overline{U}_{k_2}(m)),\quad  \overline{U}_k(m)\equiv L\frac{1-m^{k}}{k}\,.
\eqe
Note that we can rephrase the boundary as a particular solution for $\rho(z)$. It is not hard to work out that the solution for lower and upper boundary are given by
\eqa
{\rm lower}:\quad \rho(z)&=&\chi_I(z), \;\; I=[0, m]\nonumber\\
{\rm upper}:\quad \rho(z)&=&\chi_I(z), \;\; I=[m, 1]\nonumber
\eqae
where $\chi_I(z)=L$ for $z\in I$ and zero otherwise. In other words, the boundary solutions are constants along some region in $z$! 
To find the constraint equations in $a_k$ that describe the boundary, we simply combine the couplings in such a way that removes the boundary parameter.  Then the allowed space can be expressed as an inequality
\begin{align}\label{ineq1}
 \frac{L}{k_2} \left(\frac{k_1}{L}a_{k_1}\right)^\frac{k_2}{k_1}\le a_{k_2}\le  \frac{L}{k_2}\bigg(1-\left(1-\frac{k_1}{L}a_{k_1}\right)^{\frac{k_2}{k_1}}\bigg)\,.
\end{align}
This represents the allowed space for the 2D $L$ moment problem. We plot this in Figure \ref{fig:plotex3}, for $k_1=1,k_2=2$ and for $L=1,2,\infty$, observing the $L=\infty$ limit reproduces the projective bounds of the usual moment problem, namely $0\le \frac{a_{k_2}}{a_{k_1}}\le 1$.

\begin{figure}[H] 
   \centering
    \includegraphics[height=2.3in]{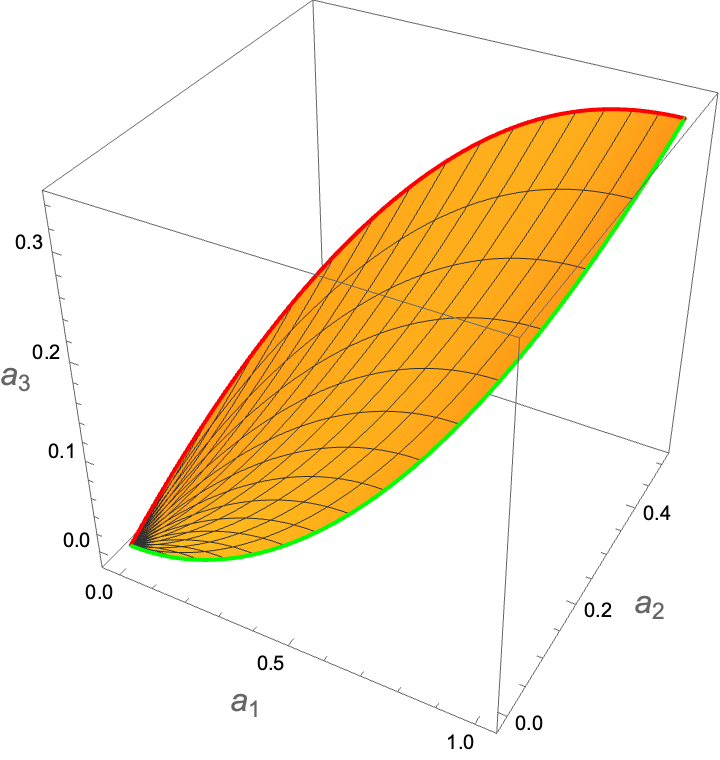} 
        \includegraphics[height=2.3in]{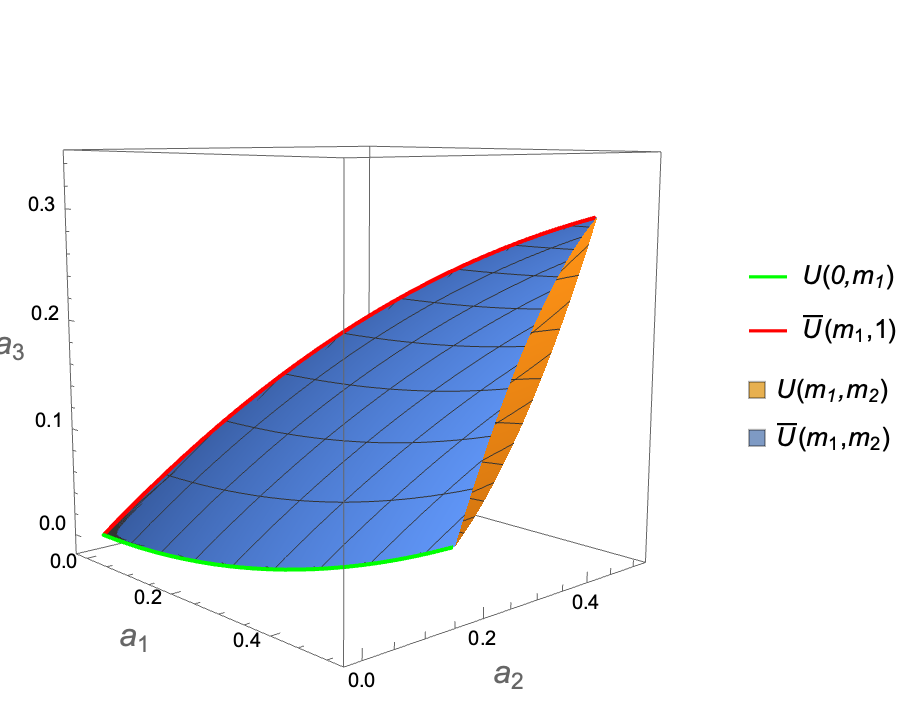} 
\caption{Boundaries for $(a_1,a_2,a_3)$ space, $L=1$. The left figure shows only the lower boundary, corresponding to eq.(\ref{bdrg}). The right figure contains both lower and upper boundaries, and has been cropped at $a_0=0.5$ to make both boundaries visible.}
   \label{fig:plotex4}
\end{figure}

\noindent \textbf{Three dimensions}
In 3D the same procedure gives the boundary as a parametric surface. Taking the continuous limit of a vertex ${\bf V}_{i_1,i_2}$ we find the boundary is
\eqa\label{bdrg}
(a_{k_1}, a_{k_2}, a_{k_3})&=&( U_{k_1}(m_{1}, m_{2}), U_{k_2}(m_{1}, m_{2}),  U_{k_3}(m_{1}, m_{2})),\nonumber\\
U_k(m_{1},m_{2})&\equiv&\frac{L}{k}\left(m_{2}^{k}{-}m_{1}^{k}\right)\,,
\eqae
while for the upper  boundary one has $\overline{U}_k(m_{1},m_{2})\equiv\frac{L}{k}\left(1{-}m_{2}^{k}{+}m_{1}^{k}\right)$. The boundary distribution is given by 
\eqa
{\rm Lower}:\quad \rho(z)&=&\chi_{I}(z),\;\; I=[m_1, m_2], \nonumber\\
{\rm Upper}:\quad \rho(z)&=&\chi_{I_1}(z){+}\chi_{I_2}(z) ,\;\; I_1=[0, m_1],  \;\;  I_2=[m_2, 1],\nonumber\,.
\eqae
For a space $(a_1,a_2,a_3)$ the two boundaries lead to the inequalities:  
\eqa 
\frac{a_1^4+12 L^2 a_2^2}{12 L^2a_1}\le a_3\le\frac{a_1^4-4 La_1^3+6L^2 a_1^2+12 L^2 a_2^2-12L^3 a_2}{12 L^2(a_1-L)}\,.
\eqae
The explicit region is shown in Figure \ref{fig:plotex4}. The generalization to higher dimensions can be readily obtained from eqs.(\ref{evenpol}) and (\ref{oddpol}), by replacing all terms ${\bf V}_{i_1,i_2}\rightarrow {\bf U}(m_1,m_2)$. Explicitly, we obtain 
\eqa\label{evenU}
\nonumber \textrm{lower: } U_k&=&\frac{L}{k}\sum_{i=1}^{D-1}(-1)^{i+1} m_i^{k}\,\\
\textrm{upper: } \overline{U}_k&= &\frac{L}{k}\left(1-\sum_{i=1}^{D-1}(-1)^{i+1} m_i^{k}\right)\,.
\eqae

Note that the inequalities above are non-projective, i.e. they involve polynomials of $a_k$ with different degrees. This is different from the usual moment problem where the constraints are given by the positivity of Hankel matrices, which are homogenous in the degree of couplings. However, through a set of non-linear (exponential) maps  to new couplings, the constraints can again be expressed through the positivity of Hankel matrices. In fact this is how the solution to the $L$-moment problem is usually given in the literature, which we review in Appendix \ref{aspectsL}. For our purpose, the parametric form of the boundary will be more useful, as we will perform further Minkowski sums for the general problem.



\section{Minkowski sums of single $L$-moments}\label{sec3}
We now return to the our general problem 
\eq\label{a}
a_{k,q}=\sum_\ell  v_{\ell,k,q}\int_0^1 \rho_\ell(z)z^{k-1} , \quad 0\le \rho_\ell\le L\,.
\eqe
We will solve this problem by treating the $a_{k,q}$ as yet another Minkowski sum. We can write the above as
\eq
\begin{pmatrix}
a_{k_1,q_1}\\
a_{k_1,q_2}\\
\vdots\\
a_{k_D,q_D}
\end{pmatrix}=\sum_\ell \begin{pmatrix}
a_{k_1,q_1}^{(\ell)}\\
a_{k_2,q_2}^{(\ell)}\\
\vdots\\
a_{k_D,q_D}^{(\ell)}
\end{pmatrix}=\sum_\ell \begin{pmatrix}
a_{k_1}^{(\ell)}v_{\ell,k_1,q_1}\\
a_{k_2}^{(\ell)}v_{\ell,k_2,q_2}\\
\vdots\\
a_{k_D}^{(\ell)}v_{\ell,k_D,q_D}
\end{pmatrix}\,.
\eqe
where for each spin, the $a_{k,q}^{(\ell)}$ is identified as the solution to the $L$-moment problem $a_k=(a_{k_1},a_{k_2},\ldots,a_{k_D})$ , with each component $a_{k_i}$ rescaled by some number $v_{\ell, k_i, q_i}$. The full geometry is then given by the Minkowski sum over all spins of such geometries.

As discussed in the beginning of the previous section, to compute the boundary of the Minkowski sum, one first computes the Minkowski sum of the boundaries. The final boundary can be found by extremizing this result, most simply using Lagrange multipliers, as we will discuss shortly. Alternatively we can revert the $L$-moment problem to its discrete version, and simply consider the whole problem as a double Minkowski sum of segments. We will explore this view point in detail in appendix~\ref{DiscreteM}. Here we will present the result from the Lagrange method. 

We first solve the problem for any 2D space $(a_{k_1,q_1},a_{k_2,q_2})$, for any $v_{\ell,k,q}$. This will be sufficient to derive optimal bounds on any 2D space of couplings without imposing null constraints, or bounds on any single coupling with one null constraint.  

To implement more null constraints, or to obtain obtain the space of more couplings, would require us to solve the geometry in higher dimensions. In this case we only obtain partial results. First, we can easily derive boundaries when only a a finite number of spins are relevant. When the sum over infinite spins cannot be avoided, we are only able to find closed form expressions for spaces $(a_{k_1,q_1},a_{k_2,q_2},\ldots)$ satisfying $k_{i+1}{-}k_i{=}\Delta_k$ , $q_{i+1}{-}q_i{=}\Delta_q$, and when $v_{\ell,k,q}{=}v_{\ell,q}{=}g(\ell)f(\ell)^q$, where $g(\ell)$ and $f(\ell)$ can be general polynomials of $\ell$.

For the entire space of couplings, the problem can be formulated in terms of envelopes of various moment constraints. This is more amendable to numerical approaches which we explore in appendix~\ref{appgend}.

\subsection{Extremizing the Minkowski sum}
In this section we briefly explain how the boundary of a Minkowski sum can be obtained by extremizing. Let us use the example of the disks in Figure \ref{Minkfig2} to introduce this method. We can describe the boundaries of disks A and B as
\eqa
\nonumber \partial A(\theta_A)&=&(a_x,a_y)+r_A(\textrm{sin}\theta_A, \textrm{cos}\theta_A)\,,\\
\partial B(\theta_B)&=&(b_x,b_y)+r_B( \textrm{sin}\theta_B, \textrm{cos}\theta_B)\,,
\eqae
The Minkowski sum of boundaries $\partial A(\theta_A)+\partial B(\theta_B)$ is then a 2D region, with the two components given by 
\eqa
f_x(\theta_A, \theta_B)&=&(a_x{+}b_x){+}r_A\textrm{sin}\,\theta_A{+}r_B\textrm{sin}\,\theta_B \nonumber\\
f_y(\theta_A, \theta_B)&=&(a_y{+}b_y){+}r_A\textrm{cos}\,\theta_A{+}r_B\textrm{cos}\,\theta_B\,.
\eqae
To find the boundary of this region we can consider extremizing in one coordinate while the other is held fixed. This gives a constraint equation that will allow us to solve one parameter in terms of the remaining, giving the co-dimension one boundaries of the Minkowski sum $\partial A(\theta_A)+\partial B(\theta_B)$.  More precisely, fixing the $x$-coordinate while extremizing in $y$, one solves
\eq
\left(\begin{array}{c} \partial_{\theta_A} f_x(\theta_A, \theta_B) \\  \partial_{\theta_B} f_x(\theta_A, \theta_B) \end{array}\right)=\lambda\left(\begin{array}{c} \partial_{\theta_A} f_y(\theta_A, \theta_B) \\  \partial_{\theta_B} f_y(\theta_A, \theta_B) \end{array}\right)\,,
\eqe
where $\lambda$ is the Lagrangian multiplier. Note that this is simply saying that the two component vectors are proportional and hence 
\eq
{\rm det} \left(\begin{array}{cc} \partial_{\theta_A} f_x(\theta_A, \theta_B) & \partial_{\theta_A} f_y(\theta_A, \theta_B) \\  \partial_{\theta_B} f_x(\theta_A, \theta_B) & \partial_{\theta_B} f_y(\theta_A, \theta_B) \end{array}\right)=0\;\;\rightarrow \;\;\cos\theta_A \sin\theta_B=\cos\theta_B \sin\theta_A\,.
\eqe
We find two solutions: $\theta_A=\theta_B$ and $\theta_A=\theta_B+\pi$. These precisely correspond to the outer and inner boundaries in Figure \ref{Minkfig2}, given by 
\eq
\partial(\partial A+\partial B)=(a_x+b_x, a_y+b_y)+(r_A\pm r_B)( \textrm{sin}\phi_A, \textrm{cos}\phi_A)\,.
\eqe
This generalizes to higher dimensions. Consider the combination of two $D$-dimensional regions. We find the boundary of their Minkowski sum by taking the Minkowski sum of the boundary $(f_{x_1}, f_{x_2},\cdots,f_{x_D})$, where $f_{x_i}$ is a function of $2(D{-}1)$ boundary variables, $(\theta_1, \theta_2, \cdots, \theta_{2D{-}2})$. Holding $(f_{x_2},\cdots,f_{x_D})$ fixed and extremizing $f_{x_1}$ we solve 
\eq
\left(\begin{array}{c} \partial_{\theta_1} f_{x_1}  \\   \partial_{\theta_2} f_{x_1} \\   \vdots \\ \partial_{\theta_{2(D{-}1)}} f_{x_1} \end{array}\right)=\sum_{i=2}^{D}\lambda_i \left(\begin{array}{c} \partial_{\theta_1} f_{x_i}  \\   \partial_{\theta_2} f_{x_i} \\   \vdots \\ \partial_{\theta_{2(D{-}1)}} f_{x_i} \end{array}\right)\,.
\eqe
This gives us $2D{-}2$ constraints on $D{-}1$ Lagrangian multipliers $\lambda_i$ and  $2(D{-}1)$ boundary variables $\theta_i$, giving us $D{-}1$-dimensional boundaries.

Finally we comment that since the Lagrange method relies on the variation of the individual boundaries, it will miss boundaries coming from the end points (or boundary of boundaries). These end point contributions must be analyzed by hand, which we will demonstrate through explicit examples.

\subsection{Two dimensions with $k_1{\neq}k_2$}\label{sec2d2}
\begin{figure}[H] 
   \centering
    \includegraphics[height=2.0in]{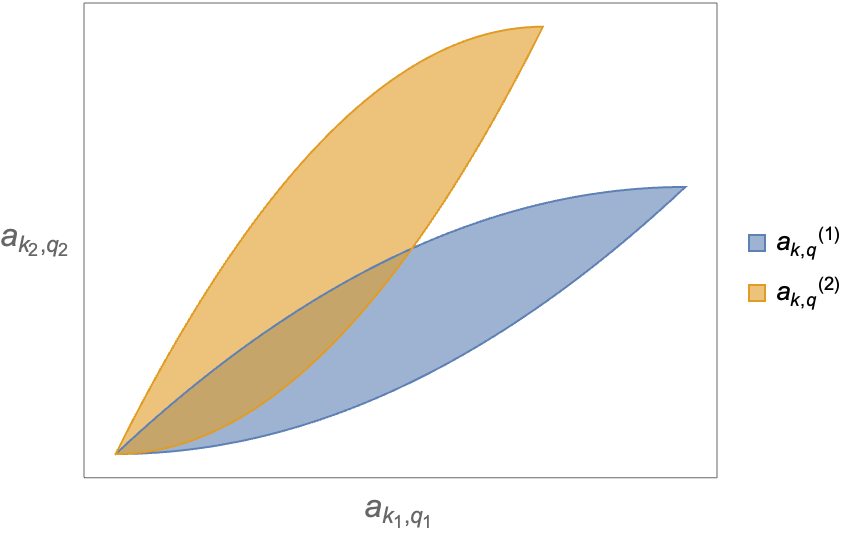} 
   \caption{The allowed space for two single $L$-moments $a_{k,q}^{(\ell)}$, rescaled by different $v_{\ell,k,q}$. The moment $a_{k,q}$ is the Minkowski sum of such geometrical objects.}
   \label{fig:plotM00}
\end{figure}

Let us consider first the sum of two moments
\eq\label{eqdo}
a_{k,q}=v_{\ell_1,k,q}a_{k}^{(\ell_1)}+v_{\ell_2,k,q}a_{k}^{(\ell_1)}=a_{k,q}^{(\ell_1)}+a_{k,q}^{(\ell_2)}\,,
\eqe
where $a_{k}^{(\ell)}$ takes the form of eq.(\ref{Lcoupling}). First, we note that $a_{k,q}^{(\ell)}$ simply represents a single $L$-moment $a_{k}$, with each component rescaled by some number $v_{\ell,k,q}$. This implies the boundaries of $a_{k,q}^{(\ell)}$ are also the original boundaries of $a_k$ that we found in Section \ref{sect2L}, with each component rescaled. We will be considering arbitrary two couplings, $(a_{k_1,q_1}, a_{k_2,q_2})$. The boundary for the $L$-moment problem in a 2D space $(a_{k_1}, a_{k_2})$ is given by the two curves in eqs.(\ref{pareq}) and (\ref{parequ}). After the rescaling, we now have: 
\eqa
\label{eq311}\textrm{lower bdy:}&&\quad  (a_{k_1,q_1}^{(\ell)}, a_{k_2,q_2}^{(\ell)})=L\bigg(\frac{v_{\ell,k_1,q_1}}{k_1} m^{k_1}, \frac{v_{\ell,k_2,q_2}}{k_2} m^{k_2}\bigg)\nonumber\,,\\
\label{eq312}\textrm{upper bdy:}&&\quad  (a_{k_1,q_1}^{(\ell)}, a_{k_2,q_2}^{(\ell)})=L\bigg(\frac{v_{\ell,k_1,q_1}}{k_1}\left(1{-} m^{k_1}\right),\frac{v_{\ell, k_2, q_2}}{k_2}\left(1- m^{k_2}\right)\bigg)\,.
\eqae
Now eq.(\ref{eqdo}) requires us to compute the Minkowski sum of such boundaries for different $\ell$, which we sketch in Figure \ref{fig:plotM00}. Note that in principle, we simply need to order the points on the boundaries of $a_{k,q}^{(\ell)}$, and we can obtain the 2D boundary using the result of the previous section. However for the problem in hand, the proper ordering of points on multiple moments is a cumbersome task. Instead, we will take the boundaries of each $a_{k,q}^{(\ell)}$ and extremize using the Lagrange multiplier as discussed previously. If we assume that  all $v_{\ell,k,q}$s have the same sign, then each moment is oriented in the same ``quadrant" and the lower boundary of  the Minkowski sum  $a_{k,q}$ is contained in the sum of lower boundaries of $a_{k,q}^{(\ell)}$, and similarly for the upper boundary. If some of the $v_{\ell,k,q}$ have different sign, then special care is needed. We will return to this issue at the end of this section. 

Using eq.(\ref{eq311}) we therefore have that the lower boundary we are after is contained in the 2D region 
\eq\label{starteq2}
a_{k,q}(m_1,m_2)=\frac{L}{k}\left(v_{\ell_1,k_,q}m_1^{k_1}+v_{\ell_2,k_,q}m_2^{k_1}\right),\quad m_1,m_2\in[0,1]\,.
\eqe
We can now employ the Lagrange method to find the relation between $m_1$ and $m_2$ that corresponds to a boundary. The Lagrange equation reads
\eq
\grad_{m_1,m_2}a_{k_1,q_1}=\alpha \grad_{m_1,m_2}a_{k_2,q_2}\,,
\eqe
which implies a remarkably simple relation between the parameters
\eq\label{lagrs}
\frac{m_1}{m_2}=\left(\frac{v_{\ell_1,k_1,q_1}}{v_{\ell_1,k_2,q_2}}\frac{v_{\ell_2,k_2,q_2}}{v_{\ell_2,k_1,q_1}}\right)^{\frac{1}{k_2-k_1}}\equiv r_{1,2}\,.
\eqe
Note that $r_{i,j}=r_{j,i}^{-1}$. If the vectors
\eq
{\bf v}_{\ell}=\left(\begin{array}{c} v_{\ell,k_1,q_1} \\ v_{\ell,k_2,q_2}\end{array}\right)\,,
\eqe
are ordered with respect to spin-$\ell$, then  $r_{i,i-1}\le 1$. We can use this to write $m_{2}=r_{2,1} m_{1}$,  which plugging into eq.(\ref{starteq2}) leads to a boundary curve 
\eq\label{eqbdr1}
a_{k,q}=\frac{L}{k}m_1^{k}\left(v_{\ell_1,k,q}+r_{2,1}^{k} v_{\ell_2,k,q}\right)\,,
\eqe
valid for $m_1\in[0,1]$.

\begin{figure}[H] 
   \centering
    \includegraphics[height=2.0in]{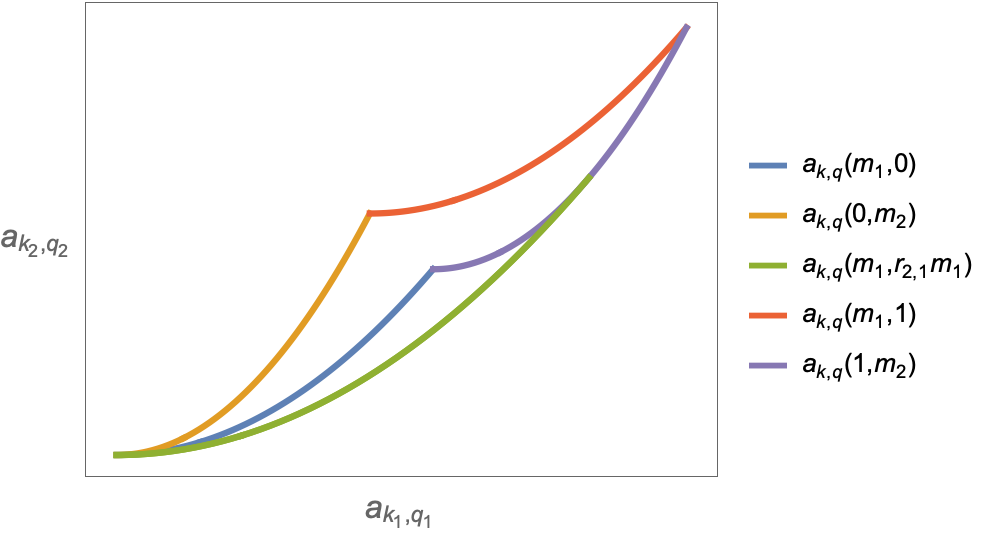} 
   \caption{The five possible boundaries of the Minkowski sum of two lower boundaries, corresponding to yellow and blue. The true lower boundary is given by the green curve, obtained by extremizing via the Lagrange method, continued by a section of the purple curve.}
   \label{fig:plotM01}
\end{figure}

Note that if $m_1, m_2$ are unbounded, then these are the only boundaries. Since they are bounded by $0$ and $1$, there could exist other boundaries that correspond to fixing one of the parameters to either 0 or 1. There are four possible choices
\eq
(m_1,m_2)=\{(m ,0),(m ,1),(0,m),(1,m)\}\,.
\eqe
Plugging these into eq.(\ref{starteq2}) we obtain four curves. We plot these together with the curve in eq.(\ref{eqbdr1}) in Figure \ref{fig:plotM01}. 
We observe that the lower boundary is in fact formed of two sections, corresponding to $a_{k,q}(m,r_{2,1}m)$ and $a_{k,q}(1,m)$
\eqa\label{lbdya}
\nonumber \textrm{lower bdy 1: }&& a_{k,q}(m,r_{2,1}m)=\frac{L}{k}m^{k}\left(v_{\ell_1,k,q}+r_{2,1}^{k} v_{\ell_2,k,q}\right),\quad m\in[0,1]\\
\textrm{lower bdy 2: }&& a_{k,q}(1,m)=\frac{L}{k}\left(v_{\ell_1,k,q}+m^{k} v_{\ell_2,k,q}\right),\quad m\in[r_{2,1},1]\,.
\eqae
The lower bound for $m$ in the second section can be understood as follows. The extremized solution is given by the sum of two boundaries in eq.(\ref{starteq2}) with the relation $m_{2}=r_{2,1} m_{1}$. The second section starts when $m_1$ reaches $1$, and at that point we have $m_2=r_{2,1}m_1=r_{2,1}$.

\begin{figure}[H] 
   \centering
    \includegraphics[height=2.0in]{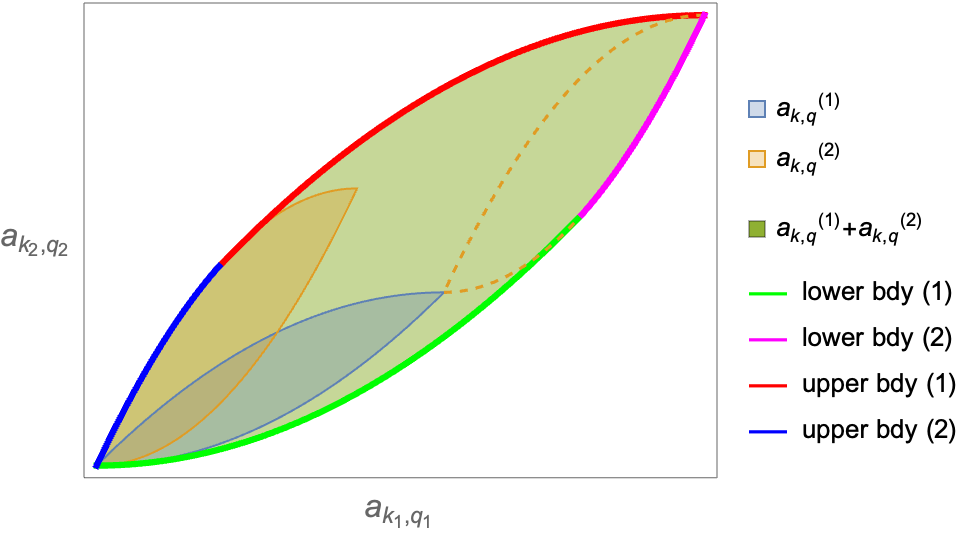} 
   \caption{The structure of a Minkowski sum of two $L$-moments. The lower and upper boundaries are each composed of two sections, corresponding to eqs.(\ref{lbdya}) and (\ref{ubdya}).}
   \label{fig:plotM0}
\end{figure}

Next we can perform the same analysis for the upper boundary. In this case the region spanned by the sum of the two upper boundaries is 
\eq
\overline{a}_{k,q}(m_1,m_2)=\frac{L}{k}\left(v_{\ell_1,k_,q_1}(1-m_1^{k_1})+v_{\ell_2,k_,q_1}(1-m_2^{k_1})\right)\,,
\eqe
and we find
\eqa\label{ubdya}
\nonumber \textrm{upper bdy 1: }&& \overline{a}_{k,q}(m,r_{2,1}m),\quad m\in[0,1]\\
\textrm{upper bdy 2: }&& \overline{a}_{k,q}(1,m),\quad m\in[r_{2,1},1]\,.
\eqae
In conclusion, we find the region given by the sum of two moments in Figure \ref{fig:plotM0}. Note that as before, the upper boundary is simply a flipped lower boundary, with endpoints matching the (opposite) endpoints of the lower boundary.

For $N$ moments, the above result generalizes by induction. We have the lower boundary contained in the Minkowski sum
\eq
a_{k,q}=\frac{L}{k}\sum_{i=1}^N v_{\ell_i,k,q} m_i^{k}\,.
\eqe
Repeating the arguments from above, we find the lower boundary of a 2D space $(a_{k_1,q_1},a_{k_2,q_2})$ is composed of $N$ sections, where for a section $(j)$, with $1\le j\le N$
\eqa\label{solfinal}
\nonumber i<j&:&\quad m_i=1\,,\\
i\ge j&:&\quad m_i=r_{i,j}m,\quad m\in[r_{j,j-1},1]\,.
\eqae
A section $(j)$ corresponds to the curve
 \eq\label{eqbdrj}
a_{k,q}^{(j)}=\frac{L}{k}\left(G_{k,q}^{(j-1)}+m^{k}v_{\ell_j,k,q} F_{k,q}^{(j)}\right),\quad m\in[r_{j,j-1},1]\,,
\eqe
where $G$ and   $F$ are purely numerical factors that depend only on the spin content and the $(k_1,k_2,q_1,q_2)$ of the space being considered
\eqa\label{FF}
G_{k,q}^{(j)}=\sum_{i=1}^{j}v_{\ell_i,k,q}\,,\quad F_{k,q}^{(j)}=\sum_{i=j}^{N} \frac{v_{\ell_i,k,q}}{v_{\ell_j,k,q}} r_{i,j}^{k}\,.
\eqae
Note that $r_{i,j}$ depends on $(k_1, q_1, k_2, q_2)$, so in fact we have $F^{(j)}_{k_1,q_1}=F^{(j)}_{k_2,q_2}$.

Finally, we also have the upper boundary, which is simply obtained by plugging in the same solution in eq.(\ref{solfinal}) in the sum of individual upper boundaries
\eq
a_{k,q}=\frac{L}{k}\sum_{i=1}^N v_{\ell_i,k,q} \left(1-m_i^{k}\right)\,.
\eqe

This leads to an expression for section $(j)$ of the upper boundary
 \eq\label{eqbdrju}
\overline{a}_{k,q}^{(j)}=\frac{L}{k}\left(\overline{G}_{k,q}^{(j)}-m^{k}v_{\ell_j,k,q} F_{k,q}^{(j)}\right),\quad m\in[r_{j,j-1},1]\,,
\eqe
where $\overline{G}_{k,q}^{(j)}=\sum_{i=j}^N v_{\ell,k,q}$. 

In some situations it will be helpful to rewrite the upper boundary such that counting the sections starts from the origin. If we instead order the spins according to decreasing slope, then we can  rewrite eq.(\ref{eqbdrju}) as
\eq\label{eqbdrju2}
\overline{a}_{k,q}^{(j)}=\frac{L}{k}\left(G_{k,q}^{(j)}-m^k v_{j,k} \overline{F}_{k,q}^{(j)}\right)\,,
\eqe
where
\eq
\overline{F}^{(j)}=\sum_{i=1}^{j} \frac{v_{i,k}}{v_{j,k}} r_{i,j}^k\,.
\eqe

\paragraph{Convergence for infinite spin limit}
In general a physical theory will contain an infinite number of spins, so we must take the Minkowski sum to infinity, $N\rightarrow \infty$. One then needs to worry if the sum in $F_{k,q}$ given by eq.(\ref{FF}) converges. If the sum diverges, then we only get a trivial projective boundary, such as $a_{k_2}\ge 0$. Convergence will therefore be necessary to obtain non projective bounds.

\begin{figure}[H] 
   \centering
    \includegraphics[height=2.6in]{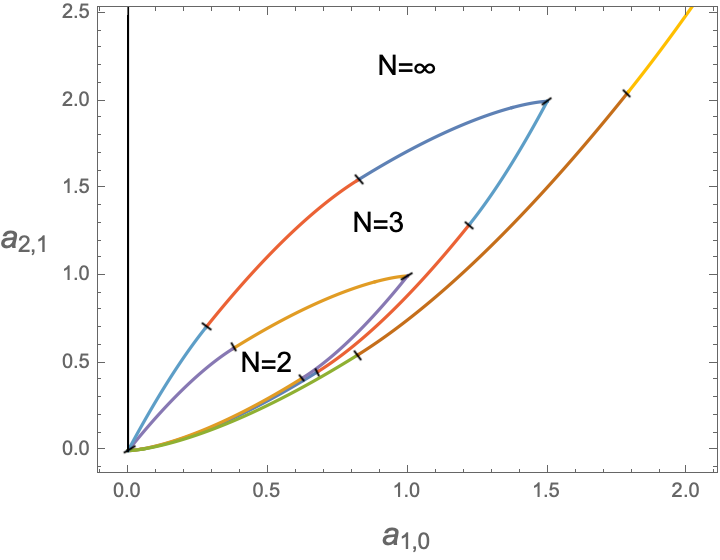} 
   \caption{Boundaries for $(a_{2,0},a_{3,1})$ space, for $N=2,3,\infty$, with $L=1$.   For $N$ spins the lower and upper boundaries each have $N$ sections or facets. In this example, in the infinite spin limit the upper boundary is pushed to infinity, and only the lower boundary remains non-trivial. The green and brown sections of the lower boundary correspond to  eqs.(\ref{sect1}) and (\ref{sectb}) respectively.}
   \label{fig:ploteA3c}
\end{figure}

Assuming a simple case with $v_{\ell,k,q}=\ell^q$, the convergence is easy to determine. We have
\eq
r_{i,1}=\left(\frac{1}{\ell}\right)^\frac{q_2-q_1}{k_2-k_1},\quad  F_{k,q}^{(1)}=\sum_{\ell=1}^{\infty} \ell^{\frac{q_1{-}q_2}{k_2{-}k_1}k{+}q}=\zeta\left(\frac{q_2{-}q_1}{k_2{-}k_1}k{-}q\right)\,,
\eqe
which converges for 
\eq
\frac{q_2{-}q_1}{k_2{-}k_1}k{-}q>1\,.
\eqe
For example, for the space $(a_{1,0},a_{2,0})$ we find $r_{i,1}=1$ and the sum diverges
\eq
F_{1,0}^{(1)}=\sum_{\ell=1}^\infty 1=\infty\,.
\eqe
This means there are no non-projective bounds for this 2D space for $v_{\ell,k,q}=\ell^q$. However, for the space $(a_{2,0},a_{3,1})$ we instead get $r_{i,1}=\ell^{-1}$ and the sum is finite
\eq
F_{2,0}^{(1)}=\sum_{\ell=1}^\infty \frac{1}{\ell^2}=\frac{\pi^2}{6}\,.
\eqe
Plugging this in eq.(\ref{eqbdrj}), we obtain the first section of the infinite spin configuration as:
\eq\label{sect1}
(1):\quad a_{2,0}=L\frac{\pi^2}{12}m^2,\quad a_{3,1}=L\frac{\pi^2}{18}m^3, \quad m\in[0,1]\,.
\eqe
The second section can similarly be obtained, using $F_{2,0}^{(2)}{=}4(F_{2,0}^{(1)}{-}1)$
\eq\label{sectb}
(2):\quad a_{2,0}=\frac{L}{2}\left(1+4\left(\frac{\pi ^2}{6}-1\right) m^2\right),\quad a_{3,1}=\frac{L}{3}\left(1+8 \left(\frac{\pi ^2}{6}-1\right) m^3\right),\quad m\in[\frac{1}{2},1]\,.
\eqe
We plot this in Figure \ref{fig:ploteA3c}, also comparing with finite spin. Note that in this example the upper boundary given by eq.(\ref{eqbdrju2}) is a vertical line starting at the origin, since the first section of the upper boundary contains $v_{\ell_N,k,q}\sim N^q$, so has it has infinite slope. Generically this need not be true, as it is possible for the maximum slope among all spins to be finite, even as we take $N\rightarrow \infty$.

\paragraph{Non-positive $v_{\ell,k,q}$}

As mentioned previously when taking the Minkowski sum, if the $v_{\ell,k,q}$s have non-uniform signs, then special care is needed to identify the correct combination of upper/lower boundaries of the individual moments to extremize. For simplicity we consider the example where we have three moments with $v_{\ell, k_1,q_1}>0$ and $\textrm{sgn}[v_{\ell, k_2, q_2}]=\{-,0,+\}$, where $\textrm{sgn}[x]{=}0$ means $x{=}0$. We will refer to such moments as type $(+,-)$, $(+,0)$ and $(+,+)$ respectively. This particular case will be relevant in the physical problem. Other cases can be treated in a similar fashion.

\begin{figure}[H] 
   \centering
    \includegraphics[height=2.4in]{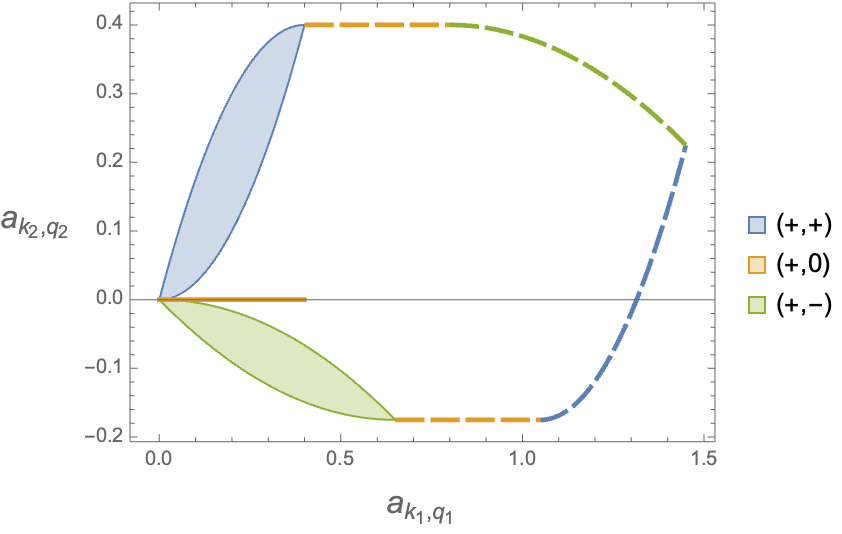} 
   \caption{The Minkowski sum of three distinct rescaled moments $a_{k,q}^{(\ell)}$, with $v_{\ell, k_1,q_1}>0$ and $\textrm{sgn}[v_{\ell, k_2, q_2}]=\{-,0,+\}$. The boundary of the sum  is obtained by placing the individual boundaries according to their slope. }
   \label{fig:plotM67}
\end{figure}

When $v_{\ell,k_2,q_2}<0$, this corresponds to the moment $a_{k,q}^{(\ell)}=(v_{\ell,k_1,q_1}a_{k_1},|v_{\ell,k_2,q_2}|a_{k_2})$ reflected about the ($a_{k_1, q_1}$) (horizontal) axis. When $v_{\ell,k_2,q_2}=0$, the moment is given by $a_{k,q}^{(\ell)}=(v_{\ell,k_1,q_1}a_{k_1},0)$, so it is just horizontal segment of length $\frac{L}{k_1}v_{\ell,k_1,q_1}$.  These are illustrated in Figure \ref{fig:plotM67}.
The three different ``lower boundaries" for each independent moment are given by
\eq
\begin{array}{ll}
\displaystyle a_{k,q}^{(+-)}(m)=Lv_{\ell_-,k,q}\frac{(1-m^{k})}{k}, & m\in[0,1]\,,\\
 \displaystyle a_{k,q}^{(+0)}(m)=L v_{\ell_0,k,q} \frac{m^k}{k},& m\in[0,1]\,,\\
\displaystyle a_{k,q}^{(++)}(m)=Lv_{\ell_+,k,q}\frac{m^{k}}{k},&m\in[0,1]\,.
\end{array}
\eqe
In the first case for $(+-)$, due to the fact that $v_{\ell, k_2, q_2}<0$, the lower boundary is the upper boundary of the original moment. Furthermore, the origin now corresponds to $m=1$, while its other endpoint to $m=0$. For the second case with $v_{\ell, k_2, q_2}=0$ the moment itself is a $1D$ segment, and we use the original lower boundary to parameterize it. Their Minkowski sum can be written as:
\eq\label{+-}
a_{k,q}(m_{+-},m_{+0},m_{++})=a_{k,q}^{(+-)}(m_{+-})+a_{k,q}^{(+0)}(m_{+0})+a_{k,q}^{(++)}(m_{++})\,.
\eqe
Since the slope of each boundary has no overlap, one does not need to perform extremization and the boundary of the Minkowski sum is simply given by placing the three boundaries in order, as illustrated in Figure \ref{fig:plotM67}. This means the lower boundary is given by
\eqa\label{+-b}
\textrm{Section 1: }&&\nonumber a_{k,q}(m,0,0),\quad m\in[0,1]\,,\\
\textrm{Section 2: }&&\nonumber a_{k,q}(0,m,0),\quad m\in[0,1]\,,\\
\textrm{Section 3: }&& a_{k,q}(0,1,m), \quad m\in[0,1]\,.
\eqae
 The upper boundary can be obtained in an identical manner. If we had several moments of type $(+,+)$, their respective boundaries would again have overlap in slope, and the result would need to be obtained by extremizing in the ranges of slopes where there is overlap. Then the boundary would be given by eq.(\ref{eqbdrj}) we previously derived, replacing $a_{k,q}^{(++)}$ in the above expressions.

\subsection{Higher dimensional spaces}\label{sect33}
\begin{figure}[H] 
   \centering
   \includegraphics[height=2.1in]{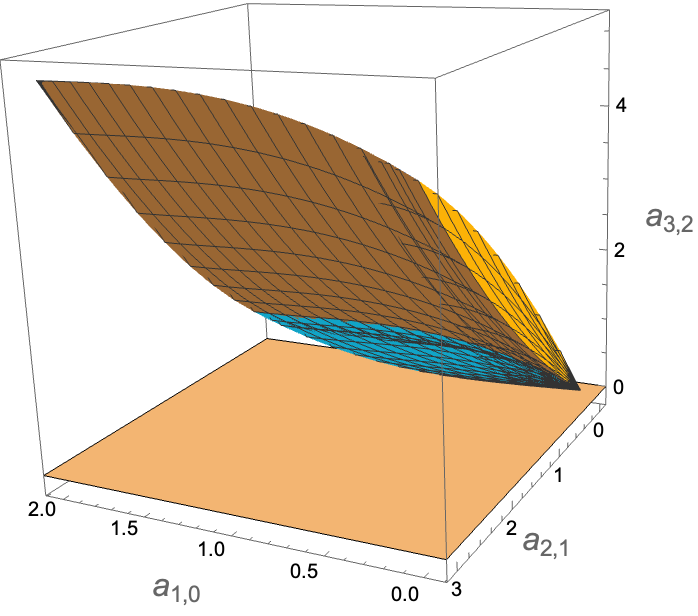} 
      \includegraphics[height=2.4in]{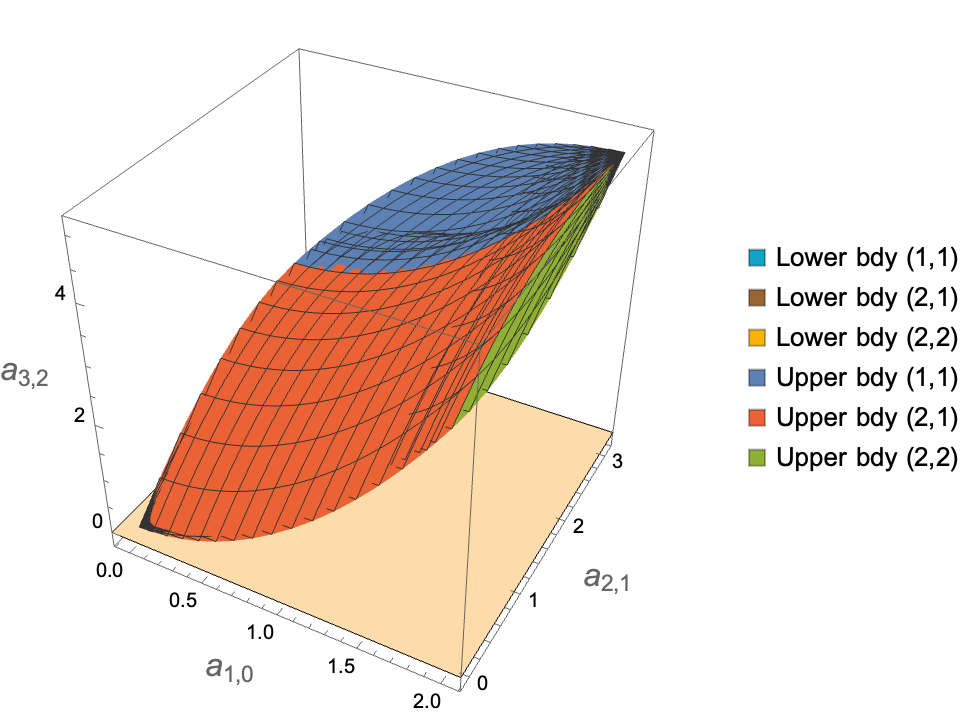} 
   \caption{Boundaries for $(a_{1,0},a_{2,1},a_{3,2}$) space, assuming two spins $\ell_1=1$, $\ell_2=2$, $v_{\ell,k,q}=\ell^q$ and $L=1$}
   \label{fig:3dex}
\end{figure}

Let us now attempt to solve the same problem in higher dimensions. As in 2D, for $v_{\ell,k,q}>0$, in higher dimensions the lower/upper boundary of the sum will be contained in the sum of lower/upper boundaries of rescaled individual moments $a_{k,q}^{(\ell)}$. Consider a $3D$ space $(a_{k_1,q_1}, a_{k_2,q_2}, a_{k_3,q_3})$ and two spins for simplicity. The Minkowski sum of the lower boundaries in eq.(\ref{bdrg}) is given by
\eqa\label{gendist}
a_{k,q}(m_1,n_1,m_2,n_2)&=&\frac{L}{k}\left( v_{\ell,k}\left(n_1^k-m_1^k\right)+ v_{\ell_2,k}\left(n_2^k-m_2^k\right)\right)\,.
\eqae
for $0\le m_i\le n_i\le 1$. We must compute the boundary of the Minkowski sum of these boundaries, which is again easily accomplished by the Lagrange method.
Unlike the 2D case, in general the solution to the Lagrange equation is more complicated. Even for the simplest 3D example $(a_{1,1},a_{2,2},a_{3,3})$ we find
\eqa\label{ugl}
\left.\begin{array}{c}m_2, \\n_2\end{array}\right.&=&\frac{\left(m_1{+}n_1\right) v_{\ell_1,1} v_{\ell_1,3} v_{\ell_2,2}{\mp}\sqrt{v_{\ell_1,1} v_{\ell_1,3} \left(\left(m_1{+}n_1\right){}^2 v_{\ell_1,1} v_{\ell_1,3} v_{\ell_2,2}^2{-}4 m_1 n_1 v_{\ell_1,2}^2 v_{\ell_2,1} v_{\ell_2,3}\right)}}{2 v_{\ell_1,1} v_{\ell_1,2} v_{\ell_2,3}}\,,\nonumber\\
\eqae
where for simplicity we defined $v_{\ell,k}=v_{\ell,k,q}$.   As before, this represents just one section of the 2D boundary. To obtain the others we have to set the various parameters to either 0 or 1, and invoke continuity of the boundary to arrange them in the correct position. To illustrate a simpler example of a 3D geometry, let us consider a particular case, when $v_{\ell,k,q}=\ell^q$, $\ell_1=1$ and $\ell_2=2$, and a particular space $(a_{1,0},a_{2,1},a_{3,2})$. In this case eq.(\ref{ugl}) drastically simplifies, and we find 
\eq\label{simp}
\frac{n_2}{n_1}=\frac{m_2}{m_1}=\frac{\ell_1}{\ell_2}=\frac{1}{2}\,.
\eqe
Checking for the remaining boundaries when $m_i,n_i=\{0,1\}$, we find a total of three sections for the lower boundary, given by
\eqa
\begin{array}{lll}
\nonumber \text{Section $(1,1)$ :}&\quad a_{k,q}(m_1,n_1,\frac{1}{2}m_1,\frac{1}{2}n_1),\quad 0\le m_1\le n_1\le 1\\
\nonumber \text{Section $(2,1)$ :}& \quad a_{k,q}(1,n_1,m_2,\frac{1}{2}n_1),\quad 1/2\le m_2\le 1,\ 0\le n_1\le 1\\
\text{Section $(2,2)$ :}&\quad  a_{k,q}(1,1  ,m_2,n_2),\quad 1/2\le m_2\le n_2\le 1
\end{array}
\eqae
and similar for the upper boundary. We illustrate the result in Figure \ref{fig:3dex}.

While the particular choices for the space and the $v_{\ell,k,q}$ lead to the simple solutions from above, more general configurations lead to similar boundary structures.

However, in general  cases, where we also have an infinite number of spins, the parameters $m_1,n_1$ cannot be separated from the sum over spins due to the square root in eq.(\ref{ugl}), and so we cannot obtain a closed form expression for the complete boundary. However, this approach is still useful in situations when relevant boundaries only contain a finite number of spins. We will encounter such a case in our physical problem in the following section.

This is not the end of the story for higher dimensions, since quite remarkably, we can obtain results similar to eq.(\ref{simp}) even for slightly more general cases. Specifically, for $v_{\ell,k,q}=g(\ell)f(\ell)^q$, where $g$ and $f$ are any polynomials in $\ell$, and for any $D$-dimensional space of the form $k_{i+1}{-}k_i=\Delta_k$ and $q_{i+1}{-}q_i=\Delta_q$, with $(\Delta_k,\Delta_q)$ fixed, we obtain
\eq
\frac{n_2}{n_1}=\frac{m_2}{m_1}=\left(\frac{f(\ell_1)}{f(\ell_2)}\right)^\frac{\Delta_q}{\Delta_k}=r_{2,1}\,.
\eqe
In this case the sum over infinite spins can again be computed in closed form. The freedom in choosing $f$ will also allow to extend the result beyond such diagonal type spaces. We discuss this further in Appendix \ref{appgend} as a promising future direction. 

 \section{The non-projective EFThedron}\label{sec4}
 We now return to the physical setting. Let us begin with the dispersive representation of the EFT couplings eq.(\ref{ref1})
 \eq
g_{k,q}=\sum_{\ell} v_{\ell,k,q}  \int_0^1 \rho_\ell(z) z^{k{-}1}  dz, \quad 0\le \rho_\ell(z)\le2\,,
\eqe
where $v_{\ell,k,q}=16(2\ell{+}1)\lambda_{\ell,k,q}$.  
 
Couplings are subjected to ``null" constraints, which originate from permutation invariance of the amplitude. These constraints can be easily found by requiring the amplitude to be expandable in terms of $\sigma_2=s^2{+}t^2{+}u^2$ and $\sigma_3=s t u$. Since the dispersion relation is valid for $k,k{-}q\ge 2$, this means the first null constraint is at order $k=4$, so we must have that $\sigma_2^2{=}4(s^4{+}2s^3 t{+}3s^2 t^2)\propto s^4 g_{4,0}{+}s^3t g_{4,1}{+}s^2 t^2 g_{4,2}{+}\ldots$, leading to 
\eq\label{nullk4}
g_{4,0}=\frac{g_{4,1}}{2}=\frac{g_{4,2}}{3}\,.
\eqe
Since $\lambda_{\ell,4,0}=\frac{\lambda_{\ell,4,1}}{2}$ from eq.(\ref{lambdadef}), the first equality trivially holds. However, $\lambda_{\ell,4,2}=6 - 4 \mathcal{J}^2 + \mathcal{J}^4/2$, where $\mathcal{J}^2=\ell(\ell+1)$, so the second equality leads to the null constraint 
\eq
n_4: 2g_{4,2}-3g_{4,1}=0\,.
\eqe 
Similarly for $k=5$ and $6$, we have:
\eq
n_5: g_{5,3}{-}2g_{5,1}=0, \quad n_6:  g_{6,4}{-}g_{6,2}=0\,.
\eqe
Geometrically these null constraints define  hyperplanes in the space of couplings. Thus imposing a set of null constraints is equivalent to intersecting the space of couplings with a collection of hyperplanes.

We will first consider the equal $k$ space, i.e. $(g_{k,q_1}, g_{k,q_2}, \cdots)$ which we showed in Section \ref{sec2} is a polytope. Its intersection with the null plane at the same order $k$ will then give bounds on coefficients. More generally, to impose null conditions of different $k$ order, it will be useful to consider the linear combinations of couplings that correspond to null constraints as one of the coordinates in our space. For example, we define
\eqa\label{null1}
n_4&=&\sum_{\ell} (2v_{\ell,4,2}{-}3v_{\ell,4,1}) \int_0^1 \rho_\ell(z)   z^{3} dz \nonumber\\
\nonumber &=&\sum_{\ell}16(2\ell+1) (\mathcal{J}^4-8\mathcal{J}^2) \int_0^1 \rho_\ell(z)   z^{3} dz\\
&=&\sum_{\ell} u_{\ell,4}\int_0^1\rho_\ell(z)z^{3}dz\,.
\eqae
 Then to impose the null constraint on a coupling $g_{k,q}$ we simply consider the geometry in $(g_{k,q},n_4)$ space, and intersect with the line $n_4{=}0$. Importantly, since we have derived the boundary for any 2D space in Section \ref{sec3}, this approach indeed allows us to find optimal bounds on any single coupling assuming one null plane.  We also discuss partial results for a space of two couplings subject to one null constraint.

\subsection{Bounding $g_{k,q}$ with $k$-null constraint}\label{eqk}
The space of equal $k$ couplings is a polytope described by the determinant constraints in eq.(\ref{gendimpol}). We can then easily intersect any space at order $k$ by the corresponding null constraints at the same order. This result is of course weaker than imposing for example the first null constraint at $k=4$, but we find it a useful exercise.

\begin{figure}[H] 
   \centering
   \includegraphics[width=4in]{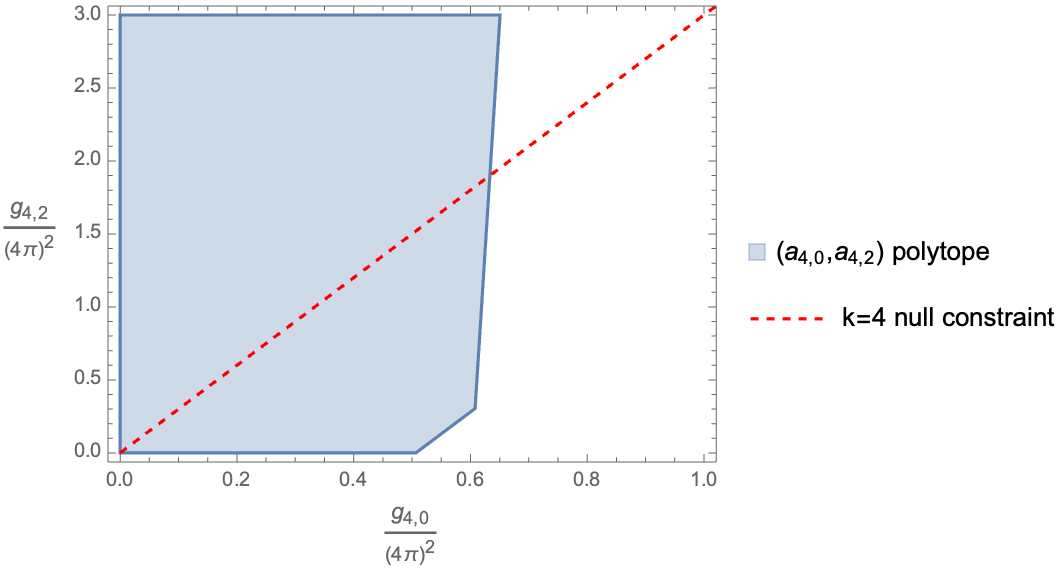} 
   \caption{$g_{4,0}, g_{4,2}$  space with $2g_{4,2}-3g_{4,1}$ null plane}
   \label{fig:plotz1}
\end{figure}

\paragraph{$k=4$ bounds}
As the first null constraint is at $k=4$, we consider the space $(g_{4,0}, g_{4,1}, g_{4,2})$ subjected to the null constraint $n_4=2g_{4,2}-3g_{4,1}=0$. We find explicitly
\eq\label{k4lam}
\lambda_{\ell,4,0}=2,\quad \lambda_{\ell,4,1}=4,\quad \lambda_{\ell,4,2}=6 - 4 \mathcal{J}^2 + \mathcal{J}^4/2\,.\quad 
\eqe
This implies $2g_{4,0}=g_{4,1}$ identically, so the space for the couplings $(g_{4,0}, g_{4,1}, g_{4,2})$ is in fact just a 2D polytope. We can then find the boundaries of this polytope in  $(g_{4,0},g_{4,2})$ space. To satisfy cyclic ordering of vectors ${\bf v}_\ell=(v_{\ell,4,0},v_{\ell,4,2})$, the spins must be ordered as $\{\ell_1,\ell_2,\ell_3,\ldots\}=\{2,0,4,\ldots\}$. Now we can impose the polytope conditions
\eq
P_i=\begin{pmatrix} g_{4,0}-V_{1,i |4,0}& v_{\ell_i|4,0}\\ g_{4,2}-V_{1,i |4,2}& v_{\ell_i|4,2}
\end{pmatrix}\ge 0\,.
\eqe
We do not need to impose the upper boundary conditions $\overline{P}_i\ge 0$, since they always contain a divergent infinite sum, leading to a trivial boundary. Going up to only $i=3$  we obtain the polytope in Figure \ref{fig:plotz1}, which is sufficient to intersect with the $k=4$ null plane. 
We find that $g_{4,0}$ is bounded as:
\eq\label{fing40}
0\le  \frac{{g}_{4,0}}{(4\pi)^2} \lessapprox\frac{0.633}{M^8}\,.
\eqe

\paragraph{$k=5$ bounds}
At $k=5$ the space is similarly simple. We find 
\begin{align}
\lambda_{\ell,5,0}=0,\quad \lambda_{\ell,5,1}=\frac{1}{2}\lambda_{\ell,5,2}= {-}5 {+} 2 \mathcal{J}^2, \quad \lambda_{\ell,5,3}=\frac{1}{36} \left({-}360 {+} 294 \mathcal{J}^2 {-} 43 \mathcal{J}^4 {+} 2 \mathcal{J}^6\right)\,.
\end{align}
Since there are just two independent couplings, the relevant space is again a 2D polytope. However, these vectors span more than a half plane. The vector corresponding to $\ell{=}0$ is ${\bf v}_0{=}({-}5,{-}10)$, so is located in the 3rd quadrant. As discussed in subsection \ref{noncyc}, one can simply compute the Minkowski sum by taking ${\bf v}_0\rightarrow -{\bf v}_0$, order the spins and impose the polytope conditions using this new vector, and finally translate the boundary thus obtained by ${\bf v}_0$. Intersecting with the null plane $g_{5,3}-2g_{5,1}=0$ we obtain a bound
\eq
- \frac{0.202}{M^{10}} \lessapprox \frac{{g}_{5,1}}{(4\pi)^2} \lessapprox \frac{2.269}{M^{10}}\,.
 \eqe
\paragraph{$k=6$ bounds}
At $k=6$ the $\lambda$ are given by
\eqa\label{vk6}
\nonumber \lambda_{\ell,6,0}&=&\frac{1}{3}\lambda_{\ell,6,1}=2, \quad \lambda_{\ell,6,2}=15 - 6 \mathcal{J}^2 + \mathcal{J}^4/2,\quad  \lambda_{\ell,6,3}=-5\lambda_{\ell,6,0}+2\lambda_{\ell,6,2},\\
\lambda_{\ell,6,4}&=& \frac{1}{288} (4320{-}4176 \mathcal{J}^2 + 732 \mathcal{J}^4 {-} 44 \mathcal{J}^6 {+} \mathcal{J}^8)\,.
\eqae
There are just three independent couplings, so again we can choose a subspace, for example spanned by $(g_{,6,0}, g_{6,2}, g_{6,4})$. However, in this space the vectors cannot be cyclically ordered. One can tell this from the determinant $
\textrm{det}\left({\bf v}_{\ell_i}{\bf v}_{\ell_j}{\bf v}_{\ell_k}\right)$
which is not sign definite for ordered $\ell$. 

\begin{figure}[H] 
   \centering
   \includegraphics[width=4.3in]{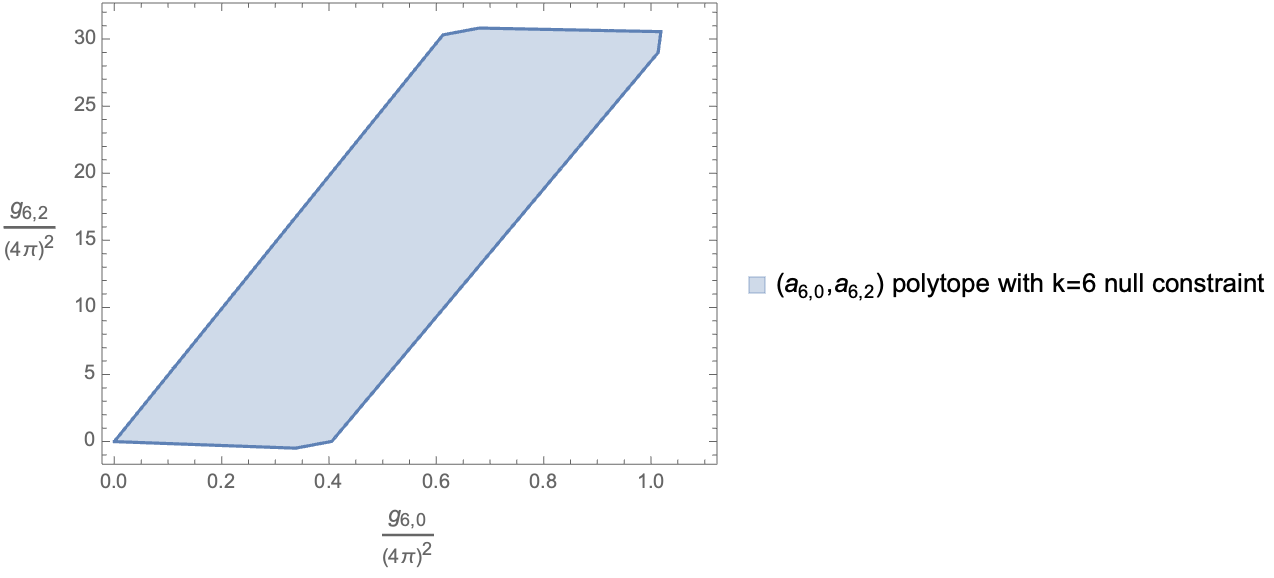} 
   \caption{$(g_{6,0}, g_{6,2})$ space, after imposing the $g_{6,4}=g_{6,2}$ null constraint on the 3D space $(g_{6,0},g_{6,2},g_{6,4})$.}
   \label{fig:plotz2}
\end{figure}

To proceed we simply first have to impose the polytope constraints $P_{i_1,i_2}$ and $\overline{P}_{i_1,i_2}$ in eq.(\ref{gendimpol}), using the general definition for $\mathcal{I}^{\pm}$ according to eq.(\ref{Idef}), as discussed at the end of Section \ref{sec22}. Since we are interested in the space resulting from intersecting with a null plane, this only needs to be done for a finite number of boundaries, ie. pairs $i_1,i_2$. Unlike the previous cases in 2D, constraints that involve the infinite sum over spins do not automatically lead  to trivial conditions. What happens instead is that the presence of infinity in the determinant of the $3\times 3$ matrix requires the positivity of the corresponding $2\times2$  minor. This simply implies that we must also impose the 2D constraints $P_i$ and $\overline{P}_i$  in order to obtain the complete 3D space. Imposing all relevant constraints, and intersecting with the $k=6$ null constraint $g_{6,4}=g_{6,2}$, we obtain  the space for $(g_{6,0},g_{6,2})$ in Figure \ref{fig:plotz2}, with  individual bounds:
\eq\label{fing60}
0\le \frac{g_{6,0}}{(4\pi)^2}\lessapprox \frac{1.01}{M^{12}}, \quad -\frac{0.49}{M^{12}}\lessapprox \frac{g_{6,2}}{(4\pi)^2}\lessapprox \frac{30.81}{M^{12}}\,.
\eqe
\subsection{Bounding $g_{k_1,q}$ with $k{\neq}k_1$-null constraint   }\label{subsect42}
We would now like to consider the effect of null constraints at some order $k_2$ on couplings $g_{k_1,q}$ where $k_2{\neq}k_1$. Since we are interested in the bound on one coupling, we will utilize the 2D geometry discussed in Section \ref{sec2d2}. That is, we will consider the two dimensional $(g_{k_1,q}, n_{k_2})$ space, and simply consider the 1D slice at $ n_{k_2}=0$. This allows us to derive analytic bounds for any coupling assuming just one null constraint. Since from experience, when considered separately, the strongest null constraint is $k=4$, we will impose this one in all further computations.

\subsubsection{Bounds on $g_{2,0}$, $g_{3,1}$ with $k{=}4$-null constraint }
\begin{figure}[H]
     \centering
     \begin{subfigure}[b]{0.35\textwidth}
         \centering
\includegraphics[width=1\textwidth]{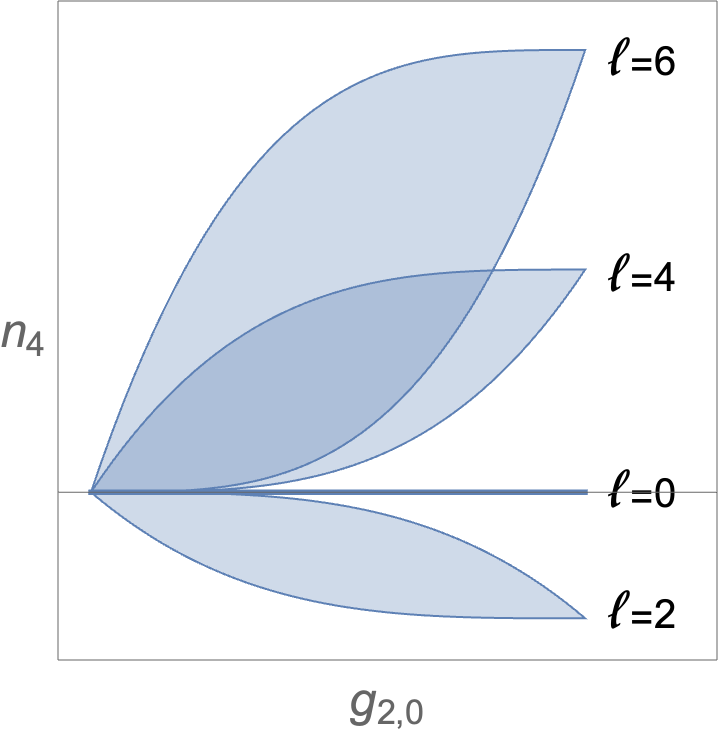}
\caption{The rescaled moments $a_{k,q}^{(\ell)}$ corresponding to spins $0$ to $6$.}
\label{fig:pa1}
\end{subfigure}
\hfill
  \begin{subfigure}[b]{0.6\textwidth}
         \centering
 \includegraphics[width=\textwidth]{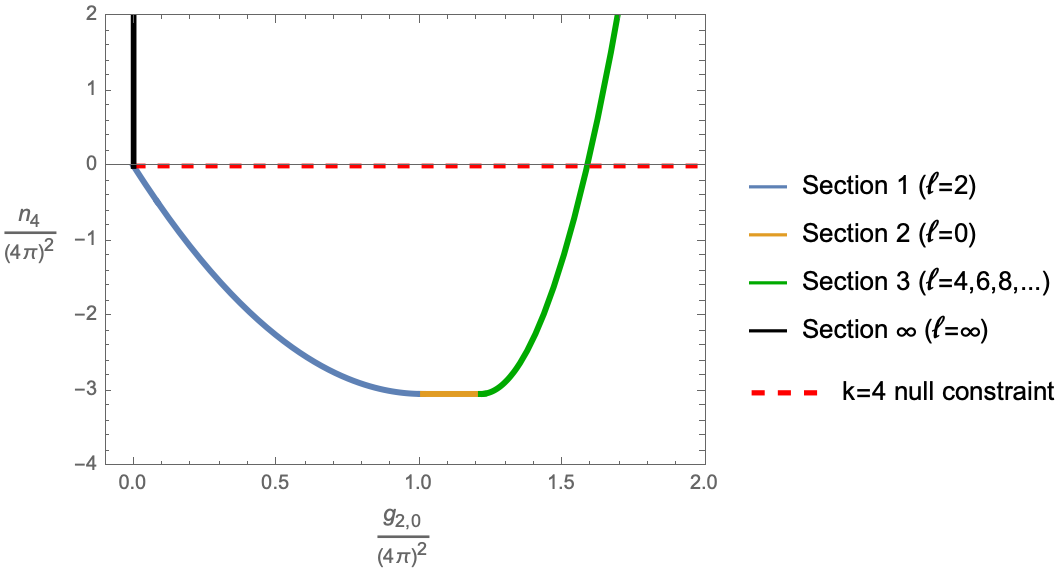}
 \caption{$(g_{2,0},n_4)$ space with four sections of the boundary shown. The upper bound on $g_{2,0}$ is follows from intersecting with the null plane.}
\label{fig:plot11a}
 \end{subfigure}
 \caption{Minkowski sums of single $L$-moments}
 \end{figure}

\paragraph{1D bounds on $g_{2,0}$}We will first obtain the necessary and sufficient conditions on $g_{2,0}$ under the $n_4$ constraint from the space ($g_{2,0}, n_4$), defined by
\eqa
\nonumber g_{2,0}&=&\sum_\ell v_{\ell,2,0}\int_0^1 \rho_\ell z dz\,,\\
n_{4}&=&\sum_\ell u_{\ell,4}\int_0^1 \rho_\ell z^3 dz\,,
\eqae
with $v_{\ell,2,0}=32(2\ell+1)$ and $u_{\ell,4}=16(2\ell+1)(\mathcal{J}^4-8\mathcal{J}^2)$, as defined in eq.(\ref{null1}). For this space the vectors $\mathbf{v}_\ell\equiv(v_{\ell,2,0}, u_{\ell,4})$ are ordered by $\{\ell_1,\ell_2,\ell_3,\ell_4,\ldots\}=\{2,0,4,6,\ldots\}$, with the respective signs of the two components given by: $(+,-)$ for $\ell_1$, $(+,0)$ for $\ell_2$, and $(+,+)$ for the remaining $\ell_i$. The individual moments are sketched in Figure \ref{fig:pa1}. 
We must compute the Minkowski sum over all such moments. The lower boundary of the space $(g_{2,0},n_4)$ will be contained in the Minkowski sum of lower boundaries for all spins, 
\eqa
\nonumber g_{2,0}(m_1,m_2,m_{i})&=&\frac{L}{2}\left( v_{\ell_1,2,0}(1-m_1^2)+v_{\ell_2,2,0} m_{2}^2+\sum_{i=3}^\infty v_{\ell_i,2,0}m_{i}^2\right)\,,\\
n_4(m_1,m_2,m_{i})&=&\frac{L}{4}\left( u_{\ell_1,4}(1-m_1^4)+u_{\ell_2,4} m_{2}^4+\sum_{i=3}^\infty u_{\ell_i,4}m_{i}^4\right)\,.
\eqae
We will now build the explicit form of this boundary. According to our previous discussion, we must arrange the boundaries such that the slope along the complete boundary is monotonically increasing or decreasing. The first section, starting at the origin, is given only by the boundary of the $\ell_1$ moment, since it has negative slope
\eqa\label{solu1}
\textrm{Section (1):} \quad  \nonumber g_{2,0}(m,0,0)&=&\frac{L}{2}v_{\ell_1,2,0}(1-m^2)={160}(1-m^2) \,,\\
n_4(m,0,0)&=&\frac{L}{4} u_{\ell_1,4}(1-m^4)=-{960}(1-m^4)\,,
\eqae
valid for $m$ ranging from 1, at the origin, to 0, at the end-point of this section. In the second section only $\ell_2$ contributes, and  for the boundary to be continuous we must have
\eqa\label{solu2}
\textrm{Section (2):}\quad \nonumber g_{2,0}(0,m,0)&=&\frac{L}{2}v_{\ell_1,2,0}+\frac{L}{2} v_{\ell_2,2,0}m_2^2 =32(5+m^2)\,, \\
n_4(0,m,0)&=&\frac{L}{4}  u_{\ell_1,4}+\frac{L}{4} u_{\ell_2,4}m^4=-{960}\,,
\eqae
valid for $m\in[0,1]$. Next we have the third section, where all spins $\ell_{i\ge 3}$ have positive slope and can contribute. Using our previous result eq.(\ref{eqbdrj}) obtained by extremizing with the Lagrange method, we know that in this section the parameters for all contributing moments will be related by $m_i=r_{i,3}m_3$, for $i\ge 3$, where 
\eq
r_{i,3}=\left(\frac{u_{\ell_3,4}}{v_{\ell_3,2,0}}\frac{v_{\ell_i,2,0}}{u_{\ell_i,4}}\right)^\frac{1}{2}=\left(\frac{240}{\mathcal{J}^4-8\mathcal{J}^2}\right)^\frac{1}{2}\,.
\eqe
So for this section we obtain 
\begin{align}\label{solu3}
\textrm{Section (3):}\quad \nonumber g_{2,0}(0,1,r_{i,3}m)&=\frac{L}{2}v_{\ell_1,2,0}+\frac{L}{2}v_{\ell_2,2,0}+\frac{L}{2}v_{\ell_3,2,0} Fm^2 =96(2+3 F m^2)\,, \\
n_4(0,1,r_{i,3}m)&=\frac{L}{4}u_{\ell_1,4}+\frac{L}{4}u_{\ell_2,4}+ \frac{L}{4}u_{\ell_3,4} Fm^4=960(-1+36  F m^4)\,,
\end{align}
valid for $m\in[0,1]$, and where
\eqa
F&=&\sum_{i\ge 3}^\infty \frac{v_{\ell_i,2,0}}{v_{\ell_3,2,0}}r_{i,3}^2=\frac{80}{3}\sum_{\ell\ge4,even}^\infty \frac{2\ell+1}{(\ell^2+\ell)^2-8(\ell^2+\ell)}\sim1.49\,.
\eqae
Finally, starting from the origin in the opposite direction, the first section would be given by the upper boundary of the $\ell=\infty$ moment, which is simply a vertical line. This simply implies $g_{2,0}\ge 0$.  We obtain the boundary shown in Figure \ref{fig:plot11a}, with the sections of the lower boundary corresponding to the parametric curves given by eqs.(\ref{solu1}), (\ref{solu2}), and (\ref{solu3}). 

We can now check that $n_4=0$ intersects the third section. This gives an optimal upper bound on $g_{2,0}$
\eq\label{fing20}
\frac{g_{2,0}}{(4\pi)^2}\le \frac{48 (4 + \sqrt{F})}{(4\pi)^2M^4}\approx  \frac{1.58}{M^4}\,,
\eqe
We find this analytically matches the result of \cite{Caron-Huot:2020cmc} for $g_2=g_{2,0}/2$.

\begin{figure}[H] 
   \centering
      \includegraphics[height=2.2in]{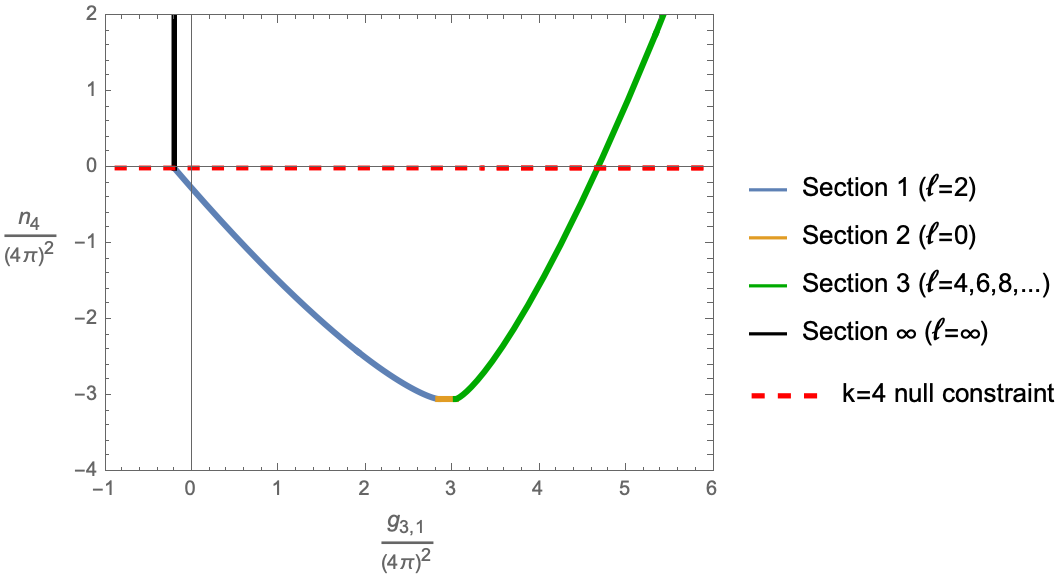}  
   \caption{$(g_{3,1},n_4)$ space with four sections of the boundary. The lower and upper bounds on $g_{3,1}$ follow from intersecting with the null plane $n_4=0$.}
   \label{fig:plot11b}
\end{figure}

\paragraph{1D bounds on $g_{3,1}$} For the coupling $g_{3,1}$ we find $\lambda_{\ell,3,1}=2\mathcal{J}^2-3$, so again there are three types of moments: $(+,-)$ for $\ell=2$, $(-,0)$ for $\ell=0$, and $(+,+)$ for for $\ell\ge4$. The boundary structure is almost identical to the previous case, the only difference being the segment corresponding to $\ell=0$ is now in the negative direction. We can find the boundary in this configuration by reflecting this segment about the origin, solving the problem as before, and then translating the whole boundary by $\frac{L}{3}(v_{0,3,1},0)$. This procedure gives the boundary structure shown in Figure \ref{fig:plot11b}. All sections can be obtained from the previous computation for $g_{2,0}$ by simply setting $k_1=3$ and $q_1=1$, including in the $r$ factor and implicitly in the infinite sum $F$, which we find still converges, now to a value of $\sim 1.29$. The intersection with the null plane gives: 
\eq\label{fing31}
-\frac{0.20}{M^6}\lessapprox  \frac{{g}_{3,1}}{(4\pi)^2}\lessapprox   \frac{4.67}{M^6}\,,
\eqe
We also notice that the lower bound is in fact independent of the null constraint, and simply follows from adding just the negative contributions to $g_{3,1}$, which in this case only come from $\ell=0$.

\subsubsection{2D theory space for $(g_{2,0}, g_{3,1})$ with $k{=}4$-null constraint }
\begin{figure}[H] 
   \centering
      \includegraphics[width=5.5in]{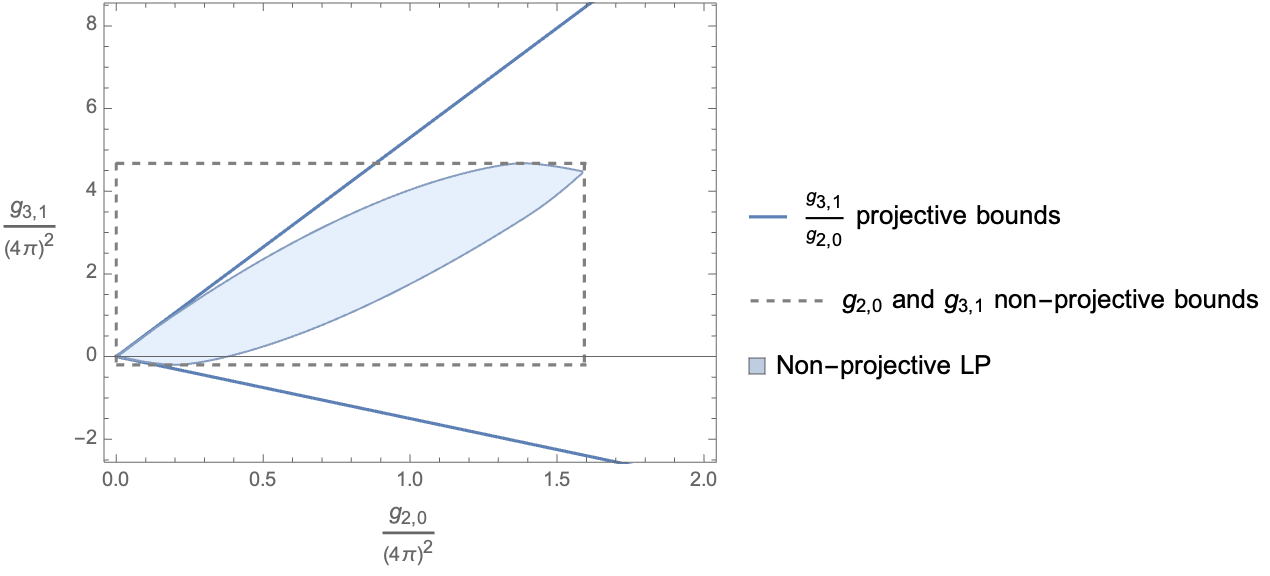} 
   \caption{$(g_{2,0},g_{3,1})$ after imposing $k=4$ null constraint, showing projective and non-projective bounds.}
   \label{fig:plot12bb}
\end{figure}

It is interesting to also consider the theory space for couplings $(g_{2,0},g_{3,1})$,  subjected to the $k=4$ null constraint. We can first collect the individual bounds derived in the previous section, and projective bounds derived in ~\cite{Caron-Huot:2020cmc,Chiang:2021ziz}. We illustrate this in Figure \ref{fig:plot12bb}, together with the numerical result due to LP. Note that once we obtain the non-projective bound for $g_{2,0}$, we could use the projective bounds on ratios $\frac{g_{3,1}}{g_{2,0}}$ to obtain non-projective bounds on $g_{3,1}$. As evident from the graph, the latter is weaker than the actual $g_{3,1}$ bound.

We can ask how much of the LP boundaries in Figure \ref{fig:plot12bb} can be obtained analytically though the geometry we developed so far. To obtain this complete space directly, would require us to first build the space $(g_{2,0},g_{3,1},n_4)$. As discussed in Section \ref{sect33}, for such a space we find a complicated solution to the Lagrange equation, making the infinite sum of spins difficult. However, from previous analysis of the EFThedron, it was observed that in general at a finite order in $k$, only the geometry associated with low spins intersects with the null plane. Thus one might expect the same to hold true for the non-projective geometry. As we will see, indeed the upper boundary in Figure \ref{fig:plot12bb} computed via LP can be identified as boundaries of the Minkowski sum of low spin moments. For the lower boundary, we observe that in all previous cases the lower bounds hold independently of the null constraint. We can use this as motivation to simply import the lower boundary from the 2D space $(g_{2,0},g_{3,1})$ in the absence of any null constraint. We expect this to be a weaker bound, but surprisingly find it is very close to LP results. 
 
\begin{figure}[H] 
   \centering
      \includegraphics[width=4.5in]{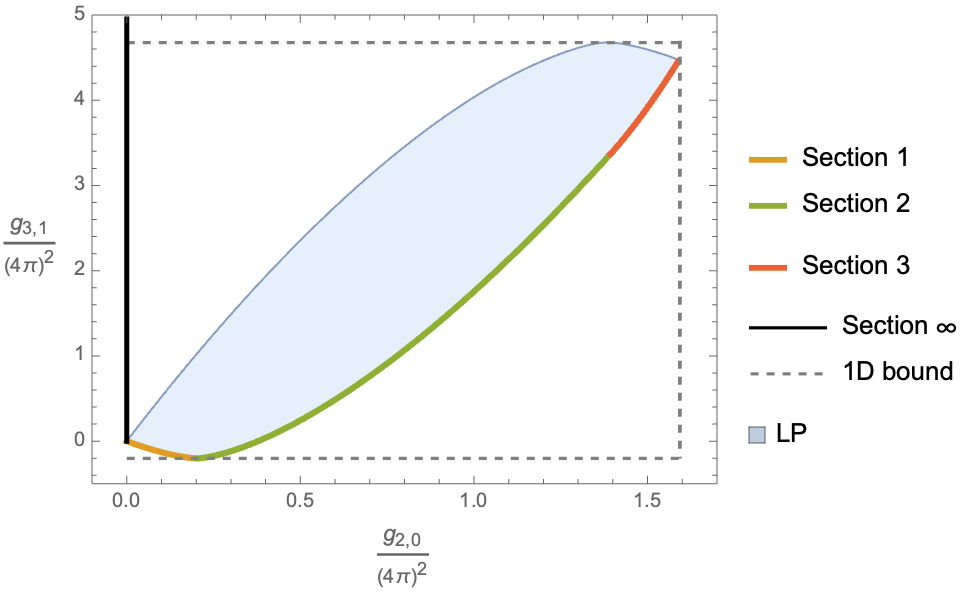} 
   \caption{$(g_{2,0},g_{3,1})$ space. Analytic results for the boundaries emphasized are derived in the absence of any null constraint, while the numerical LP results are obtained assuming the $k=4$ null constraint. The three sections of the lower boundary nevertheless seem to closely match the numerical results.}
   \label{fig:gzv2}
\end{figure}

\paragraph{Lower boundary}
First, we directly consider the space $(g_{2,0},g_{3,1})$ (without any null constraint). We have just two types of moments, $(+,-)$ for $\ell=0$ and $(+,+)$ for $\ell\ge 2$. The first three sections are given by
\begin{align}
\nonumber& \textrm{ Section 1: }g_{2,0},g_{3,1}:\ g_{k,q}=\frac{L}{k}v_{\ell_1,k,q} (1 - m^k),\quad m\in[0,1]\,,\\
&\nonumber \textrm{ Section 2: }g_{2,0},g_{3,1}:\ g_{k,q}=\frac{L}{k}\left(v_{\ell_1,k,q}+m^kv_{\ell_2,k,q} F^{(1)} \right),\quad m\in[0,1]\,,\\
&\textrm{ Section 3: }g_{2,0},g_{3,1}:\ g_{k,q}=\frac{L}{k}\left(v_{\ell_1,k,q}+v_{\ell_2,k,q}+m^k v_{\ell_3,k,q} F^{(2)} \right),\quad m\in[9/37,1]\,,
\end{align}
where $F^{(1)}\approx1.175$ and $F^{(2)}\approx1.643$. We illustrate this in Figure \ref{fig:gzv2}, finding it is in fact indiscernible from numerical results due to LP, at least to the precision we tested.
Section 3 reaches the upper bound for $g_{2,0}$, so we can terminate the boundary at that point.

\paragraph{Upper boundary}
The situation is different however for the upper boundary. It turns out the boundary from the Minkowski sum of a finite number of low spins approximates the LP boundaries very well. Let us consider first the sum of just spins $\ell=2, 4$. The Minkowski sum of lower boundaries is given by
\eqa
\nonumber g_{2,0},g_{3,1}:\ g_{k,q}&=&\frac{L}{k}\left(v_{\ell_2,k,q}(1-n_2^k+m_2^k)+v_{\ell_3,k,q}(n_3^k-m^k_3)\right)\,, \\
n_4&=&\frac{L}{4}\left(u_{\ell_2,4}(1-n_2^4+m_2^4)+u_{\ell_3,4}(n_3^4-m^4_3)\right)\,,
\eqae
and there will be six boundaries in total, which can be easily obtained by the Lagrange method as before. The boundary of interest that would correspond to an upper bound for $(g_{2,0},g_{3,1})$ is given by $m_2=0$, with $m_3$ fixed by the Lagrange method to be 
\eqa\label{complic}
m_3&=&\frac{n_2 v_{\ell_3,2,0} \left(n_2 u_{\ell_2,4} v_{\ell_3,3,1}-n_3 v_{\ell_2,3,1} u_{\ell_3,4}\right)}{u_{\ell_3,4} \left(n_2 v_{\ell_2,3,1} v_{\ell_3,2,0}-n_3 v_{\ell_2,2,0} v_{\ell_3,3,1}\right)}\,.
\eqae

We can repeat the exercise now also adding spin $\ell=6$, where the new constraints on $m_4,n_4$ give
\eqa
  \begin{array}{l}m_4,\\n_4\end{array}&=&\frac{v_{\ell_4,3,1} \left(n_2^2 u_{\ell_2,4} v_{\ell_3,2,0}-n_3^2 v_{\ell_2,2,0} u_{\ell_3,4}\right)\mp\sqrt{R}}{2 u_{\ell_4,4} \left(n_2 v_{\ell_2,3,1} v_{\ell_3,2,0}-n_3 v_{\ell_2,2,0} v_{\ell_3,3,1}\right)}\,,
\eqae
where
\begin{align}
\nonumber R&=v_{\ell_4,3,1}^2 \left(n_2^2 u_{\ell_2,4} v_{\ell_3,2,0}-n_3^2 v_{\ell_2,2,0} u_{\ell_3,4}\right){}^2\\
&-4 n_2 n_3 v_{\ell_4,2,0} u_{\ell_4,4} \left(n_2 v_{\ell_2,3,1} v_{\ell_3,2,0}-n_3 v_{\ell_2,2,0} v_{\ell_3,3,1}\right) \left(n_2 u_{\ell_2,4} v_{\ell_3,3,1}-n_3 v_{\ell_2,3,1} u_{\ell_3,4}\right)\,.
\end{align}

\begin{figure}[H] 
   \centering
      \includegraphics[width=3.4in]{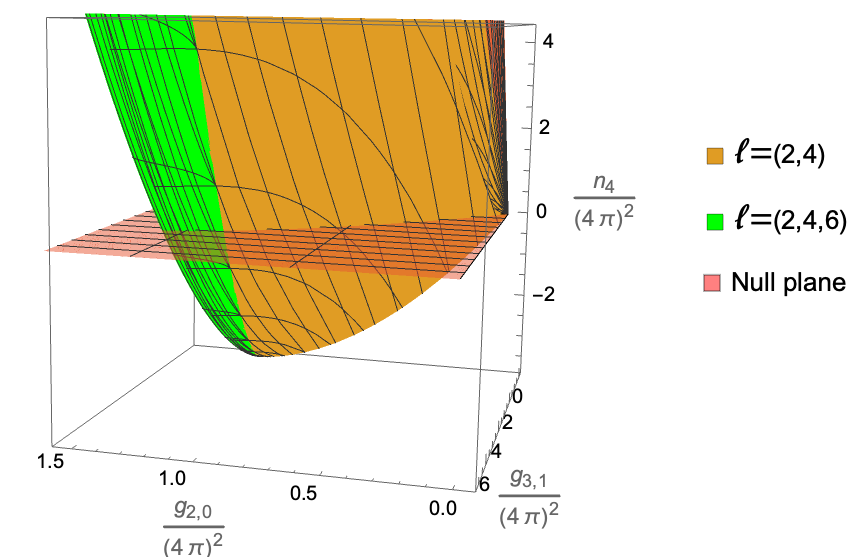} 
   \caption{$(g_{2,0},g_{3,1},n_4)$ space with $n_4=0$ null plane. The two sections of the boundary are given by only a finite number of spins contributing to the Minkowski sum.}
   \label{fig:plot3d}
\end{figure}

Plotting these surfaces in Figure \ref{fig:plot3d}, we observe the null plane first intersects the section corresponding to the sum of spins $(2, 4)$, next the one corresponding to spins $(2, 4, 6)$, and so on. This allows us to find the intersection analytically for any such section, since it always contains a sum over a finite number of spins.

\begin{figure}[H] 
   \centering
      \includegraphics[width=5.9in]{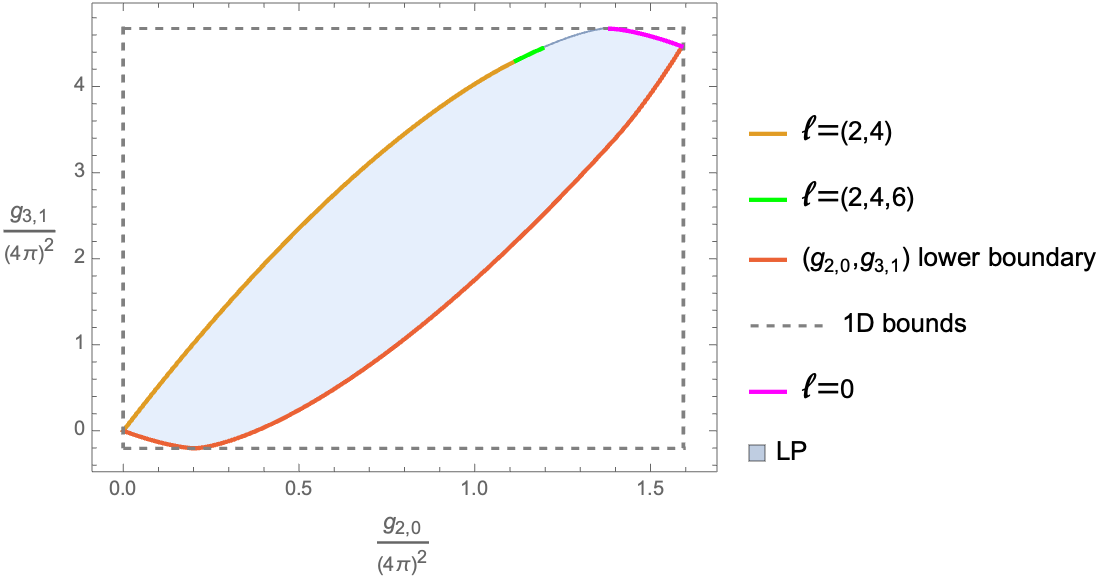} 
   \caption{$(g_{2,0},g_{3,1})$ space with $k=4$ null constraint imposed. The gap in the boundary corresponds to an infinite number of sections, each including an increasing number of higher spins.}
   \label{fig:gz}
\end{figure}

To find the intersection of section $(2, 4)$ with the null plane we must solve $n_4(m_2,n_2,m_3,n_3)=0$, with $m_2=0$ and $m_3$ given by eq.(\ref{complic}). This leads to a degree 8 polynomial in either $n_2$ or $n_3$, but only two solutions are real-valued, and only one satisfies $0\le m_i,n_i\le 1$. Plugging this remaining solution into $g_{2,0},g_{3,1}$ we find a boundary shown in Figure \ref{fig:gz}. This boundary is exact, but only valid until the next section $(2,4,6)$, whose intersection can be computed in a similar way.

Finally, we can obtain another exact piece of the upper boundary, which corresponds to a contribution given only by the $\ell=0$ moment. We can place it by matching it to the 1D bounds on $g_{2,0}$ and $g_{3,1}$ that we know exactly from eqs.(\ref{fing20}) and (\ref{fing31}). We obtain
\eq
(g_{2,0},g_{3,1}):\ g_{k,q}(m)=A_{k,q}-\frac{L}{k}v_{\ell_1,k,q}  m^k\,,
\eqe
with $g_{2,0}(1)=g_{2,0}^{\textrm{upper bound}}$ and $g_{3,1}(0)=g_{3,1}^{\textrm{upper bound}}$, allowing us to fix $A_{k,q}$.

Putting all the boundaries together, we obtain Figure \ref{fig:gz}. The remaining gap can be filled by successively adding more and more spins to the 3D space, and intersecting with the null constraint.  

\begin{figure}[H] 
   \centering
      \includegraphics[width=4.in]{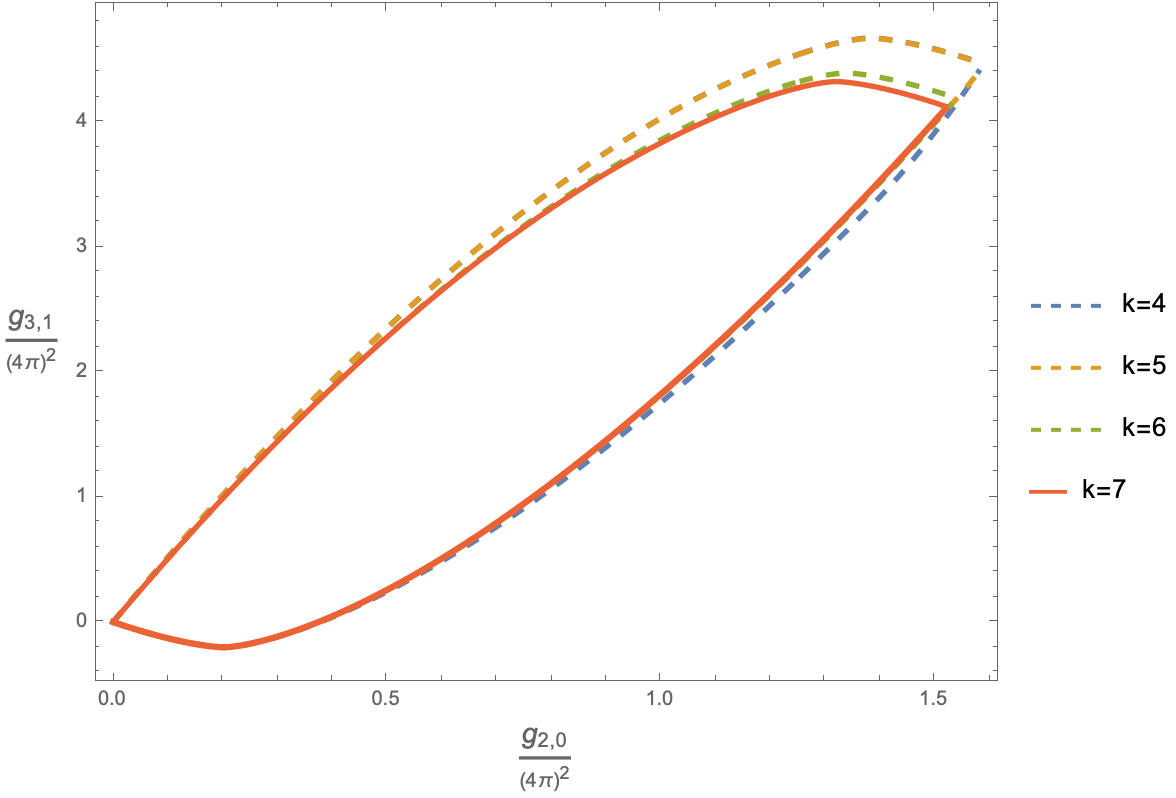} 
   \caption{$(g_{2,0},g_{3,1})$ space with an increasing number of null constraints, obtained from linear programming. The boundary converges quickly after $k=7$ null constraints.}
   \label{fig:LP}
\end{figure}

\paragraph{Higher order null constraints from LP}
The linear programming setup introduced in \cite{Chiang:2022jep} allows us to compute the effect of several null constraints. We plot the result including null constraints up to $k=7$ in Figure \ref{fig:LP}, observing the boundaries already start converging around this order.

\subsubsection{Bounds for arbitrary $k$ with $k=4$ null constraint}
\begin{figure}[H] 
   \centering
      \includegraphics[height=1.9in]{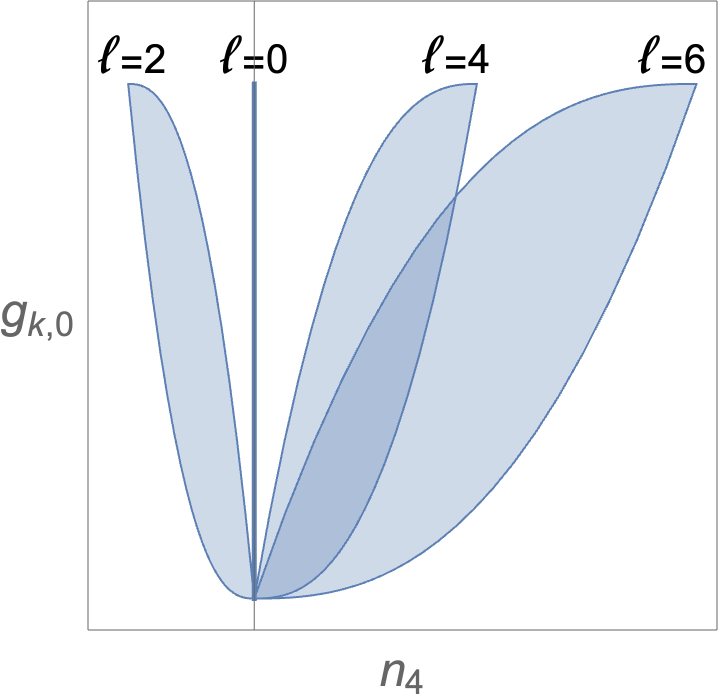} 
      \includegraphics[height=1.9in]{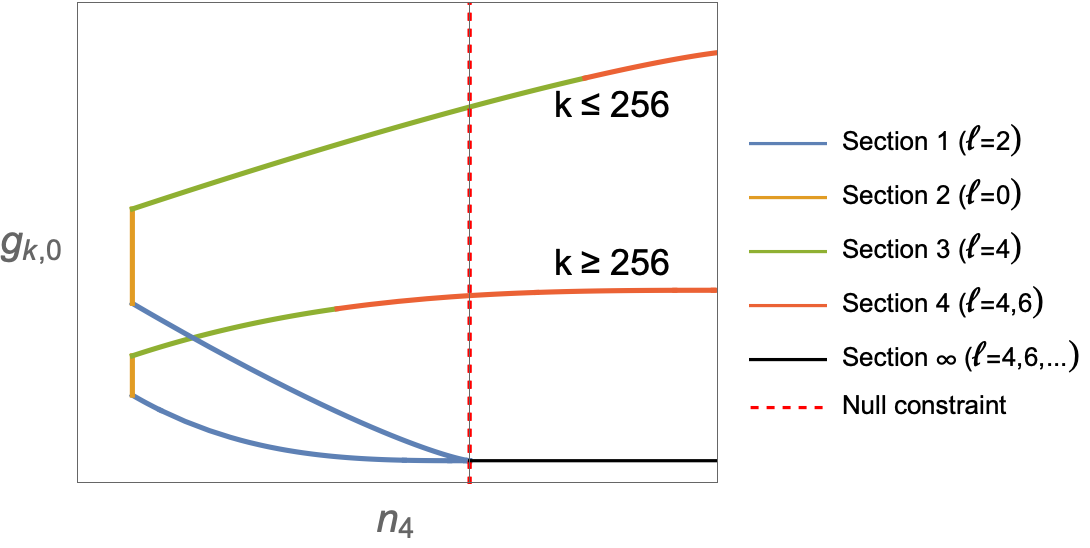} 
   \caption{Left: The first few individual $L$-moments that need to be summed for a space $(g_{k,0},n_4)$, for $k>4$. Right: The boundary of the sum over all spins, together with the $n_4=0$ null plane. As $k$ increases the boundary shrinks, and the null plane intersects a different section.}
   \label{fig:ptt}
\end{figure}

We can easily derive an upper bound due to the $k=4$ null constraint for any coupling of the form $g_{k,0}$ for $k=even$, since  $\lambda_{\ell,k,0}=2$, and so the overall boundary structure does not change as we increase $k$. We must therefore  find the boundaries for a space $(n_4,g_{k,0})$. We choose this ordering to maintain our convention that $k_2\ge k_1$. For any $k$ there will be three types of moments. For $\ell=2$ we have $(-,+)$, for $\ell=0$ we have $(0,+)$, and for all the spins $\ell\ge 4$ we have $(+,+)$. These are shown in Figure \ref{fig:ptt}, together with the boundary structure of their sum. Note that unlike the previous two cases for $g_{2,0}$ and $g_{3,1}$, the sum of upper boundaries for spins $\ell\ge 4$ no longer gives just a trivial vertical line.  Because the slope for each spin is given by $\theta_\ell=\frac{v_{\ell,k,0}}{u_{\ell,4}}=\frac{2}{\mathcal{J}^4-8\mathcal{J}^2}$, we actually have that largest slope is finite and corresponds to $\ell_3$, with  $\theta_{\ell_3}\ge \theta_{\ell_4}\ge \ldots\ge \theta_\infty=0$. We can therefore use eq.(\ref{eqbdrju2}) for this boundary. 

The Minkowski sum of all boundaries is
\eqa
\nonumber g_{k,0}(m_1,m_2,m_{i})&=&\frac{L}{k}\left( v_{\ell_1,k,0}m_1^k+v_{\ell_2,k,0} m_{2}^k+\sum_{i=3}^\infty v_{\ell_i,k,0}(1-m_{i}^k)\right)\,,\\
n_4(m_1,m_2,m_{i})&=&\frac{L}{4}\left( u_{\ell_1,4}m_1^4+u_{\ell_2,u} m_{2}^4+\sum_{i=3}^\infty u_{\ell_i,4}(1-m_{i}^4)\right)\,,
\eqae
Arranging the boundaries in order, starting at the origin, the first two sections are given by choosing
\eqa
\nonumber \textrm{Section 1: }  &&g_{k,0}(m,0,0),\quad n_{4}(m,0,0)\,, \\
\textrm{Section 2: }  && g_{k,0}(1,m,0),\quad n_4(1,m,0)\,.
\eqae
The next two sections are given by
\begin{align}
\textrm{Section 3: }\nonumber g_{k,0}&=\frac{L}{k}\left(v_{\ell_2,k,0}+v_{\ell_1,k,0}+v_{\ell_3,k,0}(1-m^k)\right)\,, \\
 n_4&=\frac{L}{4}\left(u_{\ell_2,4}+u_{\ell_1,4}+u_{\ell_3,u}(1-m^4)\right)\,,
 \end{align}
 for $m\in[r_{3,4},1]$ and 
 \begin{align}
\nonumber \textrm{Section 4: }g_{k,0}&=\frac{L}{k}\left(v_{\ell_2,k,0}+v_{\ell_1,k,0}+\left(v_{\ell_3,k,0}(1-\left(mr_{2,3}\right)^k)+ v_{\ell_4,k,0}(1-m^k)\right)\right)\,,\\
n_4&=\frac{L}{4}\left(u_{\ell_2,4}+u_{\ell_1,4}+\left(u_{\ell_3,4}(1-\left(mr_{2,3}\right)^4)+ u_{\ell_4,4}(1-m^4)\right)\right)\,,
\end{align}
for $m\in[r_{4,5},1]$.

The intersection with the null plane will occur in some section $j\ge 3$, depending on $k$.  For instance, for $4\le k\le 256$ the intersection is in section $j=3$, while for $258\le k\le 1876$ the intersection is in section $ j=4$, and so on, as shown in Figure \ref{fig:ptt}. For $4\le k\le 256$ we find an upper bound
\eq\label{finbound}
 \frac{g_{k,0}}{(4\pi)^2}\le\frac{2}{k} \left(30-18\left(\frac{35}{36}\right)^\frac{k}{4}\right)\frac{1}{\pi ^2 M^{2k}}\,.
\eqe
This upper bound matches that of $g_{4,0}$ in eq.(\ref{fing40}) that we found using the polytope constraints, as it must. For $g_{6,0}$ we obtain an upper bound $\sim 0.430$, stronger than the bound obtained in eq.(\ref{fing60}) using the $k=6$ null constraint. This result can be compared to purely projective bounds, which would only imply $g_{k,0}\le g_{2,0}$.

\begin{figure}
\centering
\begin{minipage}{.47\textwidth}
  \centering
  \includegraphics[width=\linewidth]{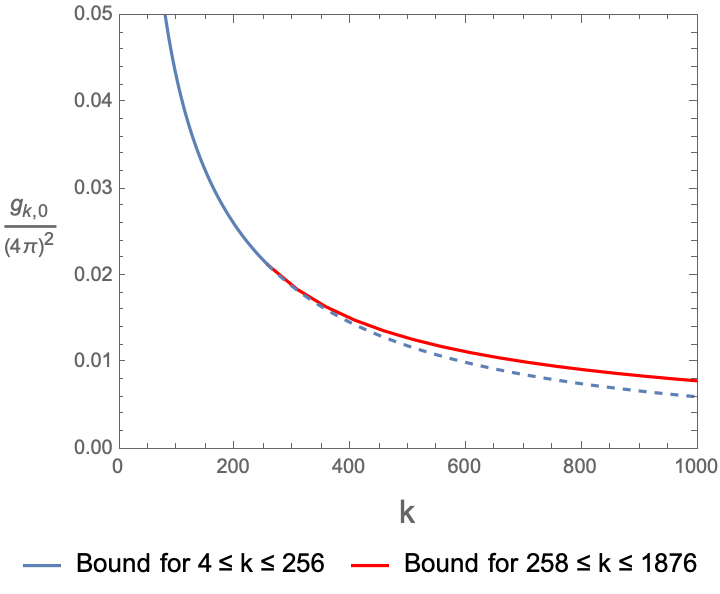}
  \captionof{figure}{Upper bound on $g_{k,0}$, as a function of $k$, with $k=4$ null constraint. Blue corresponds to eq.(\ref{finbound}), valid for $k\le256$, while red is valid for $258\le k\le 1876$; both exhibit $1/k$ dropoff.}
  \label{fig:test1}
\end{minipage}%
\hfill
\begin{minipage}{.47\textwidth}
  \centering
  \includegraphics[width=0.98\linewidth]{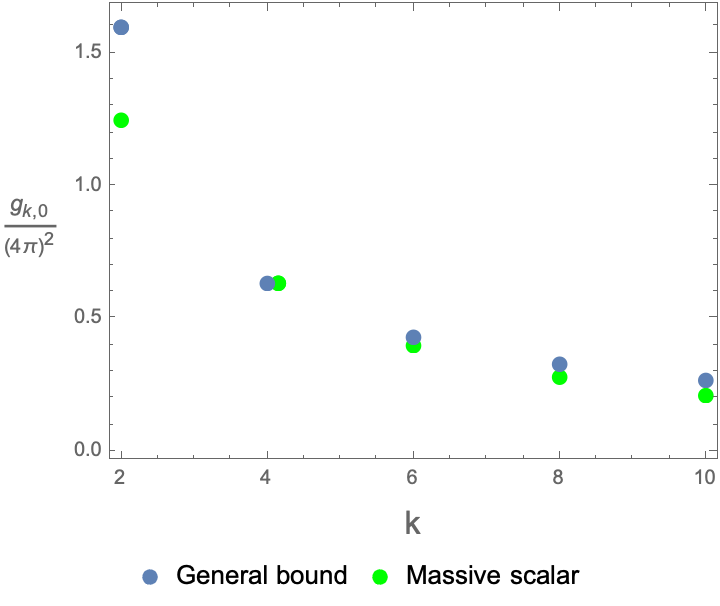}
  \captionof{figure}{Eq.(\ref{finbound}) compared to the massive scalar in eq.(\ref{scalarbox}), after normalizing by matching at $g_{4,0}$; the scalar amplitude exhibits a stronger, approximately $1/k!$ dropoff.}
  \label{fig:test2}
\end{minipage}
\end{figure}

For higher $k$, when the null plane intersects the section $j=4$, the modification is relatively small. We compare the two in Figure \ref{fig:test1}. In both cases the bound roughly behaves as $\frac{1}{k}$, and we expect this behavior to continue for even higher $k$.

The same derivation can also be similarly carried out for couplings $g_{k,1}$ for $k=odd$, since these only have a slightly more complicated $\lambda_{\ell,k,q}=2\mathcal{J}^2-k$. We have verified the upper bound follows a similar  $1/k$ dropoff, but the lower bound diverges with large $k$. We can obtain the minimum bound by summing only the negative contributions when $\lambda_{\ell,k,q}<0$, that is all spins for which $\ell\le \ell^*= 1/2 (-1 + \sqrt{1 + 2 k})\sim \sqrt{k}$. This gives a lower bound
\eq
g_{k,1}\ge \frac{16}{k} \sum_{\ell<\ell^*} (2\ell+1)(2(\ell^2+\ell)-k)\,.
\eqe
For large $k$ this approximates to $-2k$. Imposing any finite number of null constraints does not affect this bound.

It is interesting to compare this behavior with an example of a physical theory. We can consider a theory with a massive scalar $m_x$, with no tree level interaction, such that the first allowed amplitude is the 1-loop box, given by
\begin{align}
\nonumber M(s, t)=\lambda^{4} & \int \frac{d^{4} \ell}{(2 \pi)^{4}} \frac{1}{\left[\ell^{2}-m_x^{2}\right]\left[\left(\ell-p_{1}\right)^{2}-m_x^{2}\right]\left[\left(\ell-p_{1}-p_{2}\right)^{2}-m_x^{2}\right]\left[\left(\ell+p_{4}\right)^{2}-m_x^{2}\right]} \\
&+p e r m(2,3,4)\,.
\end{align}
The low energy expansion can be computed by first writing the integral in Feynman parameterization, which in $D=4$ is
\eq
I(s,t)=\int_0^1 d a_i^4 \delta(1-\sum_{i} a_i)\frac{1}{\left(a_1 a_3 s+a_2 a_4 t+m_x^2\right)^2}\,,
\eqe
In terms of the EFT scale $M$, we have $M=2m_x$, since the lightest state for this process is in fact a two-particle state. Summing over all permutations, we obtain, for $k=\{0,2,4,\ldots\}$:
\eqa\label{scalarbox}
\nonumber g_{k,0}:&&\frac{2^4\tilde{\lambda}^4}{2M^4}\left\{1,\frac{2^4}{60M^4},\frac{2^8}{1890m_x^8},\frac{2^{12}}{48048m_x^{12}},\frac{2^{16}}{1093950m_x^{16}},\frac{2^{20}}{23279256m_x^{20}},\ldots\right\}\\
\eqae
We observe these couplings follow a stronger approximately $1/k!$ fall-off. For normalization we match to the upper bound we derived for $g_{4,0}\le  0.633$,  such that the remaining couplings are consistent with our bound in eq.(\ref{finbound}), as shown in Figure \ref{fig:test2}. Note that at leading orders, the scalar box is close to our numeric bounds which indicate that the bounds are close to optional. The deviation increases at large $k$.


\section{Conclusions}\label{conc}
In this paper we developed a non-projective generalization of the EFThedron, allowing us to incorporate the non-linear unitarity bound in full. This allows us to derive non-projective upper bounds for the EFT couplings, which shows that while the leading derivative couplings are of $\mathcal{O}(1)$, they are heavily suppressed at higher $k$. As a different application of this framework, we can also incorporate the low spin dominance condition, a property observed to hold for gravitational EFT \cite{Bern:2021ppb} (see also \cite{Figueroa:2022onw}). It would also be interesting to explore how to implement the upper bound on $\rho$ for impact parameter space dispersion relations in gravitational EFT \cite{Caron-Huot:2022ugt,Caron-Huot:2021rmr}.

In Section \ref{sec2} we solved the single $L$-moment problem by treating it as a Minkowski sum of segments. To our knowledge this a new proof to this problem. It would be interesting if the same approach could be used to solve the double $L$-moment problem \cite{PUTINAR1990288}, and perhaps understand new features of this much richer topic. At the same time, it would be interesting if the known solution to the double $L$-moment can be used to obtain bounds on our physical problem. This would enable the approach used in the projective case, which is easier to set up and extend to higher dimensional spaces and more null constraints.

\acknowledgments 
 L-y Chiang, Y-t Huang, and H-c Weng are supported by Taiwan Ministry of Science and Technology Grant No. 109-2112-M-002 -020 -MY3.   L Rodina is supported by Taiwan Ministry of Science and Technology Grant No. 109-2811-M-002-523.

\appendix 

\section{Aspects of the $L$-moment problem}\label{aspectsL}
In this section we discuss some other interesting aspects of the $L$-moment problem, in particular the functions $\rho$ that correspond to our boundary solutions, also known as extremal solutions, and the formulation of constraints in terms of ``exponential Hankel" matrices.
\subsection{Boundary distributions}
The boundary solutions for $a_k$ correspond to particular $\rho$, which we can easily work out. For instance, the boundary of a 2D space we found was given by
\eq
a_{k,q}=\int_0^1\rho z^{k-1}dz =L\frac{m^k}{k}\,,
\eqe
and we find the corresponding boundary $\rho$ is a step function
\eq
\chi_{I}(z)= \left\{
    \begin{array}{ll}
        L & \textrm{for }z\in I\,, \\
        0 & \textrm{for } z\notin I\,.
    \end{array}
\right.
\eqe
In general, a solution may contain a sum over several such functions,
\eq
\rho(z)=\sum_{i=1}^N  \chi_{I_i}(z)\,,
\eqe
where $I_i=[m_{2i-1},m_{2i}]$, $m_{i}\le m_{i+1}$. Just like the boundary itself, we can deduce the corresponding distributions directly from eqs.(\ref{evenpol}) and (\ref{oddpol}), depending on dimension, by simply taking ${\bf V}_{i_1,i_2}\rightarrow \chi_{[m_{i_1},m_{i_2}]}$. For $D=\textrm{even}$  we find
\eqa\label{evensola}
\nonumber \textrm{lower: }\rho(z)&=&\chi_{[0,m_{1}]}+\chi_{[m_{2},m_{3}]}+\ldots+\chi_{[{m_{D-2}},m_{D-1}]}\,,\\
\textrm{upper: }\rho(z)&=&\chi_{[m_1,m_{2}]}+\chi_{[m_{3},m_{4}]}+\ldots+\chi_{[{m_{D-1}},1]}\,,
\eqae
and  for $D=\textrm{odd}$
\eqa\label{oddsola}
\nonumber \textrm{lower: }\rho(z)&=&\chi_{[m_{1},m_2]}+\chi_{[m_{3},m_{4}]}+\ldots+\chi_{[{m_{D-2}},m_{D-1}]}\,,\\
\textrm{upper: }\rho(z)&=&\chi_{[0,m_1]}+\chi_{[m_{2},m_{3}]}+\ldots+\chi_{[{m_{D-1}},1]}\,.
\eqae

\subsection{Exponential Hankel conditions}\label{Hankel}
Similar to the projective moment problem, necessary and sufficient conditions for the $L$-moments can be expressed in terms of positive semi-definite matrices. Using the notation of \cite{PUTINAR1990288}, the conditions for the infinite dimensional problem are given in terms of two new sequences, $b$ and $c$, defined by
\eqa\label{expo1}
\nonumber \textrm{exp}\left[{-\frac{1}{L}\left(\frac{a_0}{X}{+}\frac{a_1}{X^2}{+}\cdots\right)}\right]&=&1-\left(\frac{b_0}{X}{+}\frac{b_1}{X^2}{+}\cdots\right)\,,\\
 \textrm{exp}\left[{\frac{1}{L}\left(\frac{a_0}{X}{+}\frac{a_1}{X^2}{+}\cdots\right)}\right]&=&1+\left(\frac{c_0}{X}{+}\frac{c_1}{X^2}{+}\cdots\right)\,,
\eqae
which must satisfy the usual Hankel positivity constraints, for all $n$.
\begin{enumerate}
\item $x\in (-\infty,\infty)$:
\eq\label{prob1}
H_{n}(b)\ge 0\,.
\eqe
\item $x\in [0,\infty)$:
\eq\label{prob2}
H_{n}(b),\ H_{n}^{\textrm{shift}}(b)\ge 0\,.
\eqe
\item $x\in [0,1]$:
\eq
H_{n}(b),\ H_{n}^{\textrm{shift}}(b),\ H_{n}^{\textrm{twist}}(c)\ge 0\,.
\eqe
\end{enumerate}
One can verify that for the boundary solutions we found particular subsets of the exponential Hankels indeed vanish. This implies the Hankel constraints are necessary conditions. Next we review the proof they are also sufficient conditions, in the $x\in (-\infty,\infty)$ case.

Let us begin with the exponential map
\eq\label{ExpMap2}
e^{\frac{1}{L}\left(\frac{a_0}{x}{+}\frac{a_1}{x^2}{+}\cdots \frac{a_{2m}}{x^{2m{+}1}}\right)}=1{+}\frac{b_0}{x}{+}\frac{b_1}{x^2}{+}\cdots
\eqe
Note that while there are $2m$ $a_i$s, the exponential map generates an infinite series for $b_i$s. We would like to show that if the $b_i$ satisfy Hankel positivity condition, we can use them to build a solution for the $L$-moment problem for the $\{a_i\}$s. This applies to the Hamburger,  Stieltjes and Hausdorff intervals $\mathcal{I}$. For a general discussion see~\cite{akhiezer1962some}.

So let us begin by $b_i$ being a solution to the Hamburger moment problem. Note that since the LHS of eq.(\ref{ExpMap2}) has $2m{+}1$  $\{a_i\}$s, 
this means that the first $2m{+}1$ $\{b_i\}$s on the RHS should fully determine the $\{a_i\}$s. This further suggests that $b_{2k{+}1}, b_{k{+}2},\cdots$ are also fully determined by $\{b_0, b_1,\cdots, b_{2m{+}1}\}$, and thus the $\{b_i\}$s correspond to a solution with $m{+}1$ elements in the hull:
\eq
b_i=\sum_{a=1}^{m{+}1} \tilde\rho_a y_a^i,\quad   \tilde\rho_a>0\,.
\eqe
Substituting this into the exponential map  the RHS can be rewritten as
\eq\label{LogMap}
1{+}\frac{b_0}{x}{+}\frac{b_1}{x^2}{+}\cdots=1{+}\sum_{a=1}^{m{+}1}\frac{\tilde\rho_a}{x{-}y_a}\equiv \frac{\psi(x)}{\phi(x)}\,,
\eqe
where 
\eq
\phi(x)=\prod_{a=1}^{m{+}1}(x{-}y_a), \quad \psi(x)=\prod_{a=1}^{m{+}1}(x{-}y'_a)\,,
\eqe
and $y'_1< y_1<y_2'<y_2<\cdots< y'_{m{+}1}< y_{m{+}1}$, i.e. the position of zeros for $\psi(x)$ is distributed in between the zeros of $\phi(x)$. Now consider the following density: 
\eq\label{Lrho}
\rho(u)\equiv\frac{L}{2}\left(1{-}\textrm{sign} \frac{\psi(u)}{\phi(u)}\right)\,.
\eqe
We see that $\rho(u)$ is either $L$ or $0$. Since the zeros of $\psi(x)$ and $\phi(x)$ are interspaced, the integration directly gives
\eq
\int du\frac{\rho(u)}{x{-}u}=L\left(\int_{y'_1}^{y_1}\frac{du}{x-u}{+}\int_{y'_2}^{y_2}\frac{du}{x-u}{+}\cdots\right)=L \log \frac{\psi(x)}{\phi(x)}\,.
\eqe
Now combine eq.(\ref{ExpMap2}) with eq.(\ref{LogMap}), we have  
\eq
\left(\frac{a_0}{x}{+}\frac{a_1}{x^2}{+}\cdots \frac{a_{2m}}{x^{2m{+}1}}\right)=L \log \frac{\psi(x)}{\phi(x)}=\int du\frac{\rho(u)}{x{-}u}\quad \rightarrow \quad a_k=\int du\; \rho(u) u^k\,,
\eqe
where $\rho(u)$ is defined in eq.(\ref{Lrho}). Thus we have constructed a solution to the $L$-moment problem. Note that the solution corresponds to an $m$-state solution.

\section{Discrete solution to the double moment problem}\label{DiscreteM}
\begin{figure}[H] 
   \centering
        \includegraphics[height=2in]{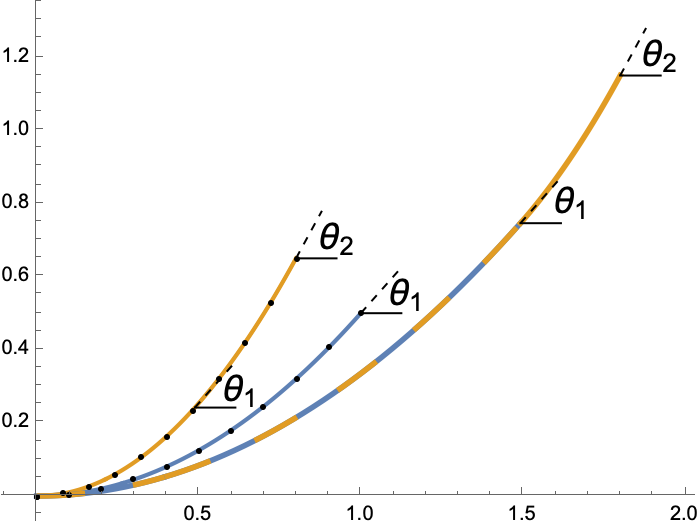} 
   \caption{Boundary of the Minkowski sum via discretization. Treating the two curves as polytopes with infinitesimal edges, the resulting polytope contains segments from both curves, in order of their slope.}
   \label{fig:plotM4b}
\end{figure}

Let us understand the results in Section \ref{sec2d2} also from the discrete perspective, instead of the Lagrange method. Our task is to find the boundary for the sum of two lower boundaries
\eq
 a_{k_1,q_1}=\frac{L}{k}v_{\ell_1,k,q}m_1^{k}+\frac{L}{k}v_{\ell_2,k,q}m_2^{k}
\eqe
If we imagine discretizing both curves into small segments, we know the resulting boundary will be given by placing the segments in order according to their slope. The slopes of segments from curve 1 range from $[0,\theta_1]$, while those of curve 2 from $[0,\theta_2]$, where $\theta_\ell=v_{\ell,k_2,q_2}/v_{\ell,k_1,q_1}$. We show this in Figure \ref{fig:plotM4b}. If we assume $\theta_2>\theta_1$, then it is clear the sum will have one section from $[0,\theta_1]$ formed with segments originating from both, placed in some particular order, and a second section from $[\theta_1,\theta_2]$ with segments just from curve 2, placed in their original order. In the continuous limit this leads to the result found via Lagrange multipliers. We work this out explicitly next.

Let us now consider a sum of an arbitrary number of moments. We assume the $v_{\ell}$ are ordered and non-negative, such that $0< \theta_{\ell_1}\le \theta_{\ell_2}\le \ldots \le \theta_{\ell_n}<  \pi/2$. This implies all spins $\ell_1,\ldots,\ell_n$ have segments with slopes up to $\theta_1$, but only spins $\ell_2,\ldots,\ell_n$  have segments with slopes from $\theta_1$ to $\theta_2$, and so on.

The discretized version of the moments, is given by
\eq
a_{k,q}^{(\ell)}=v_{\ell,k,q}\int_0^1\rho(z) z^{k-1}dz \rightarrow \sum_{i=1}^{N_\ell}\rho_{\ell,i}  v_{\ell,k,q} \frac{i^{k-1}}{N_\ell^{k}}, \quad 0\le \rho_{i,\ell}\le L\,.
\eqe
We can view each moment as a Minkowski sum of segments $s_{i}^{(\ell)}$, defined in the terms of their endpoint obtained when $\rho_{i,\ell}=L$:
\eq
s_{i,k,q}^{(\ell)}=L\,v_{\ell,k,q} \frac{i^{k-1}}{N_\ell^{k}}\,,
\eqe
with slopes 
\eq\label{slop}
\theta_{\ell,i}=\theta_\ell \frac{i^{\Delta_k}}{N^{\Delta_k}}\,.
\eqe
Note that as $i$ varies from 0 to $N_\ell$ the slopes of the segments vary from 0 to $\theta_\ell$.
Let us now consider the range $[0,\theta_1]$, where segments from all spins can appear. We wish to discretize all curves into segments $s_{i}^{(\ell)}$ with equal slope for all $\ell$. We can achieve this by tuning $N_\ell$ for each spin. To obtain segments of equal slope at equal $i$ but different $\ell$, we require
\eq
\theta_{\ell_1,i}=\theta_{\ell_2,i} \Rightarrow \frac{N_1}{N_2}=\left(\frac{\theta_1}{\theta_2}\right)^{\frac{1}{\Delta_k}}=r_{2,1}\,.
\eqe
This means each spin must be discretized in a total of $N_{\ell_i}= N_1/r_{i,1}\ge N_1$ segments. The factor $r_{i,1}$ is finite for $\ell \neq 0$, so the continuous limit remains consistent.  For a given spin $\ell_j$ we obtain a set of $N_1$ segments
\eq
s_{i,k,q}^{(\ell_j)}=L \frac{i^k}{N_1^{k}}  \ell_j^q r_{j,1}^{k}, \quad i\in[0,N_1]\,,
\eqe
with slopes
\eq\label{slope}
\theta_{\ell_j,i}=\theta_{\ell_1,i}=\theta_1 \frac{i^{\Delta_k}}{N_1^{\Delta_k}}\,,
\eqe
which unlike eq.(\ref{slop}) are now independent of spin, but still depend on $i$ as they vary from 0 to $\theta_1$. We can now compute the Minkowski sum at fixed $i$ over all spins, since the sum of segments of equal slope simply amounts to another segment of the same slope, with endpoint 
\eq
S_{i,k,q}^{(1)}=\sum_\ell s_{i,k,q}^{(\ell)}= \frac{i^k}{N_1^{k}}F_{k,q}\,,
\eqe
where 
\eq
F_{k,q}=\sum_{i=1} v_{\ell,k,q} r_{i,j}^{k}\,.
\eqe
Crucially, the new segments $S_i$ are now ordered according to $i\in[0,N_1]$, as their slope increases monotonically with $i$, from 0 to $\theta_1$. Next, we can repeat the argument for the range $[\theta_1,\theta_2]$, and we obtain a new set of vectors $S_{i,k,q}^{(2)}$, also ordered by $i$. Taking the union of  ${S}^{(1)}$ and ${S}^{(2)}$ we therefore obtain a set of segments ordered by $i\in[N_1,N_2]$, and so the boundaries of their Minkowksi sum can be obtained from   using the results in Section \ref{sec2}. For $j\in[0,N_1]$, corresponding to the range, $[0,\theta_1]$ we obtain a vertex on the boundary given by
\eq
V_{j|k,q}=\sum_{i=1}^{j}\frac{i^{k}}{N_1^{k}}F_{k,q}\,.
\eqe
In the continuous limit this becomes
\eqa
 a_{k,q}&=&\frac{L}{k}m^{k}F_{k,q}\,,
\eqae
matching what we found using the Lagrange method in eq.(\ref{eqbdrj}).

\section{Towards a solution for the complete space}\label{appgend}
In this section we present some partial results on determining the complete space for the simpler ``power" moment problem
\eq
a_{k,q}=\sum_\ell \Lambda_{\ell,q} \int_0^1 \rho_\ell(z) z^{k-1}dz\,,
\eqe
where $\Lambda_{\ell,q}=g(\ell)f(\ell)^q$, with $g$ and $f$ any positive polynomials in $\ell$.
\subsection{Diagonal spaces}
Let us first consider 3D spaces of the form $a_{k_1,q_1},a_{k_2,q_2},a_{k_3,q_3},$, where  the set of $\{k_i, q_i\}$s satisfy 
\eq\label{cond2}
k_{i+1}-k_i=\Delta_k,\quad q_{i+1}-q_i=\Delta_q\,,
\eqe
For two spins, the Minkowski sum of  lower and upper boundaries of this space is given by
\eqa\label{gendist}
\nonumber a_{k,q}&=&\frac{L}{k}\sum_\ell \Lambda_{\ell,q}\left(m_1^{(\ell)}-m_2^{(\ell)}\right)\,,\\
\overline{a}_{k,q}&=&\frac{L}{k}\sum_\ell \Lambda_{\ell,q} \left(1-m_1^{(\ell)}+m_2^{(\ell)}\right)\,,
\eqae
for $1\ge m_1^{(\ell)}\ge m_2^{(\ell)}\ge 0$. We must compute the boundary of the Minkowski sum of these boundaries, which is again easily accomplished by the Lagrange multiplier method, also keeping in mind to check the boundaries of parametric space.

For such case, applying the Lagrange method we find: 
\eq\label{ert}
\frac{m_1^{(\ell_i)}}{m_1^{(\ell_j)}}=\frac{m_2^{(\ell_i)}}{m_2^{(\ell_j)}}=r_{j,i}=\left(\frac{f(\ell_i)}{f(\ell_j)}\right)^\frac{\Delta_q}{\Delta_k}\,.
\eqe
This reduces the Minkowski sum to just two independent parameters, as required to span the $2$ dimensional boundary. Expressing for instance all parameters as $m_i^{(\ell_j)}=m_i^{(\ell_1)} r_{j,1}=m_i r_{j,1}$, the lower boundary is given by 
\eq
a_{k,q}=\frac{L}{k}\left((m_{1})^{k}-(m_2)^{k}\right)\sum_i \Lambda_{\ell,q} r_{i,1}^{k} \,.
\eqe
which is valid for $1\ge m_1\ge m_2\ge 0$, and similarly for the upper boundary.
The sum over infinite spins is now separate from the parameters, and so can be evaluated numerically. In higher dimensions, the sum of boundaries is given by
\eqa\label{gendd}
\nonumber a_{k,q}&=&\frac{L}{k}\sum_\ell \Lambda_{\ell,q}\sum_{j=1}^{D-1} (-1)^{j+1}m_j^{(\ell)}\,,\\
\overline{a}_{k,q}&=&\frac{L}{k}\sum_\ell \Lambda_{\ell,q} \left(1-\sum_{j=1}^{D-1} (-1)^{j+1}m_j^{(\ell)}\right)\,.
\eqae
Remarkably, the result  eq.(\ref{ert}) holds in fact in any number of dimensions, as long as the conditions (\ref{cond2}) are satisfied, so it can be plugged in eq.(\ref{gendd}) to obtain the $D{-}1$ dimensional boundaries.

Finally we need to consider bounds of the parameter space. These will give new boundaries, which as before, can be simply deduced by imposing the upper integration limit as
\eq
m_i^{\ell_j}\rightarrow \textrm{min}[m_i^{(\ell_j)},1]\,.
\eqe
Explicitly, the boundary of a $D$-dimensional space will be labeled by $D-1$ indices $(i_1,i_2,\ldots,i_{D-1})$, with $N\ge i_1\ge i_2\ge\ldots\ge i_{D-1}\ge 1$. A value $i_j=n$ indicates $m_j^{\ell_{i<n}}=1$ and $m_j^{\ell_{i\ge n}}=r_{i,n} m_j$, with $m_j\in[r_{n,n-1},1]$.

\subsection{Approach to more general spaces}
Here we introduce a tentative approach to obtain constraints on the complete space of couplings $a_{k,q}$. This is motivated by the previous solution to the projective moment problem discussed in~\cite{Chiang:2021ziz}. Recall that a sequence $a_k$ can be expressed as a single moment
\eq\label{projeq}
a_k=\int_\mathbb{R} \rho(x) x^{k-1},\quad \rho\ge 0\,,
\eqe
if and only if the Hankel matrix is totally positive,
\eq
H=\begin{pmatrix}
a_1&a_2&a_3&\ldots\\
a_2&a_3\\
\vdots
\end{pmatrix}\ge 0\,.
\eqe
For double moments, 
\eq\label{projeq}
a_{k,q}=\int_{\mathbb{R}^2} \rho(x,y)x^{k-1} y^{q-1},\quad \rho\ge 0\,,
\eqe
the conditions are given by total positivity of the moment matrix
\eq\label{momentmat}
M=\begin{pmatrix}
a_{1,1}&a_{2,1}&a_{1,2}&a_{3,1}&a_{2,2}&\ldots\\
a_{2,1}&a_{3,1}&a_{2,2}\\
a_{1,2}&a_{2,2}\\
\vdots
\end{pmatrix}\ge 0\,.
\eqe
There is in fact a simple way to derive the moment matrix from the Hankel matrix. This can be done by considering linear combinations of the double moments that can be written as single moments. For instance we can define
\eqa
\nonumber b_1&\equiv& a_{1,1}=\int \rho \left(x+\gamma x y\right)^0\,, \\
\nonumber  b_2&\equiv& a_{2,1}+\gamma a_{2,2}=\int \rho \left(x+\gamma x y\right)^1\,, \\
 b_3&\equiv& a_{3,1}+2\gamma a_{3,2}+\gamma^2a_{3,3}=\int \rho \left(x+\gamma x y \right)^2\,,
\eqae
where $\gamma$ is some parameter. Importantly, the new couplings $b_k$ are again a sum of single moments $b_k=\int \rho  z^k$, with $z=x+\gamma x y$. Thus $b_k$ satisfy Hankel conditions, leading to constraints back on the $a_{k,q}$, such as
\eq
H=\begin{pmatrix}
b_1&b_2\\
b_2&b_3
\end{pmatrix}=\begin{pmatrix}
a_{1,1}&a_{2,1}+\gamma a_{2,2}\\
a_{2,1}+\gamma a_{2,2}&a_{3,1}+2\gamma a_{3,2}+\gamma^2 a_{3,3}
\end{pmatrix}\ge 0\,.
\eqe
Importantly, the above condition now must hold for all values of $\gamma$. Imposing this through the discriminant equation on the determinant, we recover exactly the same condition as from the $3\times 3$ moment matrix in eq.(\ref{momentmat}). We have therefore derived the double moment conditions by taking an envelope, parameterized by $\gamma$, of the single moment conditions.

Let us now attempt to do the same for the double $L$-moment problem
\eq
a_{k,q}=\sum \ell^q \int \rho(z) z^{k-1} dx\,.
\eqe 
We mirror the argument above, by taking linear combinations
\eqa
\nonumber b_{1,1}&=&a_{1,1}=\sum \int \rho(z) (z+\gamma z \ell)^0\,,\\
\nonumber b_{2,2}&=&a_{2,1}+\gamma a_{2,2}=\sum \int \rho(z) (z+\gamma z \ell)^1\,,\\
b_{3,3}&=&a_{3,1}+2\gamma a_{3,2}+\gamma^2 a_{3,3}=\sum \int \rho(z) (z+\gamma z \ell)^2\,,
\eqae
and so on, for $\gamma>0$. We have shown that diagonal spaces with $v_{\ell,k,q}=g(\ell)f(\ell)^q$ satisfy simple boundary equations, and in the case above we simply have $v_{\ell,k,q}=(\ell+\gamma)^q$. The boundary (for the first section) is therefore
\eq\label{infsum}
b_{k,k}=\frac{L}{k} (n^{k}-m^{k} )\sum_\ell  v_{\ell_i,k,q}r_{i,1}^{k}\,,
\eqe
where
\eq
r_{i,1}=\left(\frac{\ell_1+\gamma}{\ell+\gamma}\right)^\frac{\Delta q}{\Delta k}\,.
\eqe
The boundary is now a function of the parameter $\gamma$, and if the analogy with the projective problem is valid, then we expect that taking the intersection of all constraints for all values of $\gamma$ should give necessary and sufficient conditions on the $a_{k,q}$ moments. Unfortunately, it seems unlikely a closed form can be found, since $\gamma$ enters the constraints through the infinite sum in eq.(\ref{infsum}), which is no longer a purely numerical factor. One can instead simply numerically compute the envelope (for $\gamma>0$). It would also be interesting to consider taking the continuous limit of the discrete sum. In that case the sum becomes a trivial integral, and it may be possible to obtain constraints relevant to the double continuous $L$-moment problem.

\section{Low spin dominance}\label{LSD}
The solution for the non-projective polytope derived in Section \ref{sec2} can also be utilized to explore the consequence of low spin dominance (LSD), first discussed in \cite{Bern:2021ppb}. We begin with the dispersive representation for the couplings 
\eq
a_{k,q}=\sum_\ell v_{\ell,k,q}\int_0^1 \rho_\ell(z) z^{k-1} dz ,\quad \rho_\ell(z)\ge 0\,,
\eqe
and consider the mass averaged distribution
\eq
\langle \rho_\ell \rangle_k\equiv \int_0^1 \rho_{\ell}(z) z^{k-1} dz\,.
\eqe
The LSD condition is the requirement that 
\eq
\langle \rho_{\ell_1} \rangle_k\ge \alpha \langle \rho_{\ell} \rangle_k\,,
\eqe
where $\ell_1$ is the lowest spin in the spectrum, and $\alpha\ge 0$ is some parameter specifying the strength of the the LSD condition.  Since for now we only discuss equal $k$ spaces, we will drop the $k$ subscript for simplicity. Separating out the lowest spin, we can write
\eq\label{abq}
 a_{q}=\sum_\ell \langle \rho_{\ell}  \rangle v_{\ell,q}= \langle \rho_{\ell_1} \rangle v_{\ell,q}+\sum_{\ell> \ell_1} \langle \rho_{\ell} \rangle v_{\ell,q}\,.
\eqe
We now introduce couplings $\mu_q$, 
\eq
\mu_q=\sum_{\ell>\ell_1} \langle \rho_{\ell} \rangle v_{\ell,q}\,.
\eqe
which due to the LSD condition must satisfy $\langle \rho_{\ell_1} \rangle \ge \alpha \langle \rho_{\ell}\rangle $, ie. they satisfy an $L$-moment problem with $\langle\rho_\ell\rangle\le L=\frac{\langle \rho_{\ell_1}\rangle}{\alpha}$, which through eq.(\ref{abq}) will impose constraints back on $a_q$.  Since we are at equal $k$ space, the solution to this problem is given by the polytope conditions in Section \ref{sec2}. Therefore to obtain bounds on the $a_q$ couplings, we merely solve for the $\mu_q$ in terms of $a_q$
\eq
\mu_q=a_q{-}\langle \rho_{\ell_1} \rangle v_{\ell_1,q} \,,
\eqe
and impose the polytope constraints in eq.(\ref{gendimpol}) on the $\mu_q$
\eq\label{lsdcond}
P_I(\mu_q),\ \overline{P}_I(\mu_q)\ge 0\,,
\eqe
setting $L= \frac{\langle \rho_{\ell_1} \rangle}{\alpha}$. Unlike the previous problem, here the upper bound $L=\frac{\langle \rho_{\ell_1} \rangle}{\alpha}$ is not fixed beforehand, since $\langle \rho_{\ell_1} \rangle$ is not known. However, we can demand eq.(\ref{lsdcond}) to hold for any allowed values of $\langle \rho_{\ell_1} \rangle$, and this will still generate constraints.  When considering purely projective bounds, this means we simply project out $\langle \rho_{\ell_1} \rangle\ge 0$, leading to optimal constraints on $a_q$ purely in terms of $\alpha$. We can also implement the unitarity bound $\rho\le 2$, which implies $\langle \rho_{\ell} \rangle_k \le \frac{2}{k}$, by instead projecting out $0\le \langle \rho_{\ell_1} \rangle_k\le \frac{2}{k} $.

\bibliography{mybib2}{}

\providecommand{\href}[2]{#2}\begingroup\raggedright\begin{thebibliography}{10}

\bibitem{Rychkov:2016iqz}
S.~Rychkov, \emph{{EPFL Lectures on Conformal Field Theory in D\ensuremath{>}=
  3 Dimensions}}.
\newblock SpringerBriefs in Physics. 1, 2016.
\newblock 10.1007/978-3-319-43626-5.

\bibitem{Simmons-Duffin:2016gjk}
D.~Simmons-Duffin, \emph{{The Conformal Bootstrap}},  in \emph{{Theoretical
  Advanced Study Institute in Elementary Particle Physics}: {New Frontiers in
  Fields and Strings}}, pp.~1--74, 2017.
\newblock \href{http://arxiv.org/abs/1602.07982}{{\tt 1602.07982}}.
\newblock \href{http://dx.doi.org/10.1142/9789813149441_0001}{DOI}.

\bibitem{Poland:2018epd}
D.~Poland, S.~Rychkov and A.~Vichi, \emph{{The Conformal Bootstrap: Theory,
  Numerical Techniques, and Applications}},
  \href{http://dx.doi.org/10.1103/RevModPhys.91.015002}{\emph{Rev. Mod. Phys.}
  {\bf 91} (2019) 015002}, [\href{http://arxiv.org/abs/1805.04405}{{\tt
  1805.04405}}].

\bibitem{Poland:2022qrs}
D.~Poland and D.~Simmons-Duffin, \emph{{Snowmass White Paper: The Numerical
  Conformal Bootstrap}},  in \emph{{2022 Snowmass Summer Study}}, 3, 2022.
\newblock \href{http://arxiv.org/abs/2203.08117}{{\tt 2203.08117}}.

\bibitem{Caracciolo:2009bx}
F.~Caracciolo and V.~S. Rychkov, \emph{{Rigorous Limits on the Interaction
  Strength in Quantum Field Theory}},
  \href{http://dx.doi.org/10.1103/PhysRevD.81.085037}{\emph{Phys. Rev. D} {\bf
  81} (2010) 085037}, [\href{http://arxiv.org/abs/0912.2726}{{\tt 0912.2726}}].

\bibitem{Manohar:1983md}
A.~Manohar and H.~Georgi, \emph{{Chiral Quarks and the Nonrelativistic Quark
  Model}}, \href{http://dx.doi.org/10.1016/0550-3213(84)90231-1}{\emph{Nucl.
  Phys. B} {\bf 234} (1984) 189--212}.

\bibitem{Caron-Huot:2020cmc}
S.~Caron-Huot and V.~Van~Duong, \emph{{Extremal Effective Field Theories}},
  \href{http://dx.doi.org/10.1007/JHEP05(2021)280}{\emph{JHEP} {\bf 05} (2021)
  280}, [\href{http://arxiv.org/abs/2011.02957}{{\tt 2011.02957}}].

\bibitem{Adams:2006sv}
A.~Adams, N.~Arkani-Hamed, S.~Dubovsky, A.~Nicolis and R.~Rattazzi,
  \emph{{Causality, analyticity and an IR obstruction to UV completion}},
  \href{http://dx.doi.org/10.1088/1126-6708/2006/10/014}{\emph{JHEP} {\bf 10}
  (2006) 014}, [\href{http://arxiv.org/abs/hep-th/0602178}{{\tt
  hep-th/0602178}}].

\bibitem{Pham:1985cr}
T.~N. Pham and T.~N. Truong, \emph{{Evaluation of the Derivative Quartic Terms
  of the Meson Chiral Lagrangian From Forward Dispersion Relation}},
  \href{http://dx.doi.org/10.1103/PhysRevD.31.3027}{\emph{Phys. Rev. D} {\bf
  31} (1985) 3027}.

\bibitem{Ananthanarayan:1994hf}
B.~Ananthanarayan, D.~Toublan and G.~Wanders, \emph{{Consistency of the chiral
  pion pion scattering amplitudes with axiomatic constraints}},
  \href{http://dx.doi.org/10.1103/PhysRevD.51.1093}{\emph{Phys. Rev. D} {\bf
  51} (1995) 1093--1100}, [\href{http://arxiv.org/abs/hep-ph/9410302}{{\tt
  hep-ph/9410302}}].

\bibitem{Bellazzini:2014waa}
B.~Bellazzini, L.~Martucci and R.~Torre, \emph{{Symmetries, Sum Rules and
  Constraints on Effective Field Theories}},
  \href{http://dx.doi.org/10.1007/JHEP09(2014)100}{\emph{JHEP} {\bf 09} (2014)
  100}, [\href{http://arxiv.org/abs/1405.2960}{{\tt 1405.2960}}].

\bibitem{Bellazzini:2016xrt}
B.~Bellazzini, \emph{{Softness and amplitudes\textquoteright{} positivity for
  spinning particles}},
  \href{http://dx.doi.org/10.1007/JHEP02(2017)034}{\emph{JHEP} {\bf 02} (2017)
  034}, [\href{http://arxiv.org/abs/1605.06111}{{\tt 1605.06111}}].

\bibitem{deRham:2017avq}
C.~de~Rham, S.~Melville, A.~J. Tolley and S.-Y. Zhou, \emph{{Positivity bounds
  for scalar field theories}},
  \href{http://dx.doi.org/10.1103/PhysRevD.96.081702}{\emph{Phys. Rev. D} {\bf
  96} (2017) 081702}, [\href{http://arxiv.org/abs/1702.06134}{{\tt
  1702.06134}}].

\bibitem{deRham:2017zjm}
C.~de~Rham, S.~Melville, A.~J. Tolley and S.-Y. Zhou, \emph{{UV complete me:
  Positivity Bounds for Particles with Spin}},
  \href{http://dx.doi.org/10.1007/JHEP03(2018)011}{\emph{JHEP} {\bf 03} (2018)
  011}, [\href{http://arxiv.org/abs/1706.02712}{{\tt 1706.02712}}].

\bibitem{deRham:2018qqo}
C.~de~Rham, S.~Melville, A.~J. Tolley and S.-Y. Zhou, \emph{{Positivity Bounds
  for Massive Spin-1 and Spin-2 Fields}},
  \href{http://dx.doi.org/10.1007/JHEP03(2019)182}{\emph{JHEP} {\bf 03} (2019)
  182}, [\href{http://arxiv.org/abs/1804.10624}{{\tt 1804.10624}}].

\bibitem{Bellazzini:2019bzh}
B.~Bellazzini, F.~Riva, J.~Serra and F.~Sgarlata, \emph{{Massive Higher Spins:
  Effective Theory and Consistency}},
  \href{http://dx.doi.org/10.1007/JHEP10(2019)189}{\emph{JHEP} {\bf 10} (2019)
  189}, [\href{http://arxiv.org/abs/1903.08664}{{\tt 1903.08664}}].

\bibitem{Alberte:2019zhd}
L.~Alberte, C.~de~Rham, A.~Momeni, J.~Rumbutis and A.~J. Tolley,
  \emph{{Positivity Constraints on Interacting Pseudo-Linear Spin-2 Fields}},
  \href{http://dx.doi.org/10.1007/JHEP07(2020)121}{\emph{JHEP} {\bf 07} (2020)
  121}, [\href{http://arxiv.org/abs/1912.10018}{{\tt 1912.10018}}].

\bibitem{Bellazzini:2020cot}
B.~Bellazzini, J.~Elias~Mir\'o, R.~Rattazzi, M.~Riembau and F.~Riva,
  \emph{{Positive moments for scattering amplitudes}},
  \href{http://dx.doi.org/10.1103/PhysRevD.104.036006}{\emph{Phys. Rev. D} {\bf
  104} (2021) 036006}, [\href{http://arxiv.org/abs/2011.00037}{{\tt
  2011.00037}}].

\bibitem{Tolley:2020gtv}
A.~J. Tolley, Z.-Y. Wang and S.-Y. Zhou, \emph{{New positivity bounds from full
  crossing symmetry}},
  \href{http://dx.doi.org/10.1007/JHEP05(2021)255}{\emph{JHEP} {\bf 05} (2021)
  255}, [\href{http://arxiv.org/abs/2011.02400}{{\tt 2011.02400}}].

\bibitem{Arkani-Hamed:2020blm}
N.~Arkani-Hamed, T.-C. Huang and Y.-T. Huang, \emph{{The EFT-Hedron}},
  \href{http://dx.doi.org/10.1007/JHEP05(2021)259}{\emph{JHEP} {\bf 05} (2021)
  259}, [\href{http://arxiv.org/abs/2012.15849}{{\tt 2012.15849}}].

\bibitem{Green:2019tpt}
M.~B. Green and C.~Wen, \emph{{Superstring amplitudes, unitarily, and Hankel
  determinants of multiple zeta values}},
  \href{http://dx.doi.org/10.1007/JHEP11(2019)079}{\emph{JHEP} {\bf 11} (2019)
  079}, [\href{http://arxiv.org/abs/1908.08426}{{\tt 1908.08426}}].

\bibitem{Huang:2020nqy}
Y.-t. Huang, J.-Y. Liu, L.~Rodina and Y.~Wang, \emph{{Carving out the Space of
  Open-String S-matrix}},
  \href{http://dx.doi.org/10.1007/JHEP04(2021)195}{\emph{JHEP} {\bf 04} (2021)
  195}, [\href{http://arxiv.org/abs/2008.02293}{{\tt 2008.02293}}].

\bibitem{Sinha:2020win}
A.~Sinha and A.~Zahed, \emph{{Crossing Symmetric Dispersion Relations in
  Quantum Field Theories}},
  \href{http://dx.doi.org/10.1103/PhysRevLett.126.181601}{\emph{Phys. Rev.
  Lett.} {\bf 126} (2021) 181601}, [\href{http://arxiv.org/abs/2012.04877}{{\tt
  2012.04877}}].

\bibitem{Wang:2020xlt}
Z.-Y. Wang, C.~Zhang and S.-Y. Zhou, \emph{{Generalized elastic positivity
  bounds on interacting massive spin-2 theories}},
  \href{http://dx.doi.org/10.1007/JHEP04(2021)217}{\emph{JHEP} {\bf 04} (2021)
  217}, [\href{http://arxiv.org/abs/2011.05190}{{\tt 2011.05190}}].

\bibitem{Trott:2020ebl}
T.~Trott, \emph{{Causality, unitarity and symmetry in effective field theory}},
  \href{http://dx.doi.org/10.1007/JHEP07(2021)143}{\emph{JHEP} {\bf 07} (2021)
  143}, [\href{http://arxiv.org/abs/2011.10058}{{\tt 2011.10058}}].

\bibitem{Wang:2020jxr}
Y.-J. Wang, F.-K. Guo, C.~Zhang and S.-Y. Zhou, \emph{{Generalized positivity
  bounds on chiral perturbation theory}},
  \href{http://dx.doi.org/10.1007/JHEP07(2020)214}{\emph{JHEP} {\bf 07} (2020)
  214}, [\href{http://arxiv.org/abs/2004.03992}{{\tt 2004.03992}}].

\bibitem{Hebbar:2020ukp}
A.~Hebbar, D.~Karateev and J.~Penedones, \emph{{Spinning S-matrix Bootstrap in
  4d}},  \href{http://arxiv.org/abs/2011.11708}{{\tt 2011.11708}}.

\bibitem{EliasMiro:2021nul}
J.~Elias~Mir\'o and A.~Guerrieri, \emph{{Dual EFT bootstrap: QCD flux tubes}},
  \href{http://dx.doi.org/10.1007/JHEP10(2021)126}{\emph{JHEP} {\bf 10} (2021)
  126}, [\href{http://arxiv.org/abs/2106.07957}{{\tt 2106.07957}}].

\bibitem{Alberte:2021dnj}
L.~Alberte, C.~de~Rham, S.~Jaitly and A.~J. Tolley, \emph{{Reverse
  Bootstrapping: IR lessons for UV physics}},
  \href{http://arxiv.org/abs/2111.09226}{{\tt 2111.09226}}.

\bibitem{Chowdhury:2021ynh}
S.~D. Chowdhury, K.~Ghosh, P.~Haldar, P.~Raman and A.~Sinha, \emph{{Crossing
  Symmetric Spinning S-matrix Bootstrap: EFT bounds}},
  \href{http://arxiv.org/abs/2112.11755}{{\tt 2112.11755}}.

\bibitem{Raman:2021pkf}
P.~Raman and A.~Sinha, \emph{{QFT, EFT and GFT}},
  \href{http://dx.doi.org/10.1007/JHEP12(2021)203}{\emph{JHEP} {\bf 12} (2021)
  203}, [\href{http://arxiv.org/abs/2107.06559}{{\tt 2107.06559}}].

\bibitem{Haldar:2021rri}
P.~Haldar, A.~Sinha and A.~Zahed, \emph{{Quantum field theory and the
  Bieberbach conjecture}},
  \href{http://dx.doi.org/10.21468/SciPostPhys.11.1.002}{\emph{SciPost Phys.}
  {\bf 11} (2021) 002}, [\href{http://arxiv.org/abs/2103.12108}{{\tt
  2103.12108}}].

\bibitem{Zahed:2021fkp}
A.~Zahed, \emph{{Positivity and geometric function theory constraints on pion
  scattering}}, \href{http://dx.doi.org/10.1007/JHEP12(2021)036}{\emph{JHEP}
  {\bf 12} (2021) 036}, [\href{http://arxiv.org/abs/2108.10355}{{\tt
  2108.10355}}].

\bibitem{Du:2021byy}
Z.-Z. Du, C.~Zhang and S.-Y. Zhou, \emph{{Triple crossing positivity bounds for
  multi-field theories}},
  \href{http://dx.doi.org/10.1007/JHEP12(2021)115}{\emph{JHEP} {\bf 12} (2021)
  115}, [\href{http://arxiv.org/abs/2111.01169}{{\tt 2111.01169}}].

\bibitem{Li:2021lpe}
X.~Li, H.~Xu, C.~Yang, C.~Zhang and S.-Y. Zhou, \emph{{Positivity in Multifield
  Effective Field Theories}},
  \href{http://dx.doi.org/10.1103/PhysRevLett.127.121601}{\emph{Phys. Rev.
  Lett.} {\bf 127} (2021) 121601}, [\href{http://arxiv.org/abs/2101.01191}{{\tt
  2101.01191}}].

\bibitem{Zhang:2021eeo}
C.~Zhang, \emph{{SMEFTs living on the edge: determining the UV theories from
  positivity and extremality}},  \href{http://arxiv.org/abs/2112.11665}{{\tt
  2112.11665}}.

\bibitem{Albert:2022oes}
J.~Albert and L.~Rastelli, \emph{{Bootstrapping Pions at Large $N$}},
  \href{http://arxiv.org/abs/2203.11950}{{\tt 2203.11950}}.

\bibitem{Bern:2021ppb}
Z.~Bern, D.~Kosmopoulos and A.~Zhiboedov, \emph{{Gravitational effective field
  theory islands, low-spin dominance, and the four-graviton amplitude}},
  \href{http://dx.doi.org/10.1088/1751-8121/ac0e51}{\emph{J. Phys. A} {\bf 54}
  (2021) 344002}, [\href{http://arxiv.org/abs/2103.12728}{{\tt 2103.12728}}].

\bibitem{Henriksson:2021ymi}
J.~Henriksson, B.~McPeak, F.~Russo and A.~Vichi, \emph{{Rigorous Bounds on
  Light-by-Light Scattering}},  \href{http://arxiv.org/abs/2107.13009}{{\tt
  2107.13009}}.

\bibitem{Davighi:2021osh}
J.~Davighi, S.~Melville and T.~You, \emph{{Natural selection rules: new
  positivity bounds for massive spinning particles}},
  \href{http://dx.doi.org/10.1007/JHEP02(2022)167}{\emph{JHEP} {\bf 02} (2022)
  167}, [\href{http://arxiv.org/abs/2108.06334}{{\tt 2108.06334}}].

\bibitem{Melville:2022ykg}
S.~Melville and J.~Noller, \emph{{Positivity bounds from multiple vacua and
  their cosmological consequences}},
  \href{http://arxiv.org/abs/2202.01222}{{\tt 2202.01222}}.

\bibitem{Caron-Huot:2021rmr}
S.~Caron-Huot, D.~Mazac, L.~Rastelli and D.~Simmons-Duffin, \emph{{Sharp
  boundaries for the swampland}},
  \href{http://dx.doi.org/10.1007/JHEP07(2021)110}{\emph{JHEP} {\bf 07} (2021)
  110}, [\href{http://arxiv.org/abs/2102.08951}{{\tt 2102.08951}}].

\bibitem{Caron-Huot:2022ugt}
S.~Caron-Huot, Y.-Z. Li, J.~Parra-Martinez and D.~Simmons-Duffin,
  \emph{{Causality constraints on corrections to Einstein gravity}},
  \href{http://arxiv.org/abs/2201.06602}{{\tt 2201.06602}}.

\bibitem{Henriksson:2022oeu}
J.~Henriksson, B.~McPeak, F.~Russo and A.~Vichi, \emph{{Bounding Violations of
  the Weak Gravity Conjecture}},  \href{http://arxiv.org/abs/2203.08164}{{\tt
  2203.08164}}.

\bibitem{Haring:2022cyf}
K.~H\"aring and A.~Zhiboedov, \emph{{Gravitational Regge bounds}},
  \href{http://arxiv.org/abs/2202.08280}{{\tt 2202.08280}}.

\bibitem{Kruczenski:2022lot}
M.~Kruczenski, J.~Penedones and B.~C. van Rees, \emph{{Snowmass White Paper:
  S-matrix Bootstrap}},  \href{http://arxiv.org/abs/2203.02421}{{\tt
  2203.02421}}.

\bibitem{deRham:2022hpx}
C.~de~Rham, S.~Kundu, M.~Reece, A.~J. Tolley and S.-Y. Zhou, \emph{{Snowmass
  White Paper: UV Constraints on IR Physics}},  in \emph{{2022 Snowmass Summer
  Study}}, 3, 2022.
\newblock \href{http://arxiv.org/abs/2203.06805}{{\tt 2203.06805}}.

\bibitem{Correia:2020xtr}
M.~Correia, A.~Sever and A.~Zhiboedov, \emph{{An Analytical Toolkit for the
  S-matrix Bootstrap}},  \href{http://arxiv.org/abs/2006.08221}{{\tt
  2006.08221}}.

\bibitem{Bellazzini:2021oaj}
B.~Bellazzini, M.~Riembau and F.~Riva, \emph{{The IR-Side of Positivity
  Bounds}},  \href{http://arxiv.org/abs/2112.12561}{{\tt 2112.12561}}.

\bibitem{Simmons-Duffin:2015qma}
D.~Simmons-Duffin, \emph{{A Semidefinite Program Solver for the Conformal
  Bootstrap}}, \href{http://dx.doi.org/10.1007/JHEP06(2015)174}{\emph{JHEP}
  {\bf 06} (2015) 174}, [\href{http://arxiv.org/abs/1502.02033}{{\tt
  1502.02033}}].

\bibitem{Landry:2019qug}
W.~Landry and D.~Simmons-Duffin, \emph{{Scaling the semidefinite program solver
  SDPB}},  \href{http://arxiv.org/abs/1909.09745}{{\tt 1909.09745}}.

\bibitem{Chiang:2021ziz}
L.-Y. Chiang, Y.-t. Huang, W.~Li, L.~Rodina and H.-C. Weng, \emph{{Into the
  EFThedron and UV constraints from IR consistency}},
  \href{http://arxiv.org/abs/2105.02862}{{\tt 2105.02862}}.

\bibitem{Guerrieri:2021ivu}
A.~Guerrieri, J.~Penedones and P.~Vieira, \emph{{Where Is String Theory in the
  Space of Scattering Amplitudes?}},
  \href{http://dx.doi.org/10.1103/PhysRevLett.127.081601}{\emph{Phys. Rev.
  Lett.} {\bf 127} (2021) 081601}, [\href{http://arxiv.org/abs/2102.02847}{{\tt
  2102.02847}}].

\bibitem{Figueroa:2022onw}
F.~Figueroa and P.~Tourkine, \emph{{On the unitarity and low energy expansion
  of the Coon amplitude}},  \href{http://arxiv.org/abs/2201.12331}{{\tt
  2201.12331}}.

\bibitem{Chiang:2022jep}
L.-Y. Chiang, Y.-t. Huang, W.~Li, L.~Rodina and H.-C. Weng,
  \emph{{(Non)-projective bounds on gravitational EFT}},
  \href{http://arxiv.org/abs/2201.07177}{{\tt 2201.07177}}.

\bibitem{Herrmann:2022nkh}
E.~Herrmann and J.~Trnka, \emph{{The SAGEX Review on Scattering Amplitudes,
  Chapter 7: Positive Geometry of Scattering Amplitudes}},
  \href{http://arxiv.org/abs/2203.13018}{{\tt 2203.13018}}.

\bibitem{Hausdorff:1923uf}
F.~Hausdorff, \emph{Momentprobleme f{\"u}r ein endliches intervall.},
  \href{http://dx.doi.org/10.1007/BF01175684}{\emph{Mathematische Zeitschrift}
  {\bf 16} (1923) 220--248}.

\bibitem{Kren1977TheMM}
M.~G. Krein, D.~Louvish and A.~A. Nudelman, \emph{The markov moment problem and
  extremal problems},  1977.

\bibitem{PUTINAR1990288}
M.~Putinar, \emph{The l problem of moments in two dimensions},
  \href{http://dx.doi.org/https://doi.org/10.1016/0022-1236(90)90015-D}{\emph{Journal
  of Functional Analysis} {\bf 94} (1990) 288--307}.

\bibitem{article}
D.~Freedman and P.~Diaconis, \emph{The markov moment problem and de finetti?s
  theorem: Part i},
  \href{http://dx.doi.org/10.1007/s00209-003-0633-9}{\emph{Mathematische
  Zeitschrift} {\bf 247} (05, 2004) 183--199}.

\bibitem{akhiezer1934fouriersche}
N.~Akhiezer and M.~Krein, \emph{{\"U}ber fouriersche reihen beschr{\"a}nkter
  summierbarer funktionen und ein neues extremumproblem}, {\emph{Common. Soc.
  Math., Kharkov} {\bf 9} (1934) }.

\bibitem{akhiezer1962some}
N.~I. Akhiezer and W.~Fleming, \emph{Some questions in the theory of moments},
  vol.~2.
\newblock American Mathematical Soc., 1962.

\end{thebibliography}\endgroup
\bibliographystyle{JHEP}
\end{document}